\documentclass[]{jfm}
\usepackage{graphics}
\usepackage{natbib}
\ifCUPmtlplainloaded \else
  \checkfont{eurm10}
  \iffontfound
    \IfFileExists{upmath.sty}
      {\typeout{^^JFound AMS Euler Roman fonts on the system,
                   using the 'upmath' package.^^J}%
       \usepackage{upmath}}
      {\typeout{^^JFound AMS Euler Roman fonts on the system, but you
                   dont seem to have the}%
       \typeout{'upmath' package installed. JFM.cls can take advantage
                 of these fonts,^^Jif you use 'upmath' package.^^J}%
       \providecommand\upi{\pi}%
      }
  \else
    \providecommand\upi{\pi}%
  \fi
\fi

\ifCUPmtlplainloaded \else
  \checkfont{msam10}
  \iffontfound
    \IfFileExists{amssymb.sty}
      {\typeout{^^JFound AMS Symbol fonts on the system, using the
                'amssymb' package.^^J}%
       \usepackage{amssymb}%
       \let\le=\leqslant  
         
      }{}
  \fi
\fi

\ifCUPmtlplainloaded \else
  \IfFileExists{amsbsy.sty}
    {\typeout{^^JFound the 'amsbsy' package on the system, using it.^^J}%
     \usepackage{amsbsy}}
    {\providecommand\boldsymbol[1]{\mbox{\boldmath $##1$}}}
\fi

\providecommand\bnabla{\boldsymbol{\nabla}}

\newcommand\Rey{\mbox{\textit{Re}}}  

\newsavebox{\astrutbox}
\sbox{\astrutbox}{\rule[-5pt]{0pt}{20pt}}

\def\identity{\leavevmode\hbox{\small1\kern-3.2pt\normalsize1}}%

\title[Inertial effects in spinodal decomposition]
{Inertial effects in three dimensional spinodal decomposition of a
symmetric binary fluid mixture: A lattice Boltzmann study}

\author[V. M. Kendon, M. E. Cates, J-C. Desplat, I. Pagonabarraga and P. Bladon]
{V\ls I\ls V\ls I\ls E\ls N\ns M.\ns K\ls E\ls N\ls D\ls O\ls N$^{1,2}$\thanks{Present address: Optics Section, The Blackett
Laboratory, Imperial College, London, SW7 2BW, UK.}
M\ls I\ls C\ls H\ls A\ls E\ls L\ns E.\ns C\ls A\ls T\ls E\ls S$^1$
I\ls G\ls N\ls A\ls C\ls I\ls O\ns  P\ls A\ls G\ls O\ls N\ls A\ls B\ls A\ls R\ls R\ls A\ls G\ls A$^1$
J\ls-\ls C.\ns D\ls E\ls S\ls P\ls L\ls A\ls T$^3$
\and P\ls E\ls T\ls E\ls R\ns B\ls L\ls A\ls D\ls O\ls N$^1$}

\affiliation{$^1$Department of Physics and Astronomy, University of Edinburgh,
King's Buildings, Edinburgh EH9 3JZ, UK\\[\affilskip]
$^2$Department of Physics and Applied Physics, Strathclyde University, Glasgow G4 ONG, UK\\[\affilskip]
$^3$Edinburgh Parallel Computing Centre, JCMB, King's Buildings,
Edinburgh, EH9 3JZ, UK}

\volume{??}
\pagerange{1--??}
\date{?? and in revised form ??}
\begin{document}

\maketitle

\begin{abstract}
The late-stage demixing following spinodal decomposition of a three-dimensional symmetric binary fluid mixture is studied numerically, using a thermodynamicaly consistent lattice Boltzmann method.
We combine results from simulations with different numerical parameters to obtain a unprecendented range of length and time scales when expressed in reduced physical units.
(These are the length and time units derived from fluid density, viscosity, and interfacial tension.) Using eight large ($256^3$) runs, the resulting composite graph of reduced domain size $l$ against reduced time $t$ covers  $1\lesssim l \lesssim 10^5$, $10 \lesssim t \lesssim 10^8$.
Our data is consistent with the dynamical scaling hypothesis, that $l(t)$ is a universal scaling curve.
We give the first detailed statistical analysis of fluid motion, rather than just domain evolution, in simulations of this kind, and introduce scaling plots for several quantities derived from the fluid velocity and velocity gradient fields.
Using the conventional definition of Reynolds number for this problem, $\Rey_{\phi} = l\mathrm{d}l/\mathrm{d}t$, we attain values approaching $350$.
At $\Rey_{\phi} \gtrsim 100$ (which requires $t \gtrsim 10^6$) we find clear evidence of Furukawa's inertial scaling ($l\sim t^{2/3}$), although the crossover from the viscous regime ($l\sim t$) is both broad and late ($10^2 \lesssim t \lesssim 10^6$).
Though it cannot be ruled out, we find no indication that $\Rey_{\phi}$ is self-limiting ($l\sim t^{1/2}$) at late times, as recently proposed by Grant and Elder.
Detailed study of the velocity fields confirm that, for our most inertial runs, the {\sc rms} ratio of nonlinear to viscous terms in the Navier Stokes equation, $R_2$, is of order ten, with the fluid mixture showing incipient turbulent characteristics.
However, we cannot go far enough into the inertial regime to obtain a clear length separation of domain size, Taylor microscale, and Kolmogorov scale, as would be needed to test a recent `extended' scaling theory of Kendon (in which $R_2$ is self-limiting but $\Rey_{\phi}$ not).
To obtain our results has required careful steering of several numerical control parameters so as to maintain adequate algorithmic stability, efficiency and isotropy, while eliminating unwanted residual diffusion.
(We argue that the latter affects some studies in the literature which report $l\sim t^{2/3}$ for $t \lesssim 10^4$.) We analyse the various sources of error and find them just within acceptable levels (a few percent each) in most of our datasets.
To bring these under significantly better control, or to go much further into the inertial regime, would require much larger computational resources and/or a breakthrough in algorithm design.
\end{abstract}

\section{Introduction}
\label{sec:intro}

Spinodal decomposition occurs when a fluid mixture of two species $A$ and $B$, forming a single homogeneous phase at high temperature $\mathcal{T}$, undergoes spontaneous demixing following a sudden drop in temperature (or `quench'). For suitable compositions and quenches, one enters the `spinodal' regime in which the initial homogeneous phase is locally unstable to small fluctuations. (Elsewhere one finds instead a nucleation and growth mechanism which is not the subject of this paper.)  For compositions close to 50/50, there then arises, after an early period of interdiffusion, a bicontinuous domain structure in which patches of $A$-rich and $B$-rich fluid are separated by sharply defined interfaces. The sharpness depends on the temperature drop; we assume a `deep quench' for which the interfacial thickness is, in practice, on a molecular rather than macroscopic scale. In this late-stage structure, the local compositions of the fluid patches correspond to those of the two bulk phases in coexistence; the interfacial tension approaches $\sigma$, its equilibrium value. Although locally close to equilibrium everywhere, the structure then continues to evolve so as to reduce its interfacial area. Local interfacial curvature causes stresses (equivalently, Laplace pressures) to arise, which drive fluid motion. The interface then evolves smoothly with time between isolated `pinchoff events' or topological reconnections. In principle these events reintroduce molecular physics at the short scale; however it is generally assumed that pinchoff, once initiated, occurs rapidly enough not to impede the coarsening process \citep[but see][]{jury98a,brenner97a}. Likewise it is usually assumed that at late times, the presence of thermal noise in the system is irrelevant, at least for deep quenches \citep[but see][]{gonnella99a}: the problem is thus one of deterministic, isothermal fluid motion coupled to a moving interface. Precise details of the random initial condition, which is inherited from the earlier diffusive stage, are also thought to be unimportant (assuming that no long-range correlations are initially present). 

For simplicity we address in this paper only the maximally symmetric case of two incompressible fluids with identical physical properties (shear viscosity $\eta$, density $\rho$), and also equal volume fractions, that have undergone a deep quench. With the assumptions made above, all such fluid mixtures should, in the late stages, behave in a similar manner. More precisely, 
the \textit{dynamical scaling hypothesis} is that, 
if one defines units of length and of time by
\begin{equation} L_0 \equiv \eta^2/(\rho\sigma) \;\;\;\;\;,\;\;\;\;\;T_0 \equiv \eta^3/(\rho\sigma^2) \label{defs}
\end{equation}
(which are the only such units derivable from $\eta,\rho,\sigma$)
then at late times any characteristic structural length $L(T)$ should evolve with time $T$ according to 
\begin{equation}
dL/dT = (L_0/T_0)\,\varphi(L/L_0)\label{velscaling}
\end{equation} 
where $\varphi(x)$ is the same function for all such fluids. (A specific choice of definition for $L$ is made later on, in terms of the mean domain size.) Integrating this once gives
a universal late-stage scaling
\begin{equation}
l = l(t) \label{lscaling}
\end{equation}
where we introduce `reduced physical units', 
\begin{equation}
l \equiv L/L_0\;\;\;\;\;\;,\;\;\;\;\;\; t \equiv (T- T_{int})/T_0.
\end{equation}  
Here $T_{int}$ an offset that is nonuniversal: it depends on the initial condition as fixed by the early stage diffusion processes.  
(Note that in this paper, the symbol $t$ is reserved for the reduced physical time; unscaled time is denoted $T$, and temperature $\cal T$. An overdot means time derivative in whatever units are being used.)

The form of $l(t)$ has been discussed by several authors, notably \cite{siggia79a} and \cite{furukawa85a}. Siggia argued that, for $t \ll 1$, the interfacial forces induce a creeping flow of the fluid; simple force balance in the Navier--Stokes equation then gives $l\propto t$ in this, the `viscous hydrodynamic' regime. Note that, in a creeping flow, the fluid velocity depends only on the instantaneous structure of the interface.
(This is why the nonuniversal offset is in $T$ and not $L$.) At later times, the force balance was argued by Furukawa to entail viscous and inertial effects; balancing these gives $l\propto t^{2/3}$ for $t \gg 1$, the `inertial hydrodynamic' regime. It has recently been shown by \cite{kendon99c} that Furukawa's assumption of a {\em single} characteristic length (for velocity gradients as well as interfacial structure) is inconsistent with energy conservation; her more detailed analysis nonetheless recovers $l\propto t^{2/3}$ for the domain size. Kendon's arguments, with those of Siggia and Furukawa, are discussed in \S\ \ref{sec:scaling1}, \ref{sec:scaling2}. 

For a general review of late-stage spinodal decomposition and other aspects of phase separation kinetics, see that of \cite{bray94a}. The problem is clearly intractable analytically: it involves a moving boundary with a complicated and non-constant topology whose initial condition is defined, implicitly, by the preceding, early-time diffusion. These features render it equally intractable to many numerical algorithms that might perform well for other fluid mechanics problems. Indeed, symmetrical spinodal decomposition has become a benchmark for various so-called `mesoscale' simulation techniques, developed to address the statistical dynamics of fluids with microstructure. The results from different techniques can be compared, not only with each other and with experiment (with the caveat that one cannot realize exact symmetry between fluids in the laboratory), but with the predictions of the various scaling theories already mentioned.

In the present work, we study in detail the physics of spinodal decomposition for a symmetrical binary fluid using the Lattice Boltzmann (LB) technique \citep{higuera89a}, in a thermodynamically consistent form pioneered by the group of Yeomans \citep[see][]{swift96a}. 
Our work, of which a preliminary report appeared in \cite{kendon99a}, advances significantly the state of the art for simulations of spinodal decomposition, and for LB simulations of fluid mixtures. 
In any such simulation, a
balance must be struck between discretization error at small scales, and finite size errors (arising in our case from periodic boundary conditions) at large ones; this compromise is quite subtle, as we discuss below. It means that any individual simulation run can produce only around one decade of data for the $l(t)$ curve. This is true for \textit{all} first-principles simulation methods: in three dimensions there cannot be more than two decades, or at most three, separating the discretization length from the system size, before deduction of a half-decade safety margin at each end. (Three decades before such deduction is optimistic; it means simulating at least $10^9$ degrees of freedom.)

Despite this restriction, by careful scaling and combination of separate datasets for eight large ($256^3$) simulation runs, we are able to access an unprecedented range of $l$ and $t$ (five and seven decades respectively) including regions of the $l(t)$ curve not studied previously. We find nothing to contradict the universality of Equation (\ref{lscaling}), but nor can we completely rule out
violations of it.
We gain the first unambiguous evidence for a regime in which inertial effects dominate over viscous ones, and find clear evidence for $t^{2/3}$ scaling in this regime. In this and other regions of the $l(t)$ curve, we study the statistics, not only of the interfacial structure, but also of the fluid velocity. (The latter was not addressed in detail by previous simulations.) On entering the inertial hydrodynamic regime we find some evidence for breakdown of simple scaling of velocity gradients, as proposed by \cite{kendon99c}, but our data does not extend far enough into this regime to offer a meaningful test of her alternative proposals.

To obtain our new results, we have had to push the LB technique to its limits. For statistics based on the domain size, errors at the level of several percent, arising from each of several different sources (residual diffusion, lattice anisotropy etc.) remain. We do make a systematic attempt to identify and minimize the various sources of errors -- a somewhat arduous task that, our work suggests, has been neglected in several previous studies.
The errors for some of our velocity statistics (especially those for spatial derivatives of the velocity) are much larger. Nonetheless we present the data, such as it is, because it highlights several issues both in the physics of spinodal decomposition and in how simulation results should be obtained, analysed and interpreted.

The rest of this paper is organized as follows. \S\ \ref{sec:thermo}
outlines the thermodynamics of the binary fluid system, and \S\ \ref{sec:geq}
its governing equations. \S\ \ref{sec:scaling1} and \S\ \ref{sec:scaling2}
outline the simple and extended scaling analyses referred to above. \S\ \ref{sec:method} describes the LB method in the form that we use; \S\ \ref{sec:pars} describes how the simulation parameters are chosen. \S\ \ref{sec:validation} outlines a number of validation tests. \S\ \ref{sec:spin}
gives our results for the evolution of the interfacial structure, \S\ \ref{sec:velocity} those for the velocity field and \S\ \ref{sec:velocity_derivatives} those for the velocity derivatives and related quantities. \S\ \ref{sec:conc} summarizes our conclusions. Two appendices
give further information on the effects of residual fluid compressibility in the LB method and on the relation between our work and that of previous authors.

\section{Thermodynamics}
\label{sec:thermo}
Although we are interested in the late-stage demixing of two isothermal, incompressible fluids separated by sharp interfaces, the LB method resorts to a more fundamental approach, in which these interfaces are described as excitations of a thermodynamic field theory.
The central object is the Helmholtz free energy
\begin{equation}
F = E - \mathcal{TS},
\end{equation}
where $E$ is the internal energy, $\mathcal{T}$ the temperature and $\mathcal{S}$
the entropy of the system. 

\begin{figure}
\begin{minipage}{\textwidth}
    \begin{center}
    \begin{minipage}{0.44\textwidth}
        \raggedleft
        \vspace{1ex}
        \resizebox{\textwidth}{!}{\rotatebox{-90}{\includegraphics{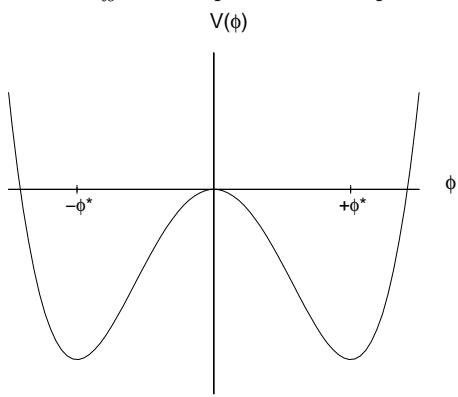}}}
    \end{minipage}
    \end{center}
    \caption{Model potential for phase separation, a symmetric
             double well, $\mathcal{V}(\phi)$.
             The equilibrium values of the order parameter are
             $\pm\phi^*$.}
    \protect\label{fig:m_hat}
\end{minipage}
\end{figure}

In a system at fixed volume $V$, and fixed contents and temperature,
equilibrium states are given by global minima of the free energy, $F$.
For a symmetric fluid mixture, $F$ is a functional of a single composition variable $\phi({\bf r})$, defined as $\phi = (n_A - n_B)/(n_A+n_B)$ where the $n$'s are number densities, and of the mean fluid density $\rho = n_A+n_B$.
(We take unit mass for $A$ and $B$ particles without loss of generality.) In the incompressible case, $\rho$ is fixed; we leave it as a parameter in what follows. 
Further restricting attention to homogenous states (so that $\phi$ is the same everywhere), we can write 
\begin{equation}
F/V = \mathcal{V}(\phi). \label{polynomial}
\end{equation}
Within mean-field theories of fluid demixing, one predicts that $\mathcal{V}$ has everywhere positive curvature at high temperatures, but becomes concave below a critical temperature $\mathcal{T}_c$. The resulting curve is as shown in Figure \ref{fig:m_hat}, with symmetric minima at $\pm \phi^*$.
Below $\mathcal{T}_c$, the free energy is therefore minimized by creating two bulk domains (of equal volume) at compositions $\pm \phi^*$ instead of a single homogenous phase with $\phi = 0$ (which is our presumed initial condition). The same phase separation occurs for any other $\phi$ between $\pm \phi^*$, but in this case the domain volumes are unequal; for sufficient asymmetry this causes depercolation. (In a depercolated, droplet structure, coarsening can only occur by diffusion so that the scaling arguments given above cease to apply. We do not address this here.)

\begin{figure}
\begin{center}
    \begin{minipage}{0.54\textwidth}
        \raggedleft
        \resizebox{\textwidth}{!}{\includegraphics{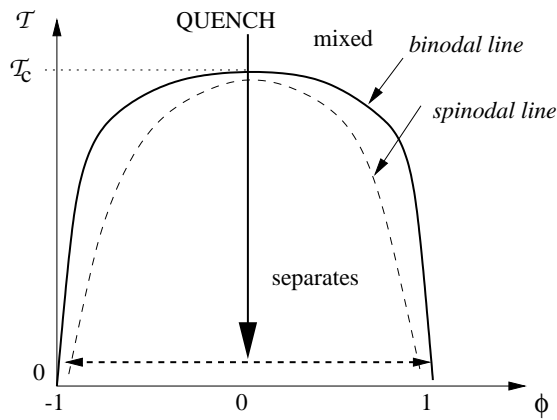}}
    \end{minipage}
    \caption {Phase diagram for spinodal decomposition.
             The order parameter is $\phi=(n_A-n_B)/(n_A+n_B)$ with
             $\rho=n_A+n_B$ the mean fluid density.
             The temperature
             axis shows the critical temperature, $\mathcal{T}_c$, 
             below which the system starts to separate, and above 
             which it remains completely mixed.}
    \protect\label{fig:phase}
\end{center}
\end{figure}

The resulting phase diagram is shown in Figure \ref{fig:phase}. Spinodal decomposition occurs 
for any quench that leaves the system beneath the spinodal line, on which $\mathrm{d}^2\mathcal{V}/\mathrm{d}\phi^2$ changes sign. Immediately after such a quench, the system is locally unstable: the free energy can be lowered, in any local neighbourhood, by creating two domains whose composition differs only infinitesimally from the initial one. (The resulting free energy density lies on a line connecting two points on $\mathcal{V}(\phi)$ at the new compositions; in the convex region, this causes a reduction in $F$.) Accordingly, infinitesimal fluctuations will grow by diffusion until there is local coexistence of domains at compositions approaching $\pm \phi^*$.

To describe quantitatively both the domains and the interfaces between them, one must specify not just $\mathcal{V}(\phi)$ but the free energy functional, $F[\phi]$. An acceptable choice is
the square gradient model \citep[see][]{bray94a}
\begin{equation}
F[\phi] = \int\!\mathrm{d}\mathbf{r}\,\left\{\mathcal{V}(\phi) +
\frac{\kappa}{2}|\bnabla\phi|^2 \right\},
\label{eq:free_gen}
\end{equation}
where $\mathcal{V}(\phi)$ is as shown
in Figure \ref{fig:m_hat}, and the term in $\kappa$ penalizes sharp gradients in composition. This ensures smooth local deviations from $\pm \phi^*$ near the interface, and provides a nonzero interfacial tension $\sigma$ which can be calculated as follows. We consider a
flat interface between two domains, introducing a coordinate normal to it, $g$. Stationarity of $F$ requires
\begin{equation}
\kappa\nabla^2\phi = \kappa\frac{\partial^2\phi}{\partial g^2}
= \frac{\mathrm{d}\mathcal{V}(\phi)}{\mathrm{d}\phi}.
\end{equation}
Integrating this once across the interface and setting $g=0$, $\phi=0$
at its centre gives
\begin{equation}
\frac{\kappa}{2}\left(\frac{\partial\phi}{\partial g}\right)^2 
= \mathcal{V}(\phi) - \mathcal{V}(\phi^*).
\label{eq:dg_dphi}
\end{equation}
The excess free energy
per unit area of interface is then given by
\begin{equation}
\sigma = \int\! dg\,\left[\frac{\kappa}{2}\left(\frac{\partial\phi}{\partial g}\right)^2 + \mathcal{V}(\phi) -  \mathcal{V}(\phi^*)\right].
\label{eq:sigma_def}
\end{equation}
By exploiting the fact that 
$\bnabla\phi\rightarrow 0$ in the bulk fluid, and
using Equation (\ref{eq:dg_dphi}), we obtain
\begin{equation}
\sigma = \int_{-\phi^*}^{+\phi^*}\,\mathrm{d}\phi\,\left(2\kappa\right)^{1/2}
         \left[\mathcal{V}(\phi) - \mathcal{V}(\phi^*) \right]^{1/2}.
\label{eq:sigma}
\end{equation}
Given a form for the potential, $\mathcal{V}(\phi)$, a value
for the interfacial tension can thus be calculated.  This is done
for the model used in our simulations in \S\ \ref{sec:method}.

We now turn to the (exchange) chemical potential, $\mu$, which describes the change in $F$ for a small local change in composition:
\begin{equation}
\mu \equiv \frac{\delta F}{\delta\phi} = \frac{\mathrm{d}\mathcal{V}}{\mathrm{d}\phi} - \kappa \nabla^2\phi.
\label{chempot}
\end{equation}
Within the LB approach, the coupling between interfacial and fluid motion arises as follows.
In the presence of a nonuniform composition, there is a thermodynamic force density $-\phi\bnabla\mu$ acting at each point on the fluid. (The two species are pulled in opposite directions by the chemical potential gradient; the net force vanishes only if $\phi = 0$.) This force density can also be 
written as the divergence of a `chemical' pressure tensor: 
\begin{equation}
\phi\bnabla\mu = \bnabla. \mbox{\boldmath{$\cal P$}}^{chem}
\label{divstress}
\end{equation}
where it is a straightforward exercise to confirm that
\begin{equation}
\mathcal{P}^{chem}_{\alpha\beta} = \delta_{\alpha\beta}\left[\phi\frac{\mathrm{d}\mathcal{V}}{\mathrm{d}\phi} - \mathcal{V} - \kappa\{\phi\nabla^2\phi + \mbox{$\frac{1}{2}$}|\bnabla\phi|^2\}\right]
+\kappa (\nabla_\alpha\phi)(\nabla_\beta\phi). \label{stress}
\end{equation}
Note that only the last term is anisotropic; the rest contributes, in effect, to the isotropic fluid pressure $P$. 
By integrating  (\ref{divstress}) across an interface, and using Eq. (\ref{stress}),
one finds that there is, in static equilibrium, a finite pressure difference across a curved interface, called the Laplace pressure:
\begin{equation}
\Delta P = \sigma \mathcal{K} \label{eq:laplace}
\end{equation}
where $\mathcal{K}$ is the interfacial curvature.

Throughout the above, our description in terms of a smooth composition variable $\phi({\bf r})$, usually known as the order parameter, assumes a
coarse-graining so that the smallest length scale under consideration
is larger than the average distance between molecules. 
In equilibrium, this coarse-graining is an almost trivial operation, but
for the dynamical description desired below, certain conditions must be met.  
On scales smaller than the coarse-graining length, the system must remain in local equilibrium, while the variations of interest
at larger scales must be slow on
the scale of the time it takes for that local equilibrium to be reached.

This does not mean that the microscopic scales can be forgotten from here
on.  Although usually a macroscopic description is sufficient
to fully understand the system, ultimately it is still the microscopic
interactions that are driving the system and determining the dynamics.
In particular, any interface between the two fluids will have a microscopic
size and structure.
It is always a possibility that the microscopic behaviour can intrude
at the macroscopic level (for example, by interfering with pinchoff) and change the results predicted by any simple
macroscopic considerations.  In particular, when using 
numerical models, care must be taken that the microscopic behaviour in
these models is admissible.

\section{Governing Equations}
\label{sec:geq}

The equation of motion for $\phi$ is taken to be a convective diffusion equation of Cahn Hilliard type \citep[see][]{bray94a,swift96a} 
\begin{equation}
\dot{\phi} + \mathbf{v}.\bnabla\phi 
= M\nabla^2\mu
= -M\nabla^2\left\{\kappa\nabla^2\phi - \frac{\partial \mathcal{V}(\phi)}{\partial\phi}\right\},
\label{eq:phi}
\end{equation}
where $M$ is an order-parameter mobility (here assumed independent of $\phi$)
that controls the strength of the diffusion, and
$\mathbf{v}(\mathbf{r})$ is the fluid velocity. 
This equation states that the order parameter responds to composition gradients by diffusion (the $M\nabla^2\mu$ term), and also changes with time because it is advected by the fluid
flow (the $\mathbf{v}.\bnabla\phi$ term).

The fluid velocity in turn obeys the Navier--Stokes equation
(NSE), which for an incompressible fluid reads
\begin{equation}
\rho\left[\dot{\mathbf{v}} + (\mathbf{v}.\bnabla)\mathbf{v}
\right] = \eta\nabla^2\mathbf{v} -\bnabla.\mbox{\boldmath{$\cal P$}}^{th}.
\label{eq:nse}
\end{equation}
Here $\mathcal{P}^{th}_{\alpha\beta}$ 
is the `thermodynamic' (or conservative) part of the pressure tensor, and
contains two pieces: an isotropic contribution $P\delta_{\alpha\beta}$, chosen to maintain constant $\rho$, and the `chemical' pressure tensor, $\mathcal{P}^{chem}_{\alpha\beta}$, defined previously in  (\ref{stress}). 
Recall that by  (\ref{divstress}), the chemical term $-\bnabla.\mbox{\boldmath{$\cal P$}}^{chem}$ can equally well be represented as a body force density $-\phi\bnabla\mu$ acting on the fluid, so that Eq. (\ref{eq:nse})
also reads
\begin{equation}
\rho\left[\dot{\mathbf{v}} + (\mathbf{v}.\bnabla)\mathbf{v}
\right] = \eta\nabla^2\mathbf{v} - \phi \bnabla \mu -\bnabla P\,.
\label{eq:nse2}
\end{equation}

Within the LB approach of \cite{swift96a}, the governing equations are solved by relaxing slightly the requirement of fluid incompressibility. We return to this in \S\ \ref{sec:method}.

\section{Simple Scaling Analyses}
\label{sec:scaling1}

The pair of coupled nonlinear differential equations, (\ref{eq:phi}) and (\ref{eq:nse}),
are intractable,
but various dimensional and scaling ideas may be used to find out how fast the domains grow
once the diffusive period is over. All these analyses assume that the interface can be characterized by a single length scale --- that is, it is basically smooth, with radii of curvature that scale as the domain size itself, which is much larger than the interfacial thickness.

Many domain-scale length measures are possible; we use $L(T)$, the inverse of the first moment of the
spherically averaged structure factor, $S(k,T)$:
\begin{equation}
\label{eq:Ldef}
L(T) = 2\upi\frac{\int S(k,T)\,\mathrm{d}k}{\int k S(k,T)\,\mathrm{d}k},
\end{equation}
where $k = |{\bf k}|$ is the modulus of the wave vector in Fourier space, and
\begin{equation}
S(k,T) \equiv 
\langle\phi(\mathbf{k},T)\phi(\mathbf{-k},T)\rangle
\label{eq:sk_def}
\end{equation}
with $\phi(\mathbf{k},T)$ the spatial Fourier transform of the order parameter. The angle brackets denote an average over a shell in $\bf k$ space
at fixed $k$. 

The aim of scaling analyses is to find the form of the time dependence of $L(T)$ by
considering the NSE (\ref{eq:nse2}), and balancing the force from the interface,$-\phi\bnabla\mu$,
against the viscous and inertial terms which
tend to oppose its motion.
The interfacial force density, 
$-\phi\bnabla\mu$, can be approximated as follows.  
The curvature, 
$\mathcal{K}$, is of order $1/L$, since $L(T)$ is roughly the 
size of the domains. This sets the scale of $\mbox{\boldmath{$\cal P$}}^{chem}$
through (\ref{eq:laplace}), as $\sigma/L$.
Likewise the gradient operator, $\bnabla$, can be approximated by $1/L(T)$ in Equation (\ref{stress}), which then reads
\begin{equation}
-\phi\bnabla\mu \simeq \frac{\sigma}{L^2}.
\label{eq:gradmuest}
\end{equation}

Now we turn to the remaining terms in the NSE (\ref{eq:nse2}).
We start by assuming that the length $L$ also controls the magnitude of $\bnabla$ as far as \textit{velocity gradients} are concerned. Approximating also the fluid velocity, $\mathbf{v}$,
by the velocity of the interface
$\dot{L}(T)$, gives for the viscous and inertial terms respectively
\begin{eqnarray}
\eta\nabla^2\mathbf{v} &\simeq & \eta\frac{\dot{L}}{L^2},\\
\rho\left[\dot{\mathbf{v}} + (\mathbf{v}.\bnabla)\mathbf{v}
\right] &\simeq & \rho\ddot{L} + \rho\frac{\dot{L}^2}{L}.
\end{eqnarray}

Under conditions in which the inertial terms are negligible, the force
from the interface will be balanced by the viscous force,
giving ${\dot{L}}/{L^2} \simeq {\sigma}/({\eta}{L^2})$.
Integrating this gives,
\begin{equation}
L \simeq \frac{\sigma}{\eta}\left(T-T_{int}\right).
\label{eq:linear1}
\end{equation}
Thus the domain size is predicted to grow linearly with time
in the region where the fluid flow is viscous dominated.
This is the result of \cite{siggia79a}.
Linear growth has been reported in experiments by, for example,
\cite{kubota92a,chen93c,hashimoto94a}, and in simulations incorporating
hydrodynamics by \cite{koga91a,puri92a,alexander93a,laradji96a,bastea97a}
and \cite{jury98a}.

To find the growth rate in the inertial region, \cite{furukawa85a} balanced instead the inertial and interfacial terms; assuming again only one relevant length, he obtained
\begin{equation}
\ddot{L} \simeq \frac{\dot{L}^2}{L} \simeq \frac{\sigma}{\rho}\frac{1}{L^2}.
\label{eq:inertial}
\end{equation}
Integrating this twice gives, for large enough $T$, 
$L^3 \simeq \sigma T^2/\rho$,
so that the domain size grows as $L\sim T^{2/3}$.
This result has not yet 
been observed experimentally (for reasons we discuss later, \S\ \ref{sec:conc}). There are a few previous
claims to see this in simulation \citep{ma92a,appert95a,lookman96a},
but none reliably establish dominance of inertial over viscous forces as we do below in \S\ \ref{sec:velocity}. 

Comparing the results of  (\ref{eq:linear1}) and (\ref{eq:inertial}) allows us to estimate a characteristic domain size, $L = L^*$, and characteristic time, $T = T^*+T_{int}$
at which the crossover from viscous to inertial scaling occurs. (To be precise, we can define $L^*,T^*$ by the interception of asymptotes on a log-log plot.) This leads to 
$L^* \simeq L_0,
T^* \simeq T_0$,
with $L_0,T_0$ defined in  \ref{defs}.
Converting to reduced physical units $l =L/L_0$, $t = (T-T_{int})/T_0$ as defined previously, and invoking the dynamical scaling hypothesis, we have respectively
\begin{eqnarray}
l &=& b_1t  \,\;\;\;\;\;\;\;\;\;\;\;\;\;\; t \ll t^*     \mbox{\hspace{0.5cm}viscous regime} \nonumber \\
l &=& b_{2/3}t^{2/3} \;\;\;\;\;\;\; t \gg t^* \mbox{\hspace{0.5cm}inertial regime},
\label{eq:l_t}
\end{eqnarray}
where $b_1,b_{2/3}$ and $t^*$ are dimensionless numbers that should be universal to all incompressible, fully symmetric, deep quenched fluid mixtures. What scaling theories cannot predict, of course, are values of the universal constants $b_1,b_{2/3},t^*$, other than to state that these are `of order unity'.

In fact our simulations show that $t^* \equiv T^*/T_0$ is between $10^4$ and $10^5$, which is `of order unity' only in a rather unhelpful sense; the implications of this are discussed in \S\ \ref{sec:conc} below. We will also find that the crossover region, between the asymptotes described by  (\ref{eq:l_t}), is several decades wide. Although
there is no explicit scaling prediction for the behaviour within this crossover region, its width means that each individual simulation dataset, either within or outside the crossover, can be well-described by a single scaling
exponent, $\alpha$, such that
\begin{equation}
l \simeq b_\alpha t^{\alpha},
\label{alphafit}
\end{equation}
where $b_1 \le b_\alpha \le b_{2/3}$, and $1\le\alpha\le 2/3$.
We use this form below, when analysing our numerical data.

\section{Extended Scaling Analysis}
\label{sec:scaling2}

In what follows we will find it useful to compare directly the relative magnitude of the 
various terms in the NSE.
Two ratios have therefore been defined,
the {\sc rms} ratio $R_1$ between the acceleration term and the viscous term,
\begin{equation}
R_1^2 \equiv \frac{\langle |\rho\dot\mathbf{v}|^2\rangle}
           {\langle |\eta\nabla^2\mathbf{v}|^2\rangle},
\label{eq:R_1t}
\end{equation}
and the {\sc rms} ratio $R_2$ between the nonlinear term and the viscous term,
\begin{equation}
R_2^2 \equiv \frac{\langle |\rho(\mathbf{v}.\bnabla)\mathbf{v}|^2\rangle}
           {\langle |\eta\nabla^2\mathbf{v}|^2\rangle}.
\label{eq:R_2t}
\end{equation}
Here $\langle\rangle$ denote spatial averages (though an ensemble average might be preferable under some conditions). These ratios obey $R_1,R_2\gg 1$ where the inertial terms dominate
and $R_1,R_2\ll 1$ where the viscous term dominates. The ratio $R_2$ we identify as the `true' Reynolds number, that is, a dimensionless measure of the relative importance of nonlinearity in the NSE.

When $R_2$,  (\ref{eq:R_2t}), is
simplified using the normal scaling assumption ($\bnabla\sim 1/L$), the result is the following 
estimate:
\begin{equation}
R_2\simeq \frac{\rho v^2/L}{\eta v/L^2} \simeq \frac{vL}{\eta/\rho}.
\label{eq:rn}
\end{equation}
Assuming that the characteristic velocity scale is $v\simeq \dot{L}$, one
finds \citep[following][]{furukawa85a,grant99a} the Reynolds number estimate
$\Rey_{\phi}$, 
\begin{equation}
R_2\simeq \Rey_{\phi} \equiv \frac{\rho}{\eta} L(T)\dot{L}
= l\dot{l} = \alpha b^2_\alpha t^{2\alpha-1}.
\label{eq:Re_phi}
\end{equation}
This `order-parameter Reynolds number' has the advantage, in simulations, that it is computable from $L(T)$ without direct access to any fluid velocity statistics.
However, $\Rey_{\phi}$ is only a good estimate of the true Reynolds number $R_2$, if the simple scaling for velocity gradients of $\bnabla \rightarrow 1/L$ does indeed hold. This assumption leads to the following paradox, as noted by \cite{grant99a}. According to  (\ref{eq:Re_phi}),
if in the inertial region $\alpha=2/3$, then
$\Rey_{\phi} \sim t^{2\alpha - 1}$ which becomes indefinitely large
as time proceeds.
Grant and Elder argued that this could not be physical, on the grounds that an infinite Reynolds number would imply turbulent remixing of domains: this would limit the domain growth to such a speed that $\Rey_{\phi}$ remained bounded at late times.  (\ref{eq:Re_phi}) then demands a lower
asymptotic growth exponent, $\alpha\le \frac{1}{2}$, as $t\to\infty$. 

However, a closer look at the scaling of all the terms in the NSE admits
an alternative resolution. \cite{kendon99c} pointed out that a minimally acceptable scaling theory
should allow, not only force balance in NSE, but a balance of terms in the global energy equation, which for an isothermal, incompressible fluid reads
\begin{equation}
\frac{\mathrm{d}\langle\rho v^2\rangle}{\mathrm{d}t} = -\varepsilon  + \varepsilon_{in} \label{energy}
\end{equation}
where $\varepsilon = \eta \langle (\nabla_{\alpha} v_{\beta})(\nabla_{\alpha} v_{\beta}) \rangle$ is the
dissipation rate, and $\varepsilon_{in}$ is the rate of energy
transfer from the interface to the fluid. Retaining the assumption that the
interface (as opposed to the velocity field) has just one characteristic
length, $\varepsilon_{in}$ is readily estimated from 
(\ref{eq:gradmuest}) as $\sigma \dot{L}/L^2$.

Applying the simple scaling for velocity gradients ($\bnabla \sim 1/L$) to each term in  (\ref{energy})
gives the following energy `balance' in the inertial regime where $L\sim T^{2/3}$:
\begin{equation}
-\rho T^{-5/3} \sim - \eta T^{-2} + \sigma T^{-5/3},
\label{energy2}
\end{equation}
where factors of $\rho,\eta,\sigma$ are retained to aid identification of the terms.
At first sight this suggests a balance of interfacial and inertial terms, with the viscous contribution negligible, at late times: this is Furukawa's assumption. However, the signs show this to be inconsistent: the kinetic energy and the energy stored in the interface are both {\em decreasing} with time, so these cannot properly be balanced against each other.

This exposes a central defect of the simple scaling analysis in the inertial regime. It is well known, of course, that even the simplest theories of fluid turbulence entail \textit{several} length scales \citep[whereas more modern `multiscaling' theories have, in effect, infinitely many,][]{frisch95a}. In the simplest,  Kolmogorov-type approach \citep[see][]{frisch95a,kolmogorov41a}, the important lengths are 
the Taylor microscale\footnote{\
The prefactors in the definitions of the Taylor and Kolmogorov microscales
differ between sources.}
\begin{equation}
\lambda\equiv(5\eta\langle v^2\rangle/\varepsilon)^{1/2},
\label{eq:microscale}
\end{equation}
characteristic of velocity gradients, 
and the Kolmogorov (dissipation) microscale, 
\begin{equation}
\lambda_d\equiv2\upi(\eta^3/\rho^3\varepsilon)^{1/4},
\label{eq:k_d}
\end{equation}
the length scale below which nothing interesting occurs. (Energy is dissipated at or above the scale $\lambda_d$.)

\cite{kendon99c} argued that in a binary fluid system where the fluid motion has become turbulent, the velocity follows the interface and scales as $\dot{L}$, but
the first and second gradients of velocity have scalings set by $\lambda$ and
$\lambda_d$ respectively, rather than by $L$. Within this simplified (Kolmogorov-level)
description,
the only scalings for these three lengths found physically admissible by Kendon for the inertial regime were
$\lambda \sim T^{1/2},\lambda_d \sim T^{5/12},L\sim T^{2/3}$.
In the NSE, this gives for the acceleration, convection, viscous and driving terms the following scalings:
\begin{equation}
\rho T^{-4/3} + \rho T^{-7/6} \sim \eta T^{-7/6} + \sigma T^{-4/3}.
\end{equation}
The predicted outcome is thus a balance between the
nonlinear and dissipative forces that is decoupled from the interfacial
motion, while interfacial stresses balance fluid acceleration.
The existence of a nonlinear/viscous balance implies an asymptotically \textit{finite} value for the
ratio of the corresponding terms in NSE, that is, a finite asymptote for the true Reynolds number $R_2$. On the other hand, since the result for the domain scale, $L\sim T^{2/3}$, survives unaltered from the simple scaling theory, the Reynolds number estimated from the order parameter, $\Rey_{\phi}$, continues to grow indefinitely. This suggestion, although speculative, 
appears to resolve the issue raised by Grant and Elder, without requiring a change in the domain scale growth law (nor any breakdown of universality of  (\ref{lscaling})).

To summarise, \cite{kendon99c} predicts a balance in which energy is first
transferred from the interface ($-\phi\bnabla\mu$) to large scale
fluid motion ($\rho\dot{\mathbf{v}}$).  The nonlinear
term ($\rho\mathbf{v}.\bnabla\mathbf{v}$) then transfers the energy from
the large scales down to smaller scales where dissipative forces
($\eta\nabla^2\mathbf{v}$) finally remove it from the system.
The resulting energy cascade thereby decouples the energy input scales from
the dissipation scales --- a familiar enough idea in turbulence theory \citep{kolmogorov41a}. 
In contrast, in the viscous hydrodynamic regime, 
the simple (one-length) scaling theory is already consistent with energy conservation, and all its results are recovered.
Kendon's predictions for scaling are summarised for ease of reference in
table \ref{table:predict}.
\begin{table}
    \begin{minipage}{\textwidth}
        \vspace{1ex}
        \begin{center}
        \begin{tabular}{@{}l@{\hspace{2em}}r@{}l@{\hspace{5em}}r@{}l@{\hspace{5em}}r@{}l@{}}
                Quantity & \multicolumn{2}{l}{viscous} &
                \multicolumn{4}{c}{inertial region}   \\[2pt]
                & \multicolumn{2}{l}{region}
		& \multicolumn{2}{l}{simple}
		& \multicolumn{2}{l}{new} \\
		& & & \multicolumn{2}{l}{scaling} 
		& \multicolumn{2}{l}{scaling} \\[5pt]
                $L(T)$ && 1 & 2&/3 & 2&/3  \\
                $\lambda=(5\eta\langle v^2\rangle/\varepsilon)^{1/2}$ && 1 & 2&/3 & \textbf{1}&\textbf{/2} \\
                $\lambda_d=2\upi(\eta^3/\rho^3\varepsilon)^{1/4}$ & 1&/2 & 1&/2 & \textbf{5}&\textbf{/12} \\
                $\mathbf{v}$ && 0 & $-$1&/3 & $-$1&/3 \\[5pt]
                $\rho\dot{\mathbf{v}}$ & $=\/\,$&$0$ & $-$4&/3 & $-$4&/3 \\
                $\rho\mathbf{v}.\bnabla\mathbf{v}$ & $=0$&, \boldmath{$-$}\boldmath{$1$} & $-$4&/3 & \textbf{$-$7}&\textbf{/6} \\
                $\eta\nabla^2\mathbf{v}$ & $-$&2 & $-$5&/3 & \textbf{$-$7}&\textbf{/6} \\
                $\phi\bnabla\mu$ & $-$&2 & $-$4&/3 & $-$4&/3 \\[5pt]
                $\Rey_{\phi}=l\dot{l}$ && 1 & 1&/3 & 1&/3 \\
                $R_1$ & $=\,\/$&$0$ & 1&/3 & \textbf{$-$1}&\textbf{/6} \\
                $R_2$ & $=\,\/$&$0$ & 1&/3 & \textbf{0} & \\
                $\varepsilon=\eta(\bnabla\mathbf{v})^2$ & $-$&2 & $-$2 && \textbf{$-$5}&\textbf{/3} \\
        \end{tabular}
        \end{center}
        \caption{Summary of predicted scaling exponents for
        the viscous and inertial regions.  The `new' scaling theory \citep{kendon99c} gives
        the same predictions as simple scaling for the viscous region, except
        for the NSE term $\rho\mathbf{v}.\bnabla\mathbf{v}$.
        Entries are powers of $T$, an entry of $0$ indicates the quantity is
        constant, while an entry of $=0$ indicates the quantity is assumed
        to be zero in the viscous approximation.
        Bold entries indicate scaling predictions that differ from
        the simple theory.  From \cite{kendon99c}.}
        \protect\label{table:predict}
    \end{minipage}
    \vspace{1ex}
\end{table}
%

\section{Numerical method}
\label{sec:method}

The model system described by  (\ref{eq:phi}) and (\ref{eq:nse})
was simulated numerically using a modular LB code called \textbf{Ludwig}, described in detail in \cite*{bladon99a}. It has both serial and parallel
versions; the parallel code uses domain decomposition and the MPI (message passing interface) 
platform.  Any cubic lattice can be used with the
\textbf{Ludwig} code; the lattice parameter is taken as unity, as is the timestep $\Delta T$, thereby defining `simulation units' of length and time.

Here we chose the D3Q15 lattice, a simple cubic arrangement in which each site communicates with its six nearest and eight third-nearest neighbours.
The fluid dynamics is, as usual \citep[see][]{higuera89a,ladd94a} encoded at each site by a distribution function $f_i$, where the subscript obeys $0\le i\le 14$. This ascribes weights to each of 15 velocities $\mathbf{c}_i$: one null, six of magnitude $1$, eight of magnitude $\sqrt 3$, with directions such that each velocity vector points toward a linked site.
In order to model the binary fluid, a second set of distribution functions,
$g_i$, is also used, following \cite{swift96a}. The $f$'s are defined such that
\begin{equation}
\sum_i f_i = \rho
\label{eq:ld_rho}
\end{equation}
where the sum is over all directions, $i$, at a single lattice point,
while for $g_i$ the same sum gives the order parameter,
\begin{equation}
\sum_i g_i = \phi.
\label{eq:ld_phi}
\end{equation}
(At this point, our algorithm departs from that of \cite{swift96a}, who would have $\phi\rho$ on the right. The two methods differ
only by terms that vanish in the incompressible limit of interest,
where $\rho \to 1$ everywhere.)

The momentum, $\rho v_{\alpha}$ (with $\alpha$ a cartesian index) is then given by
\begin{equation}
\rho v_{\alpha} = \sum_i f_i c_{i\alpha},
\label{eq:ld_mom}
\end{equation}
where $c_{i\alpha} \equiv (\mathbf{c}_i)_\alpha$. 
The full pressure tensor, $\mathcal{P}_{\alpha\beta}$, is given by
\begin{equation}
\mathcal{P}_{\alpha\beta} = \sum_i f_i c_{i\alpha} c_{i\beta}.
\label{eq:ld_stress}
\end{equation}
This expression includes not only the conservative stress $\mathcal{P}^{th}_{\alpha\beta}$ but also dissipative (viscous) contributions, and a trivial `kinetic pressure' $\rho v_\alpha v_\beta$ which arises in any fluid
moving at constant velocity $\mathbf{v}$.

The distribution functions $f_i,g_i$ obey discrete evolution equations
involving simple first-order relaxation kinetics toward a pair of equilibrium distributions:
\begin{equation}
f_i(\mathbf{r} +\mathbf{c}_i, t+1) -f_i(\mathbf{r},t) = -(f_i-f_i^{(eq)})/\tau_1
\label{eq:tau_1}
\end{equation}
\begin{equation}
g_i(\mathbf{r} +\mathbf{c}_i, t+1) -g_i(\mathbf{r},t) = -(g_i-g_i^{(eq)})/\tau_2
\label{eq:tau_2}
\end{equation}
thus defining two relaxation parameters $\tau_1,\tau_2$. In our use of the code we select $\tau_2 = 1$ which causes $g_i$ to be reset to $g_i^{(eq)}$ each time step.
The viscosity is determined by $\tau_1$, with $\eta = (2\tau_1-1)/6$ in lattice units.
The equilibrium distributions, $f_i^{(eq)}$ and $g_i^{(eq)}$, can be
derived from  (\ref{eq:ld_rho}) -- (\ref{eq:ld_mom}), along with the condition that
the order parameter is advected by the fluid,
\begin{equation}
\sum_i g_i^{(eq)} c_{i\alpha} = \phi v_{\alpha},
\end{equation}
and that the pressure tensor
and chemical potential 
at equilibrium obey
\begin{eqnarray}
\sum_i f_i^{(eq)}c_{i\alpha} c_{i\beta} &=& \mathcal{P}^{th}_{\alpha\beta} + \rho v_{
\alpha} v_{\beta} 
\label{eq:lud_P} \\
\sum_i g_i^{(eq)}c_{i\alpha} c_{i\beta} &=& \tilde M\mu\,\delta_{\alpha\beta} + \phi v_{\alpha} v_{\beta}.
\label{eq:lud_mu}
\end{eqnarray}
The parameter $\tilde M$ controls the order parameter mobility $M$ via $\tilde M \Delta t(\tau_2-1/2) = M$, so that $\tilde M = 2 M$ in our case. (Note that in \cite{kendon99a} and \cite{cates99a}, the quoted values of $M$ are in fact $\tilde M$ values and should therefore be halved to give the true order parameter mobility.) The second term on the right 
in  (\ref{eq:lud_P}) is the trivial `kinetic pressure', with an analagous term in  Eq.(\ref{eq:lud_mu}).

By expanding $f_i^{(eq)}, g_i^{(eq)}$ to second order in velocities and solving for the coefficients one obtains
\begin{equation}
f_i^{(eq)} = \rho \omega_{\nu}\left\{A_{\nu} + 3v_{\alpha} c_{i\alpha}
        + \frac{9}{2}v_{\alpha} v_{\beta}c_{i\alpha} c_{i\beta}
        - \frac{3}{2}v^2 + G_{\alpha\beta}c_{i\alpha} c_{i\beta}\right\}.
\end{equation}
Here, ${\nu}$ is an index that denotes the speed, 0, 1, or $\sqrt{3}$, and
$\omega_{\nu}$, $A_{\nu}$ and $G_{\alpha\beta}$ are constants given by
\begin{equation}
\omega_0 = 2/9;\;\;\; \omega_1 = 1/9;\;\;\; \omega_3 = 1/72,
\end{equation}
\begin{equation}
A_0 = \frac{9}{2} - \frac{7}{2} \mathrm{Tr}\mbox{\boldmath{$\cal P$}}^{th};\;\;\;
A_1 = A_3 = \frac{1}{\rho} \mathrm{Tr}\mbox{\boldmath{$\cal P$}}^{th},
\end{equation}
\begin{equation}
G_{\alpha\beta} = \frac{9}{2\rho} \mathcal{P}^{th}_{\alpha\beta} - 
\frac{3\delta_{\alpha\beta}}{2\rho} \mathrm{Tr}\mbox{\boldmath{$\cal P$}}^{th}.
\end{equation}
The equilibrium distribution for the order parameter, $g_i^{(eq)}$,
is the same as for $f_i^{(eq)}$, with {\boldmath$\mathcal{P}$}$^{th}$
replaced by $\tilde M\mu\,\mathbf{\identity}$ in the above equations.
The above results follow \cite{swift96a}, generalised to three dimensions.

To complete the model specification, one must introduce 
expressions for the pressure tensor and chemical potential derived from the free energy functional. 
In this study we chose in  (\ref{eq:free_gen}) a simple `$\phi^4$' model
for $\mathcal{V}(\phi)$:
\begin{equation}
F[\phi,\rho] = \int\! \mathrm{d}\mathbf{r}
        \left\{ \frac{A}{2} \phi^2 + \frac{B}{4} \phi^4
                + \frac{\kappa}{2} (\nabla\phi)^2
                +\frac{1}{3}\rho\ln\rho \right\}
\label{eq:free}
\end{equation}
where $A < 0$. (The term in $\rho$ is discussed below.) With this choice one finds 
$\phi^*=\pm(-A/B)^{1/2}$, and using  (\ref{eq:sigma}), (\ref{chempot}),
\begin{eqnarray}
\sigma &=&(-8\kappa A^3/9B^2)^{1/2}\label{sigmaphifour}\\
\mu &=& A\phi+B\phi^3 - \kappa \nabla^2\phi . \label{muphifour}
\end{eqnarray}
The equilibrium interfacial profile in given by
$\phi/\phi^* = \tanh(g/\xi_0)$, where $g$ is the normal coordinate introduced previously, and 
\begin{equation}
\xi_0=(-\kappa/2A)^{1/2} \label{xizero}
\end{equation} is a measure of the interfacial width.

An important addition to  (\ref{eq:free}) is the term dependent on density $\rho$, here chosen as an `ideal gas' type contribution (up to the factor 1/3). This gives a diagonal term in the thermodynamic pressure tensor, which becomes
\begin{equation}
\mathcal{P}^{th}_{\alpha\beta} = \left\{\frac{\rho}{3} + \frac{A}{2}\phi^2 + \frac{3B}{4}\phi^4
- \kappa\phi\nabla^2\phi - \frac{\kappa}{2}(\bnabla\phi)^2\right\}\delta_{\alpha\beta}
+ \kappa(\partial_{\alpha}\phi)(\partial_{\beta}\phi),
\label{eq:ld_pt}
\end{equation}
so that the thermodynamic stress obeys $\mathcal{P}^{th}_{\alpha\beta} = (\rho/3)\delta_{\alpha\beta} + \mathcal{P}^{chem}_{\alpha\beta}$. Thus in practice
$\rho/3 = P$, which is the isotropic pressure contribution normally viewed as a Lagrange multiplier for incompressibility. But in fact, our LB algorithm does not know that the fluids are meant to be incompressible; instead the ideal-gas term is relied upon to enforce incompressibility to within acceptable numerical tolerances. (This avoids a separate calculation, at each time step, of the fluid pressure $P$ and renders the algorithm local.) Compressibility errors can be minimized by increasing the coefficient of $\rho\ln\rho$ in  (\ref{eq:free}), which would require a shortened time step, or for fixed time step, by reducing the magnitudes of $A,B$, and $\kappa$ together so that an acceptable level of incompressibility is maintained. The second route is followed here.

The chosen functional,  (\ref{eq:free}), has a number of points in its favour for
numerical simulation.  The main terms in $\mathcal{P}^{th}_{\alpha\beta}$ and 
$\mu$ are simple powers of $\phi$, so are easy and quick to evaluate.
(Models involving logarithms or trigonometric functions \citep{swift96a}
pay a heavy price in computational efficiency.)  
Further, the shape of the ``$\phi^4$'' potential is fairly smooth, 
avoiding very steep gradients that might lead to inaccuracy and instability
when approximated numerically on a lattice. We nonetheless need
to evaluate spatial gradients of $\phi$; this is done using all 26 (first, second and third) nearest neighbours on the
D3Q15 lattice, so that numerically
\begin{equation}
\partial_{\alpha}\phi(\mathbf{x}) =
        \frac{\sum_i c_{i\alpha}\phi(\mathbf{r}+\mathbf{c}_i)}
        {\sum_i c_{i\alpha}c_{i\alpha}}
\label{eq:grad_def}
\end{equation}
\begin{equation}
\nabla^2\phi(\mathbf{r}) = \frac{1}{9}\left[
        \left(\sum_{i=1}^{26}\phi(\mathbf{r}+\mathbf{c}_i)\right)
        - 26\phi(\mathbf{r}) \right].
\label{eq:lap_def}
\end{equation}
Note that these are not the only possible choices and that others will only coincide when all gradients are small on the scale of the lattice. There may be considerable scope for improvement by optimising the choices made, but we leave this for future work.

As usual in LB, we have chosen a lattice with enough symmetry to ensure that 
the rotational invariance of the fluid mechanics is faithfully represented \citep{higuera89a}. However, this does not guarantee that the same holds for the thermodyamic properties. With our choice of free energy functional,  (\ref{eq:free}) and the above gradient discretisation, rotational invariance in the thermodynamic sector is recovered only when all order parameter gradients are weak, which in principle requires $\xi_0 \gg 1$. In practice, a compromise is necessary; we return to this in \S\ \ref{sec:isotropy} below.

Finally, the hydrodynamic behaviour of the LB technique requires detailed comment.
The hydrodynamic equations that correspond to LB
can be obtained by making a Chapman-Enskog expansion of the Boltzmann equations (\ref{eq:tau_1},\ref{eq:tau_2}).
If we consider the expansion for the distribution function $f$ (which relates
to the fluid momentum) we arrive at
\begin{eqnarray}
\partial_T(\rho v_{\alpha}) &+& \partial_{\alpha}(\rho v_{\alpha}v_{\beta})
= -\partial_{\beta}{\cal P}^{th}_{\alpha\beta}
  -\partial_{\beta}\left[\eta\left(\partial_{\beta} v_{\alpha}
    + \partial_{\alpha} v_{\beta}
    - \frac{2}{3}\delta_{\alpha\beta}\partial_{\gamma} v_{\gamma}\right)
    + \xi\partial_{\gamma} v_{\gamma}\delta_{\alpha\beta}\right]
\nonumber \\
&& + \frac{3\eta}{\rho}\partial_{\beta}\left[v_{\alpha}\partial_{\gamma}
 {\cal P}^{chem}_{\beta\gamma} + v_{\beta}\partial_{\gamma}
 {\cal P}^{chem}_{\alpha\gamma} + v_{\gamma}\partial_{\gamma}
 {\cal P}^{chem}_{\alpha\beta}\right]
 - \frac{3\eta}{\rho}\partial_{\beta}\partial_{\gamma}(\rho v_{\alpha}v_{\beta}v_{\gamma})
\label{eq:NS_LB}
\end{eqnarray}
where $\eta$ and  $\xi$ are the shear and second viscosities, respectively. For our single-relaxation LB scheme $\xi = 2\eta/3$.

The first line of Equation \ref{eq:NS_LB} corresponds to the standard Navier Stokes equation, and shows that, through these terms, the model 
recovers both the compressible and incompressible features of isothermal hydrodynamics. \footnote{Note that when modeling a macroscopic 
length, $L$, the ratio $\eta/(c_s \rho L)$ will be larger than in typical real fluids; due to the 
presence of the lattice, LB does not have a sharp separation between the compressible 
and incompressible time scales. In this respect, it resembles a high viscosity fluid, \cite{Hagen97a}.}

The second line in Equation \ref{eq:NS_LB} contains spurious terms, which arise partly because the enthalpic
interactions that lead to the non-ideal behaviour of 
the LB fluid (or fluid mixture) are introduced through equilibrium information only. (In a Hamiltonian system, the same 
interactions that perturb the equilibrium state away from an ideal gas would also be responsible for the dynamics.) Of the
two terms, the second one is 
not Galilean invariant but is cubic in the velocity. It will 
be negligible for small velocities (recall that the LB algorithm anyway requires fluid velocities that are small in 
lattice units). The first term can be decomposed into a Galilean-invariant part, which is a product of gradients of the 
pressure and velocity, and a non-Galilean invariant contribution. The product term is small compared to the Navier-Stokes 
terms, which are linear in such gradients, whenever these are weak; the non-Galilean invariant term is small under the same conditions.
There have been recent proposals to 
enforce Galilean invariance within compressible multiphase lattice-Boltzmann schemes, \cite{Holdych98a}, and this is a 
desirable feature of future algorithms.
Nonetheless, under circumstances where all hydrodynamic fields vary smoothly on the lattice scale, the spurious terms appearing in the second line of (\ref{eq:NS_LB}) will typically be smaller 
than the retained ones of the NSE equation in the first line. 

The evolution equation for the order parameter can be obtained analogously, by performing a 
gradient expansion of the linearized Boltzmann equation (\ref{eq:tau_2}). This leads to
\begin{equation}
\dot{\phi} + \partial_{\alpha}(\phi v_{\alpha}) 
= \left(\tau_2-\frac{1}{2}\right)\left[\nabla^2 \tilde{M}\mu
 - \partial_{\alpha}\left(\frac{\phi}{\rho}\partial_{\beta}{\cal P}^{th}_{\beta\alpha}\right)\right]
\label{eq:diff_LB}
\end{equation}
Equation (\ref{eq:diff_LB}) has the usual 
form of a convection-diffusion equation, so long as one chooses $\tilde{M}$ as a constant, except for the last 
term. As with (\ref{eq:NS_LB}), this Galilean-invariant term arises as a result of the way in which the non-ideality of the fluid mixture is introduced; like the first spurious term in (\ref{eq:NS_LB}) it contains one higher derivative than the term $\partial_{\alpha}{\mathcal{P}}^{th}_{\alpha\beta}$ that enters the Navier Stokes equation, and is expected to be small for similar reasons.
In Appendix \ref{app:hydro_modes} we confirm explicitly that, in the incompressible limit {\em only}, this spurious term does not modify the 
hydrodynamic modes of a binary mixture, at the level of a linearised expansion about a uniform quiescent fluid.

In summary, for a nominally incompressible fluid the correct fluid behaviour is recovered in the only regime where it can justifiably be expected, namely, when all the hydrodynamic fields vary slowly on the lattice length scale. Within the current LB algorithm, one also depends on having only slight fluid compressibility: this eliminates a spurious coupling between order parameter and momentum fluxes \citep[Appendix \ref{app:hydro_modes}, and][]{swift96a,Holdych98a}. Because of this it is important, with the current algorithm, to monitor closely the actual fluid flow.

\section{Parameter Steering}
\label{sec:pars}
Since it is possible in a single simulation to sample only a small piece of the $l(t)$ curve, it is necessary to work one's way along this curve via a series of different runs. This means varying $L_0$ relative to the lattice constant so that an appropriate window of \textit{reduced} length scales $l$ lies within the range of the simulation. Put differently, one must set, in lattice units, $1\ll L_0l(t) \ll \Lambda$. 

In the simulations, however, we need to choose not one parameter value ($L_0$) but seven ($A,B,\kappa,\rho,M$, and also $\tau_1,\tau_2$). 
Few of these parameters can, in practice, be set independently:
an unguided choice would typically produce a simulation
that either did nothing over sensible timescales or became unstable very rapidly. To avoid these outcomes, careful parameter steering is required. 
This was done by semi-empirical testing using small (workstation) simulation runs until satisfactory choices emerged. 
Runs on $96^3$ and $128^3$
lattices were used to confirm these before committing
the resources for $256^3$ runs.
The resulting parameters are summarized in tables
\ref{table:pars_256} and \ref{table:pars_128}. 
The guiding principles that emerged from this process, as well as the actual parameter values, are of some interest to those planning future work with LB. They are now summarised briefly, followed by a discussion (\S\ \ref{sec:validation}) of several validation exercises that were then undertaken. 

In essence, our navigation of the $l(t)$ curve involves steering three parameters ($\eta,\sigma,M$) at fixed values of the remaining four ($B/A,\kappa/A,\rho,\tau_2$).
Firstly, the (mean) density $\rho$ can be set to unity without loss of generality; we do this. Second, we set $\phi^* = 1$ by choosing $B = - A$; in terms of the original definition of the order parameter ($\phi = (n_A-n_B)/(n_A+n_B)$) this amounts to a simple rescaling of $\phi \to \phi/\phi^*$. 
Varying $\kappa$ in proportion to $A$ then gives control over the interfacial tension $\sigma$ (\ref{sigmaphifour}) while retaining a fixed interfacial width $\xi_0$ in lattice units (Eq. (\ref{xizero})); this keeps thermodynamic lattice anisotropies under control (\S\ \ref{sec:isotropy}). 

To achieve an efficient simulation, one requires the interfacial velocity $\dot L$ to be of order $0.01$ in lattice units during the main part of each run. (Any slower will exhaust resources; any faster will give compressible and inaccurate fluid motion, and, in all likelihood, numerical instability.) At each point on the $l(t)$ curve, this gives, \textit{a posteriori} a relation between $\eta$ and $\sigma$. Thus to access large $l$ one clearly requires small $L_0 = \eta^2/(\rho\sigma)$; but to avoid compressibility problems (\S\ \ref{sec:vel_comp}) this must be done by reducing viscosity rather than increasing interfacial tension.
Maintenance of numerical stability (\S\ \ref{sec:stability}) requires in fact that we \textit{decrease} $\sigma$ with decreasing $\eta$; however, these factors do not cancel in $L_0$ and a wide range of values (about six decades) can stably be achieved. Thus we were able to explore the viscous, crossover, and inertial regimes; these various regimes are delineated quantitatively in \S\ \ref{sec:l_t} below.

Setting the correct mobility $M$ is crucial throughout. Across the whole $l(t)$ curve, one has to ensure that $M$ is large enough that interfaces relax to local equilibrium on a time scale fast compared to their translational motion. But if $M$ is made too large, residual diffusion becomes a significant contributor to the coarsening rate, contaminating the data. This tradeoff can be eased in principle by going to larger system sizes than those currently available. It
could also be improved by making $M$ a function of $\phi$, setting (for example) $M = M_0(1-\phi^2)$. 
This would have the effect of giving strong diffusion only where it is needed, in the interfacial region. However, implementation of this within LB is not trivial \citep{swift96a}; specifically it is not enough to make $\tilde M$ in  (\ref{eq:lud_mu}) $\phi$-dependent\footnote{
Since in (\ref{eq:diff_LB}) $\tilde M$ enters in the form $\nabla^2(M\mu)$
rather than as
$\bnabla. (M \bnabla \mu)$ as would be required if $M$ were not constant.}.

It is not surprising that mobility is a limiting factor at large $L_0$ (viscous regime, small $l$): diffusion will always enter if the fluid flow is slow enough (high enough $\eta$). But mobility factors also come into play at the inertial end
(small $L_0$, large $l$): in physical units, the interface in this regime is unnaturally wide and to maintain it in diffusive equilibrium (and keep the algorithm stable) again requires relatively large $M$. These cause residual diffusion
which, for our system sizes, limits from above the range of $l$ accessible. 

Finally, we found that accuracy in the viscous regime (small $l$, large $L_0$)
is compromised when the viscosity becomes too large (of order unity, in lattice units). The signature of this is an apparent breakdown of energy conservation (see \S\ \ref{sec:diss_rate}). We are not sure of its origins, but note that too large a viscosity causes the dynamics of momentum diffusion and that of sound propagation (density equilibration) to mix locally. In an almost steady flows this should not matter, but in the small $l$ regime the viscous and interfacial terms in NSE are both numerically large. In principle these balance to give negligible fluid acceleration but their numerical cancellation may be imperfect. Although any such local accelerations are numerical in origin, the response to them may need to be accurate, if the global physics is to be handled correctly. We speculate that this is a limiting factor in our exploration of the $l(t)$ curve at the lower end. 

In this study, the largest
system size was $\Lambda^3 = 256^3$, although due to disk storage limitations, the results from this system size were analysed only after coarse-graining down to $128^3$.  The coarse-graining was done by averaging over blocks of eight
neighbouring lattice sites to create one coarse-grained value.
Runs at $128^3$ and $96^3$ were also done, and results for all calculated quantities were compared between $256^3$ and
$128^3$ runs with the same parameters, to identify any effects of
coarse-graining.
The main $96^3$ and $128^3,256^3$ simulations were run respectively on the EPCC
Hitachi SR-2201 machine (4 processors) and the EPCC Cray T3D (64 and 256 processors).
Follow-up studies used in some of the velocity analysis work, and for additional visualisation, were made on the CSAR Cray T3E at Manchester. 

Typical runs required, in the $256^3$ case, around 3000 T3D processor hours CPU, and $10^4$ time steps to reach the point where finite size effects set in (see \S\ \ref{sec:finitesize}). All simulation were run with
periodic boundary conditions; 
the initial configuration was always a completely mixed state, with
small random fluctuations.
For each run, the order parameter, $\phi$, and the fluid velocity vector at each lattice site were saved
periodically for later analysis.  The sampling frequency was limited
by the available disk space.  Typically, data was saved every 300
time steps giving, over a run of $10^4$ time steps, around 4Gb of data. 

\begin{table}
  \vspace{1ex}
  \begin{center}
  \begin{tabular}{lr@{\/}lr@{\/}lr@{\/}lr@{\/}lr@{\/}lr@{\/}lr@{\/}l}
    Run & $L_0$ && $T_0$ && $-\/\,$&$A,B$ && $\kappa$ && $\eta$ &
	\multicolumn{2}{c}{$\tilde M$}  && $\sigma$ \\[5pt]
    Run028 &  36&   &  935& & 0&.083 & 0&.053& 1&.41 & 0&.1 & 0&.055\\
    Run022 &  5&.95 & 71& & 0&.0625 & 0&.04 & 0&.5 & 0&.5 & 0&.042 \\
    Run033 &  5&.95 & 71& & 0&.0625 & 0&.04 & 0&.5 & 0&.2 & 0&.042 \\
    Run029 &  0&.952& 4&.54& 0&.0625 & 0&.04 & 0&.2 & 0&.3 & 0&.042 \\
    Run020 &  0&.15 & 0&.885 & 0&.00625 & 0&.004 & 0&.025 & 4&.0 & 0&.0042 \\
    Run030 &  0&.01  & 0&.016  & 0&.00625 & 0&.004 & 0&.0065 & 2&.5 & 0&.0042 \\
    Run019 &  0&.00095 & 0&.00064 & 0&.00313 & 0&.002 & 0&.0014 & 8&.0 & 0&.0021
 \\
    Run032 &  0&.0003 & 0&.00019 & 0&.00125 & 0&.0008 & 0&.0005 & 10&.0 & 0&.00083 \\
  \end{tabular}
  \end{center}
  \caption[Parameters used in $256^3$ lattice-Boltzmann runs]
          {Parameters used in $256^3$ lattice-Boltzmann runs. }
  \protect\label{table:pars_256}
  \vspace{1ex}
\end{table}
\begin{table}
  \vspace{1ex}
  \begin{center}
  \begin{tabular}{lr@{\/}lr@{\/}lr@{\/}lr@{\/}lr@{\/}lr@{\/}lr@{\/}l}
    Run & $L_0$ && $T_0$ && $-\/\,$&$A,B$ && $\kappa$ && $\eta$ &
	\multicolumn{2}{c}{$\tilde M$}  && $\sigma$ \\[5pt]
    Run010 &  381&  &25656& & 0&.125 & 0&.08 & 5&.71 & 0&.5 & 0&.084 \\
    Run026 &  36&   &  935& & 0&.083 & 0&.053& 1&.41 & 0&.25 & 0&.055 \\
    Run027 &  36&   &  935& & 0&.083 & 0&.053& 1&.41 & 0&.1 & 0&.055 \\
    Run014 &  5&.95 &   71& & 0&.0625 & 0&.04 & 0&.5 & 0&.5 & 0&.042 \\
    Run008 &  0&.952& 4&.54& 0&.0625 & 0&.04 & 0&.2 & 0&.5 & 0&.042 \\
    Run018 &  0&.15 & 0&.885 & 0&.00625 & 0&.004 & 0&.025 & 4&.0 & 0&.0042 \\
    Run015 &  0&.00095 & 0&.00064 & 0&.00313 & 0&.002 & 0&.0014 & 8&.0 & 0&.0021
 \\
    Run031 &  0&.0003 & 0&.00019 & 0&.00125 & 0&.0008 & 0&.0005 & 10&.0 & 0&.00083 \\
  \end{tabular}
  \end{center}
  \caption[Parameters used in $128^3$ lattice-Boltzmann runs]
          {Parameters used in $128^3$ lattice-Boltzmann runs.}
  \protect\label{table:pars_128}
  \vspace{1ex}
\end{table}

\section{Validation and Error Estimates}
\label{sec:validation}

\subsection{Numerical stability}
\label{sec:stability}

The LB method is not generally stable. In fact, our experience suggests that, whatever parameters are chosen, any run would eventually become unstable if continued for long enough; this is not dissimilar to some molecular dynamics algorithms, \cite{allen87a}. 
During testing, a reliable picture was acquired of the characteristic way in which this happens.  When the inaccuracies have built up to the 
point of failure, the velocities become very large over a small number
of time steps until numerical overflow causes the code to
stop running.  There seems to be no danger of taking data from a 
period when the system might be far from accurate but still apparently
running successfully, since the onset is so rapid.
Thus there are several runs among the set used for final data analysis where
the run ended prematurely due to instabilities, but the data prior
to the instability has been considered sufficiently reliable to be used.


\subsection{Anisotropy and interfacial tension}
\label{sec:isotropy}

\begin{figure}
\begin{minipage}{\textwidth}
    \raggedright
    \begin{minipage}{0.49\textwidth}
        \resizebox{\textwidth}{!}{\rotatebox{-90}{\includegraphics{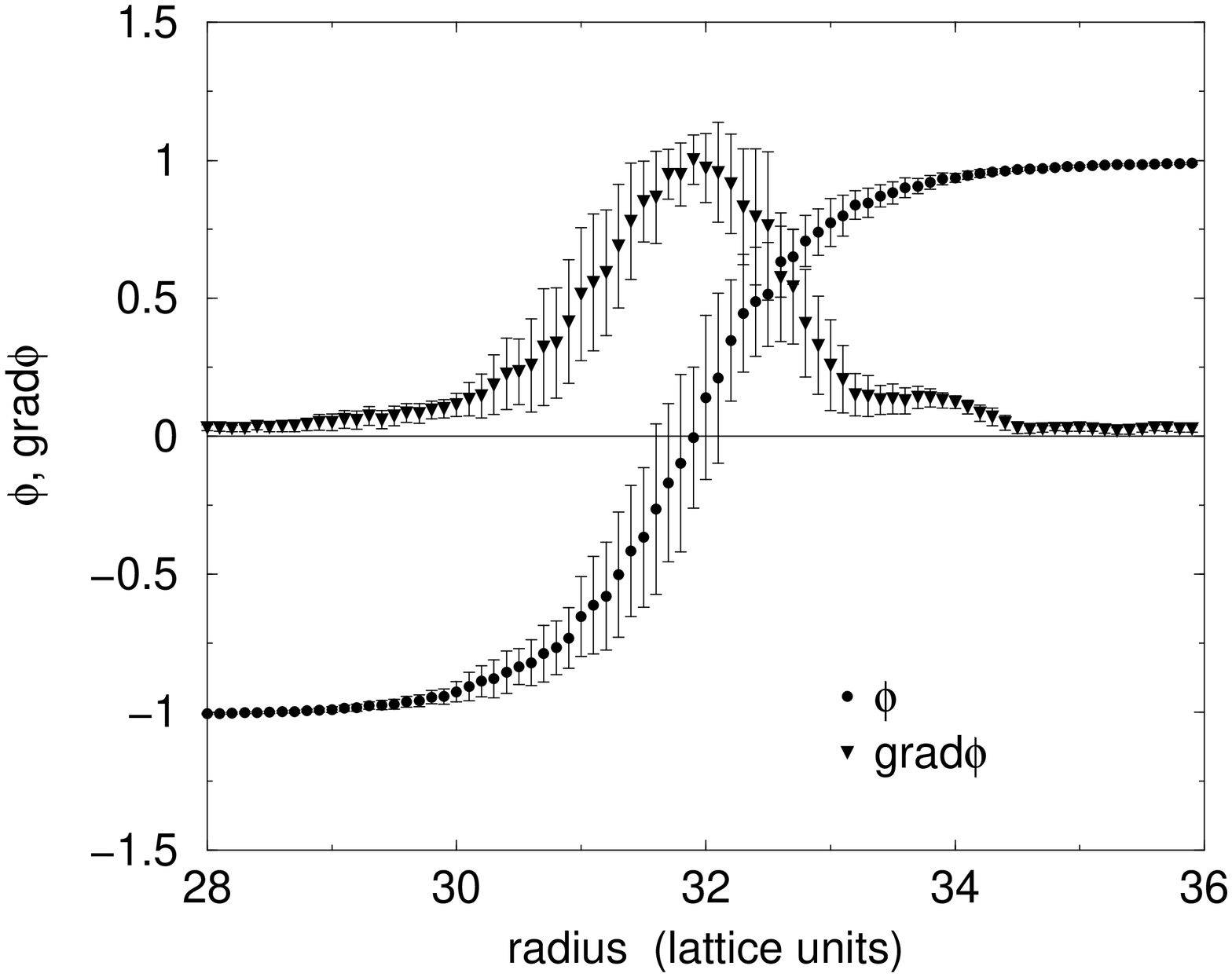}}}
    \end{minipage}
    \hfill
    \begin{minipage}{0.49\textwidth}
        \resizebox{\textwidth}{!}{\rotatebox{-90}{\includegraphics{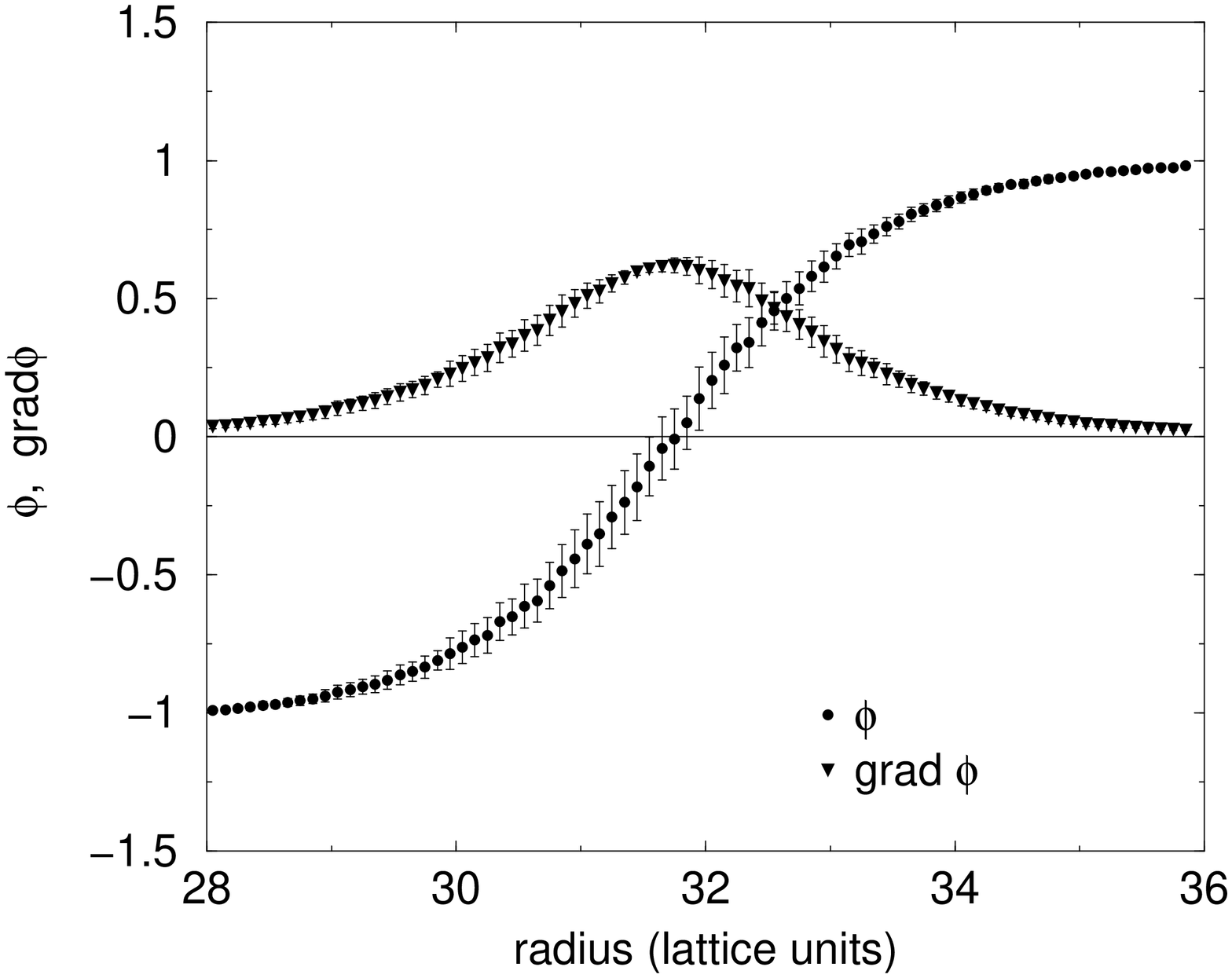}}}
    \end{minipage}
    \vspace{1ex}
    \caption[Interface profiles for spherical droplet]
            {Interface profiles ($\phi$) and gradient profile
             ($\bnabla\phi$), for spherical droplet equilibrated
             in the opposite fluid.  Data has been collected in bins of
             width 0.1 lattice spacings, and the error bars are one
             standard deviation.  The sphere radius is $32$.
             Left: interface width set by \protect$\xi_0=0.57$.
             The theoretical profile is
             \protect$\tanh(g/\xi_0)$, for a flat interface, where
             $g$ is a coordinate normal to the interface.
             Right: interface width set by \protect$\xi_0=0.88$.}
    \protect\label{graph:int_profile}
\end{minipage}
\end{figure}

The elimination of lattice anisotropy in the thermodynamic sector of the model requires $\xi_0 \gg 1$ in lattice units, to ensure that the interfacial tension $\sigma$ is independent of interface orientation. In practice this goal
must be balanced against other demands. 
To test the extent of the problem, a spherical droplet (radius 32 lattice units) 
of fluid B surrounded by fluid A was allowed to equilibrate.  The
interface profile was then measured by evaluating the mean and standard deviation of the order parameter $\phi(r)$ at various radii $r$ (binned on the scale of 0.1 lattice units) from the droplet centre.  The result is presented in figure \ref{graph:int_profile} for $\xi_0 =0.57$ and for $\xi_0 = 0.88$. 
Note that the `width'
of the interface, as judged by eye, is actually about $5\xi_0$.

A closer look at the droplet shape (not shown) in each case
reveals that the sphere has deformed slightly by squeezing along 
the Cartesian lattice directions and expanding along the diagonals.
This deformation is about 3.5\% for the narrower profile and about 1.5\%
for the wider profile. A similar test, done with a sphere of radius $31.5$, confirmed that this
deformation was not due to any tendency of the interface to lock onto
specific lattice sites but purely from anisotropy of the tension.

If all else were equal, the wider interface would be chosen. However, the
computational penalty for wider
interfaces is severe. To maintain these in local equilibrium,
the mobility $\tilde M$ must be high enough to allow diffusion 
across several $\xi_0$ on a timescale faster than fluid motion.
For the wider interface ($\xi_0 = 0.88$) the resulting residual diffusion then contaminates much of the remaining $L$ range. It was thus found necessary to
sacrifice some isotropy for efficiency, and the narrower
profile with $\xi_0=0.57$ was used for the main runs in this work. 
The resulting anisotropies are marginally detectable by eye in visualisations of the interface for the spinodal system (e.g. figure \ref{fig:vel_fields} below). We estimate that they contribute systematic errors of a few percent to the growth rate $\dot L(T)$, which is comparable to other sources of error.

The mean interfacial tension was measured for each parameter set by allowing an interface to
come to equilibrium and numerically performing the integration in
 (\ref{eq:sigma}).  Both terms were evaluated, and an
average taken over various configurations.
This gives values for the interfacial
tension, shown in table \ref{table:sigma},
that are systematically about 10--15\% smaller than the theoretical values.
(The statistical error is a few percent.)
The difference is due to the narrow interface leading to inaccuracies 
in the gradient calculations.  But as far as the simulation is concerned,
this systematic effect is removed by our using the \textit{measured} value of
the interfacial tension in subsequent calculations of $L_0$ and $T_0$.
\begin{table}
  \vspace{0.1cm}
  \begin{center}
  \begin{tabular}{@{}r@{\/}lr@{\/}lr@{\/}lr@{\/}lr@{\/}lr@{\/}l@{}}
    $-$&$A,B$ && $\kappa$ && $\eta$ & \multicolumn{2}{c}{$\tilde M$} & $\sigma\,\,$ & theory 
& $\sigma\,\,$ & measured \\[5pt]
    0&.083  & 0&.053  & 1&.41   &  0&.1 & 0&.063   & 0&.055   \\
    0&.063  & 0&.04   & 0&.5    &  0&.5 & 0&.047   & 0&.042   \\
    0&.0063 & 0&.004  & 0&.025  &  4&.0 & 0&.0047  & 0&.0042  \\
    0&.0031 & 0&.002  & 0&.0014 &  8&.0 & 0&.0024  & 0&.0021  \\
    0&.0013 & 0&.0008 & 0&.0005 & 10&.0 & 0&.00094 & 0&.00083 \\
  \end{tabular}
  \end{center}
  \caption[Interfacial tension, theoretical and measured values]
          {Interfacial tension, theoretical and measured values.}
  \protect\label{table:sigma}
  \vspace{0.1cm}
\end{table}
%

\subsection{Local equilibrium and residual diffusion}
\label{sec:diffusion}

Errors in the intereface-driven dynamics can arise if the interface is
not maintained in local equilibrium.
This was tested as follows. 
Since the bulk fluid is fully separated ($\phi=\pm1$), 
one expects $1-\langle|\phi|\rangle \propto A/V \propto 1/L$ 
where $A/V$ is the area per unit volume and angle brackets are a real-space (site) average. Within a given run, any departure from 
constancy of the product $L(1-\langle|\phi|\rangle)$
is thus an indicator that the interfaces are failing to keep up with the evolution of the surrounding fluid. (This product could have different asymptotic values in the viscous and inertial regimes, so the product need not be the same in different runs.)
At the lowest values used for the mobility $M$ (deep in the viscous regime) there was measurable deviation from constancy, from which the nonequilibrium deviations in $\sigma$ were estimated to be of order 5\%. Any deviations in the inertial regime were, however, smaller than this.

Careful checks were made to exclude residual diffusive contributions
to the coarsening process. This was done using comparator runs in which the viscosity was set to an extremely large value so that coarsening was \textit{purely} diffusive. (Such runs are depicted in figure \ref{graph:l_t_fit1} below.) From this, the diffusive coarsening rate was found as a function of domain size. Then for the full run (with fluid motion reinstated) all data was excluded for which this diffusive coarsening rate exceeded 2\% of the full rate. 
This whole procedure was repeated with a limit of 1\% instead of 2\%
on the residual diffusion.  The values of the fitted exponent $\alpha$ as per  (\ref{alphafit})
(given in the last column in table \ref{table:L_min}),
did not change beyond the estimated errors so the limit of 2\% diffusion
was taken to provide sufficient accuracy.

The result of this choice was exclusion of data with $L<L_{min} \simeq 15 - 25$ (varying somewhat between runs). Had a wider interface been used (see \S\ \ref{sec:isotropy}) then by the same criterion $L_{min}$ would be much larger giving very little usable data.

\subsection{Compressibility and small scale structure}
\label{sec:vel_comp}

\begin{figure}
\begin{minipage}{\textwidth}
    \raggedright
    \begin{minipage}{0.48\textwidth}
        \resizebox{\textwidth}{!}{\rotatebox{-90}{\includegraphics{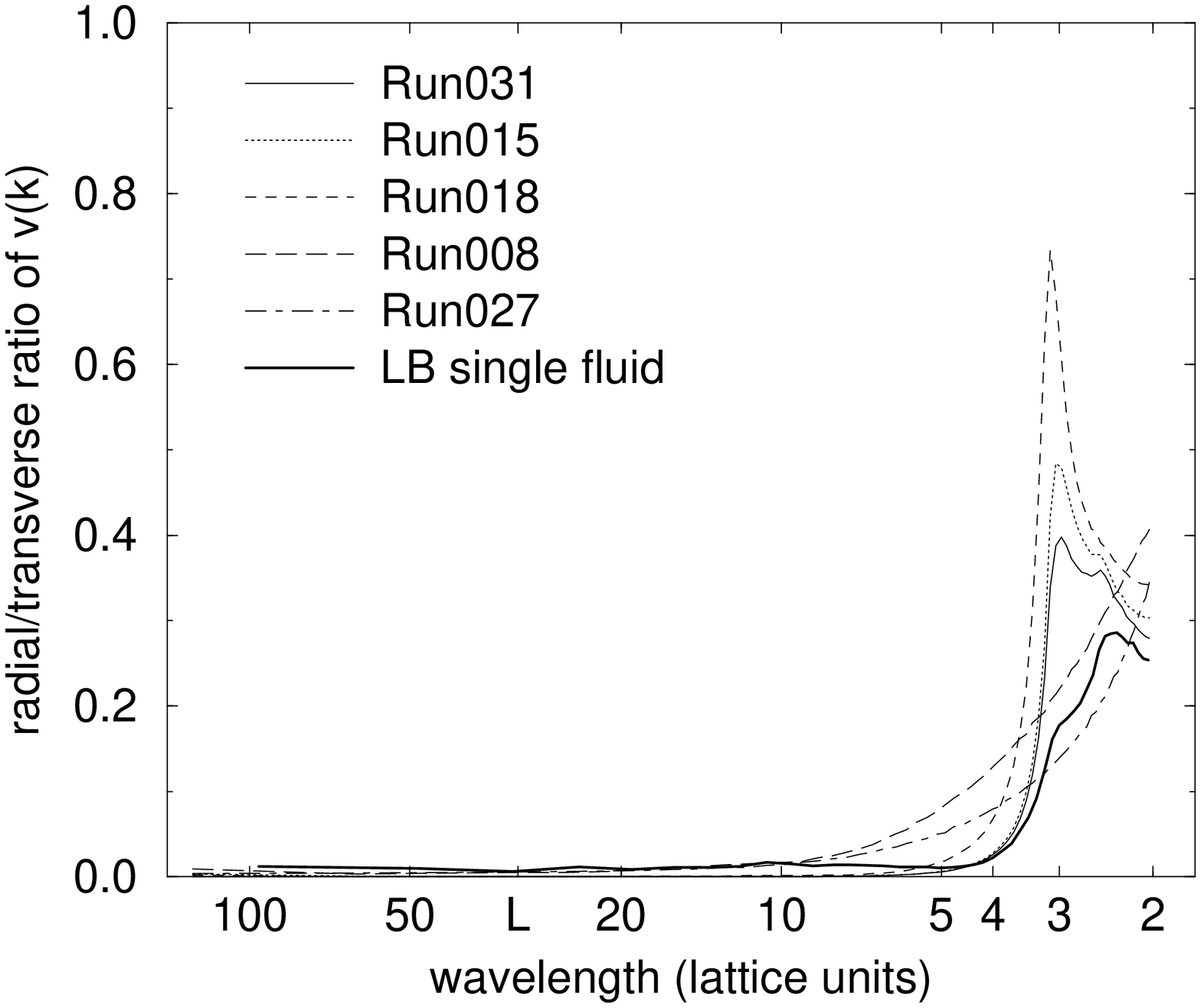}}}
    \end{minipage}
    \hfill
    \begin{minipage}{0.48\textwidth}
        \resizebox{\textwidth}{!}{\rotatebox{-90}{\includegraphics{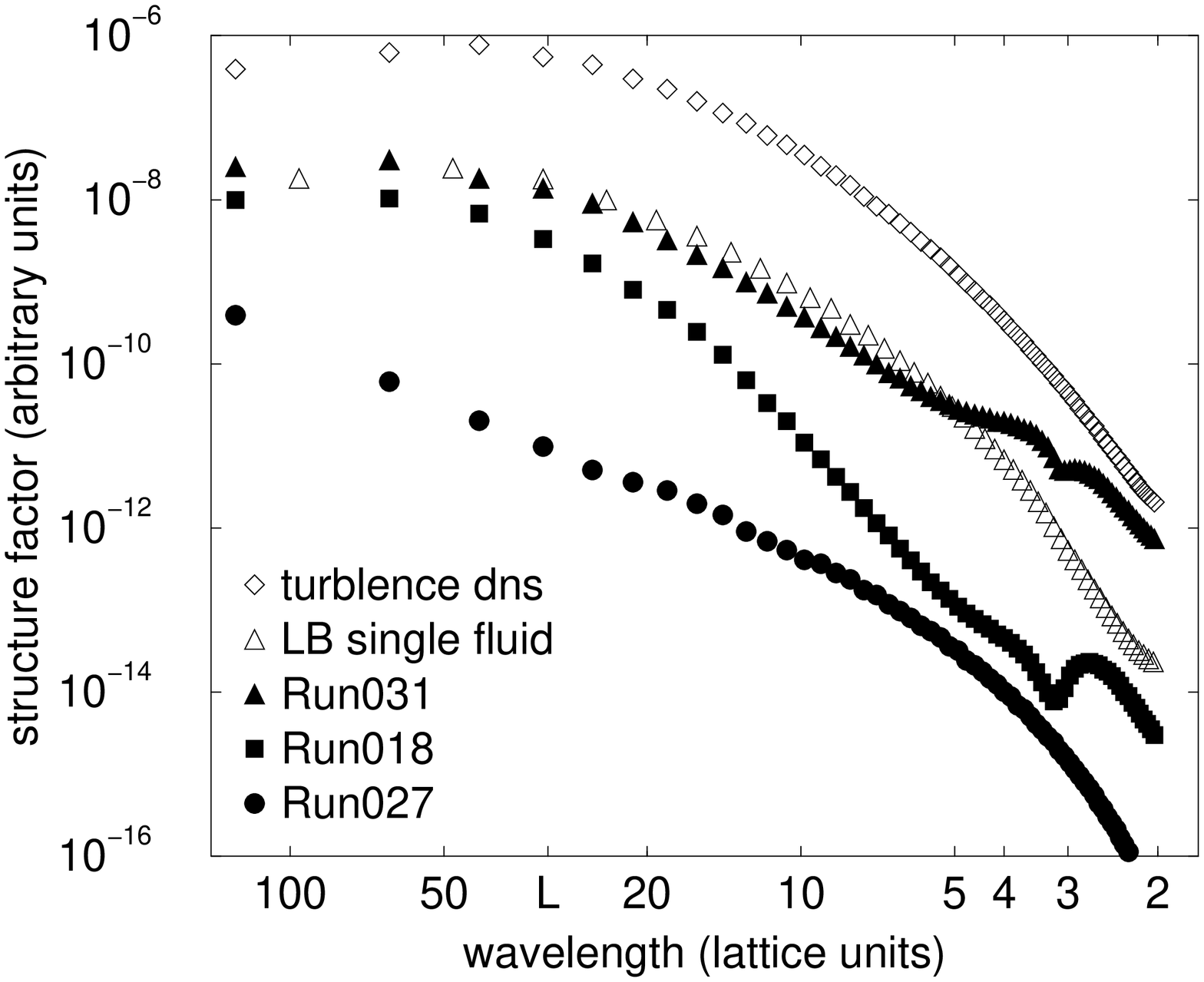}}}
    \end{minipage}
    \vspace{1ex}
    \caption{Left: ratio of the radial to transverse velocity
             components in Fourier space for various runs.
             Right: velocity structure factor showing relative magnitude
             of the Fourier space velocity at different wave vectors.
             Wavevector axis is labelled by wavelength in lattice units.}
    \protect\label{graph:vel_comp}
\end{minipage}
\end{figure}

The \textbf{Ludwig} code will only work correctly at low Mach number. This requires $\dot L \ll c$ where the sound speed is $c=3^{-1/2}$ in lattice units. Since in our simulations $\dot L$ is of order 0.01, we expect our the binary fluid mixture to remain incompressible ($\bnabla.\mathbf{v}=0$), at least at length scales larger than a few lattice sites; in Fourier space, we expect
$\mathbf{k}.\mathbf{v}(\mathbf{k})=0$ at all but high $k$.
figure \ref{graph:vel_comp} shows the {\sc rms} ratio of the radial
to the transverse velocity components in Fourier space as a function
of wavenumber, and also the spherically averaged velocity structure factor, 
$S_v(k) = \langle |\mathbf{v(k)}|^2\rangle$, for various runs. 
Also shown for comparison is single fluid turbulence\footnote{
The single fluid turbulence simulation method sets the radial
component identically to zero thus guaranteeing perfect incompressibility.},
generated using pseudo-spectral direct numerical simulation
(DNS) code by \cite{young99a},
and a LB run with a single fluid (no interface) but 
otherwise the same parameters as Run031 (inertial region).

At low wavenumbers the sytem is incompressible.  At higher wavenumbers,
there is some compressibility, whose effect varies in the different growth
regimes.  In the viscous regime, the longitudinal/transverse ratio rises with $k$, but the velocity structure factor shows that 
that all velocity components become small at high $k$ and contribute little to the overall dynamics. This is still true in the crossover region, where the compressibility
ratio is highest; a peak in $S_v(k)$ is found at a wavelengths around
3 lattice spacings.  In the inertial region, this
peak shrinks, and splits into two (at around
3.5 and 2.5 lattice spacings).  The transverse velocity component is
now larger although still an order of magnitude smaller than
the velocity at the peak of $S_v(k)$. 

Comparison with the single fluid turbulence, as simulated by both DNS and LB,
shows that these peaks in $S_v(k)$ are mainly due to the presence of the interface. Their presence only
in the crossover and inertial runs suggests that perhaps capillary waves are forming on the interface giving structure the velocity field on scales of the order of the interface width. Subsequent visualisation work showed that underdamped wavelike motion of interface is undoubtedly present at large $l$, \cite{desplat00a}, but predominantly at wavelengths much larger than the interfacial width. Another argument against the capillary wave explanation is that no similar bumps are seen in the order parameter structure factor $S(k)$ (figure \ref{graph:s_k}).

The nature of the velocity fields close to the interface certainly deserves further investigation 
\citep*[see, for example,][for related work on a different system]{theissen98a}.
Meanwhile, to have some compressibility on the length scale
of the interface itself appears unavoidable within current LB. Specifically, in the immediate vicinity of the interface the various diagonal terms in the chemical contribution to the pressure tensor,  (\ref{stress}), are individually
large, although these should nearly cancel for a slowly moving, weakly curved interface. Any numerical error here will lead to local deviations in 
the fluid density $\rho$, even if the bulk fluid motion is effectively incompressible everywhere else. On molecular physics grounds also, some coupling between density and order parameter can be expected at the interface between otherwise incompressible fluids. Such coupling is present in real physical systems, but care is needed with the current LB code where compressibility effects also bring violations of Galilean invariance (\S\ \ref{sec:method}).

\subsection{Finite size effects}
\label{sec:finitesize}

Various estimates were made of when our (periodic) boundary conditions started to significantly influence the behaviour of $L(T)$. This included several comparisons of different sized runs with the same values for other simulation parameters. On this basis, the data for the $96^3$ and $128^3$ was pruned at $L = L_{max}=\Lambda/4$ before analysis, and the $256^3$ runs terminated at this point. This criterion is much more conservative than in some previous work \citep[e.g.][]{jury98a}, and, given $L_{min}$, limits the range of $L$ accessible in a single large run to about half a decade. To balance this, averaging over different runs with the same parameter values should not then be necessary, since one has in effect $(\Lambda/L_{max})^3 = 64$ different (albeit correlated) samples being simulated within each run. 
Indeed, in the crossover and the inertial regime, we saw no sign of statistical fluctuations in the $L(T)$ plots. 

Interestingly, the same was not true for the extremely viscous runs, which showed somewhat erratic statistics (see \S\ \ref{sec:spin}). 
One possible reason for this is the presence of correlations, in the velocity field, over much larger length scales than $L(T)$, causing the local coarsening rates in different parts of the simulation to fluctuate coherently.  Long range velocity correlations are, in fact, clearly visible in the structure factor $S_v(k)$ shown in figure \ref{graph:vel_comp}. Specifically, for the most viscous run analysed (Run 027, $L_0 = 36$), $S_v(k)$ shows no sign of saturating at low $k$; instead the data suggests a power law divergence, and is consistent with $S_v(k) \sim k^{-2}$. (A theoretical argument leading to this result for the viscous regime is given in \S\ \ref{sec:s_k_v}.) In real space this translates into a long range, $1/r$ velocity correlation extending to either the system size (which is the likely case in any simulation) or some large physical length scale beyond which
the purely viscous approximation (Stokes flow) breaks down.

If this is correct, it could be practically impossible to avoid finite size effects when simulating the viscous regime. The most benign outcome is if the main effect is to correlate (rather than alter) local coarsening rates; this could be countered by averaging over a number of different runs \citep{jury98a,laradji96a}. However, this would have to be done for several system sizes before concluding that no other finite size effects were
present.
 
\section{Order parameter results}
\label{sec:spin}

We now present our results for the time evolution of the interfacial structure. These results can be extracted directly from knowledge of the order parameter using well-established procedures \citep[see][]{jury98a,appert95a,laradji96a,bastea97a}. We defer to \S\ \ref{sec:velocity} our explicit analysis of the fluid velocity field.

\subsection{Structure factor scaling}
\label{sec:s_k}

The first step in the analysis of the order parameter data was
calculation of the structure factor.
The $\phi$ field saved from the simulation runs was
processed through numerical Fourier transform routines, and the structure factor calculated as:
\begin{equation}
S(k) = \frac{1}{n_k}\sum_{k-{\upi}/{\Lambda} < |\mathbf{k}| < k+{\upi}/{\Lambda}}
       \phi(\mathbf{k}) \phi(\mathbf{-k}),
\label{eq:sk}
\end{equation}
where $\phi(\mathbf{k})$ is the Fourier transform of the order parameter, and
$n_k$ is the (actual) number of lattice sites in a shell of radius $k$
and thickness $2\upi/\Lambda$ in Fourier space (compare  (\ref{eq:sk_def})).

\begin{figure}
\begin{minipage}{\textwidth}
    \raggedright
    \begin{minipage}{0.48\textwidth}
        \resizebox{\textwidth}{!}{\rotatebox{-90}{\includegraphics{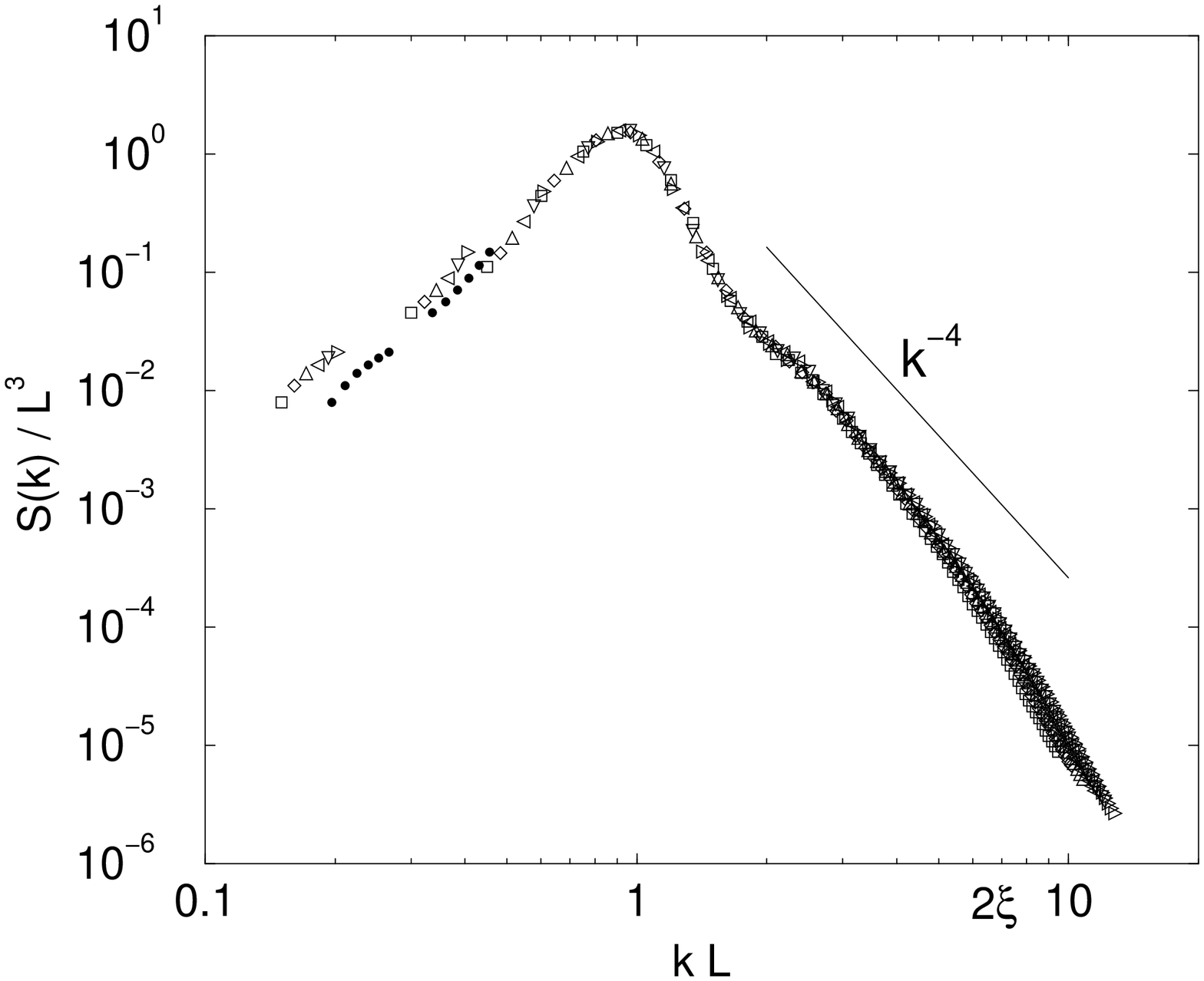}}}
    \end{minipage}
    \hfill
    \begin{minipage}{0.48\textwidth}
        \resizebox{\textwidth}{!}{\rotatebox{-90}{\includegraphics{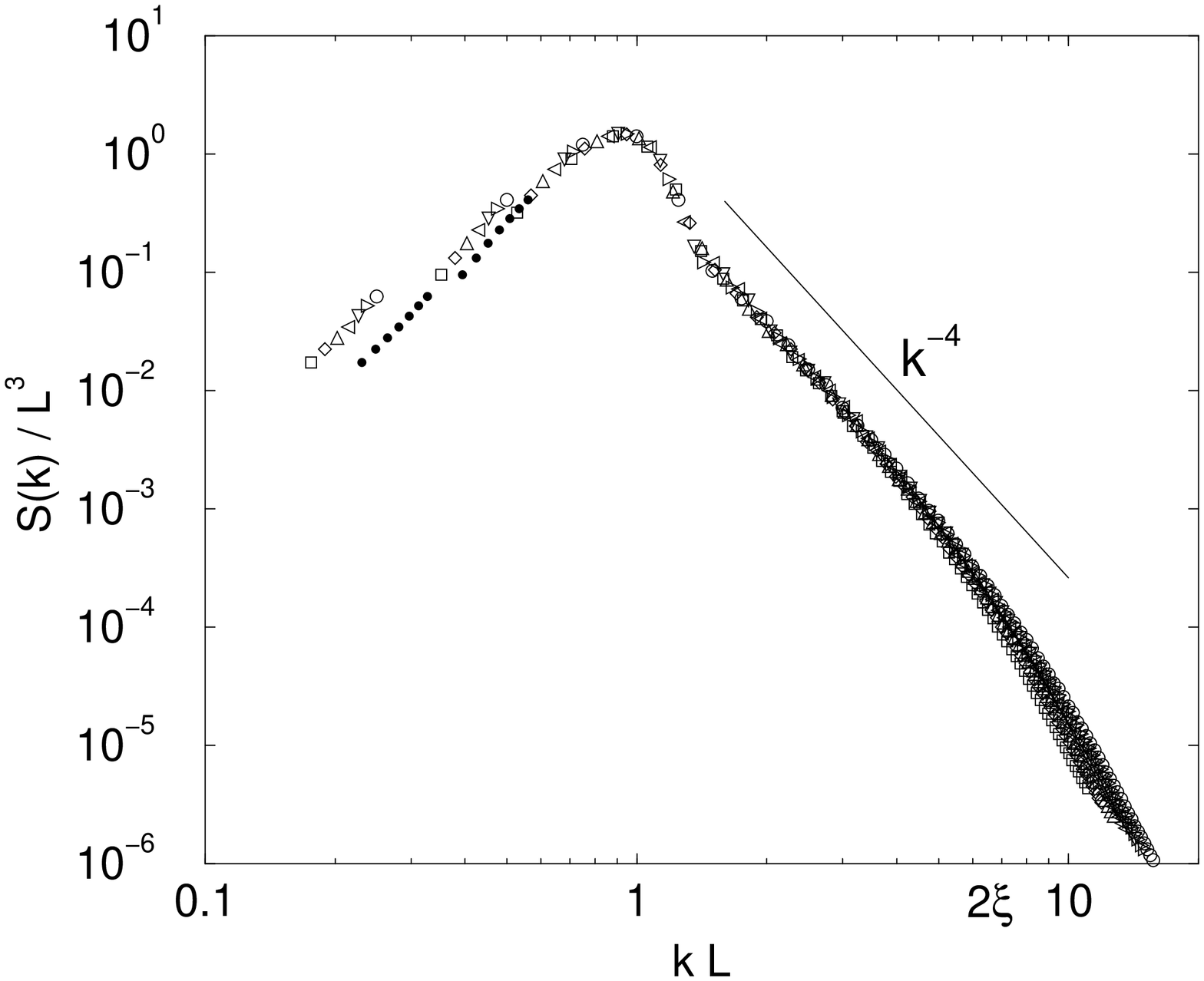}}}
    \end{minipage}
    \caption[Structure factor scaling]
             {Left: structure factor, $S(k)$, for Run028
             (viscous regime) for
             timesteps 14000 -- 19000, $L(T)= 38$ -- 52.
             Right: $S(k)$ for Run032 (inertial regime) for
             timesteps 11000 -- 17000, $L(T)= 45$ -- 64.
             Different open symbols denote different times $T$; filled circles show the same data corrected at low $k$ for discretisation effects (see text).}
    \protect\label{graph:s_k}
\end{minipage}
\end{figure}

Dynamical scaling requires that, in reduced physical units, not only the characteristic length $l(t)$ but also the statistical distribution of different interfacial structures should be the same for each $l$. In either the viscous or the inertial regime, therefore, the structure factor $S(k)$ should asymptotically collapse onto a single plot when appropriately
scaled, so that in simulation units
\begin{equation}
L^{-3} S(k) = f(k L),
\end{equation}
with a different function $f(kL)$ in each of the two limits. (More generally, dynamical scaling allows $L^{-3}S(k) = f(kL,l)$, so that the viscous and inertial asymptotes are $f(kL,0)$ and $f(kL,\infty)$ respectively.) 
figure \ref{graph:s_k} shows plots of $S(k)$ scaled in this way
for Run028 and Run032, representative of the viscous and inertial regimes
respectively. 

The collapse of the structure factor data within each run is good (figure \ref{graph:s_k}) for length scales larger than about twice the 
interface width. (The latter is marked as $2\xi$ on the graphs, with $\xi = 5 \xi_0$.) With our definition of $L$, the peak occurs at $kL$ just less than one. To the right of the peak there is a shoulder, followed by
a reasonable approximation to a $k^{-4}$ Porod tail.
(The Porod tail represents scattering from a weakly curved interface and should be found in the region $\xi\lesssim k^{-1} \lesssim L(T)$, 
see \cite{bray94a}; but between $\xi\simeq 3$ and $L_{max}=64$ there is barely room to observe it cleanly.)
The ragged sections of $S(k)$ in the low $kL$ region corresponds to the first two $k$-shells which have few $\mathbf{k}$ points and so poor statistics. The filled symbols are the same data corrected to allow for the fact that the
average value of $|\mathbf{k}|$ in such a shell differs from the nominal shell radius; the corrected result suggests no deviation from scaling even at low $k$, although the data there is less reliable than in the high wavenumber region.

The collapse \textit{between} different runs (not shown) is also good, so long as one compares runs chosen within either the viscous or the inertial regime. However, as is visible from figure \ref{graph:s_k}, the shape of $S(k)$ does evolve significantly between one regime and the other. In particular, the shoulder to the right of the peak is lower in the viscous regime than the inertial regime.  This implies that the domains are
a subtly different shape in real space, perhaps more evenly rounded in the 
linear regime since the peak is effectively a little sharper.
This may be linked to an increased number of relatively narrow necks in the inertial runs (large $l$), as first suggested by \cite{jury99a} and recently confirmed by direct visualisation of LB data, \cite{desplat00a}. Our structure factor results, taken piecewise, are compatible with those of \cite{jury99b}, \cite{appert95a}, and several other authors (see Appendix \ref{sec:other_work}). However, our study is the first to cover a wide enough parameter range to show a clear distinction, in the shape of $S(k)$, between the viscous and inertial regime.

Runs in the crossover region also show reasonable data collapse within each run, with a shape intermediate between the two shown in figure \ref{graph:s_k}, and very similar to that found by \cite{jury99b} in the same region of the $l(t)$ curve.  Note that a good collapse, within or between runs, cannot be expected \textit{a priori} in the crossover region. It arises because the $l$-dependent scaling function $f(kL,l)$ in fact evolves so slowly with $l$ that any data spanning less than a decade or two in $l$ is insensitive to the $l$ dependence. This is a consequence of the extreme breadth of the crossover region (quantified below).


\subsection{Evolution of the characteristic length scale}
\label{sec:l_t}

\begin{figure}
\begin{minipage}{\textwidth}
\begin{center}
    \resizebox{0.70\textwidth}{!}{\rotatebox{-90}{\includegraphics{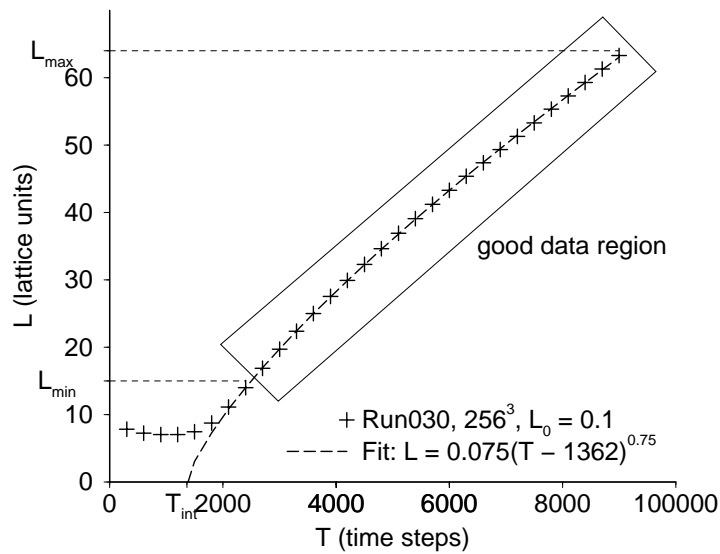}}}
\end{center}
    \caption{$L$ vs $T$ graph (unscaled) for Run030,
             illustrating the fitting procedures.}
    \protect\label{graph:L_T_030}
\end{minipage}
\end{figure}

The characteristic length scale $L(T)$, defined via
 (\ref{eq:Ldef}), has been calculated for the eight $256^3$ runs in table \ref{table:pars_256}. The order parameter data was coarse-grained to $128^3$ before analysis, but comparison with smaller runs confirmed that there was no effect of this on $L(T)$ within the `good data' range. The latter is defined as $L_{min}<L<L_{max}$, with $L_{min}$ fixed by our
criterion on residual diffusion (\S\ \ref{sec:diffusion}) and $L_{max} = 64$ as required to exclude finite size effects (\S\ \ref{sec:finitesize}). 
Figure \ref{graph:L_T_030} illustrates how the fitting was done.

\begin{figure}
\begin{minipage}{\textwidth}
    \raggedright
    \begin{minipage}{0.49\textwidth}
        \resizebox{\textwidth}{!}{\rotatebox{-90}{\includegraphics{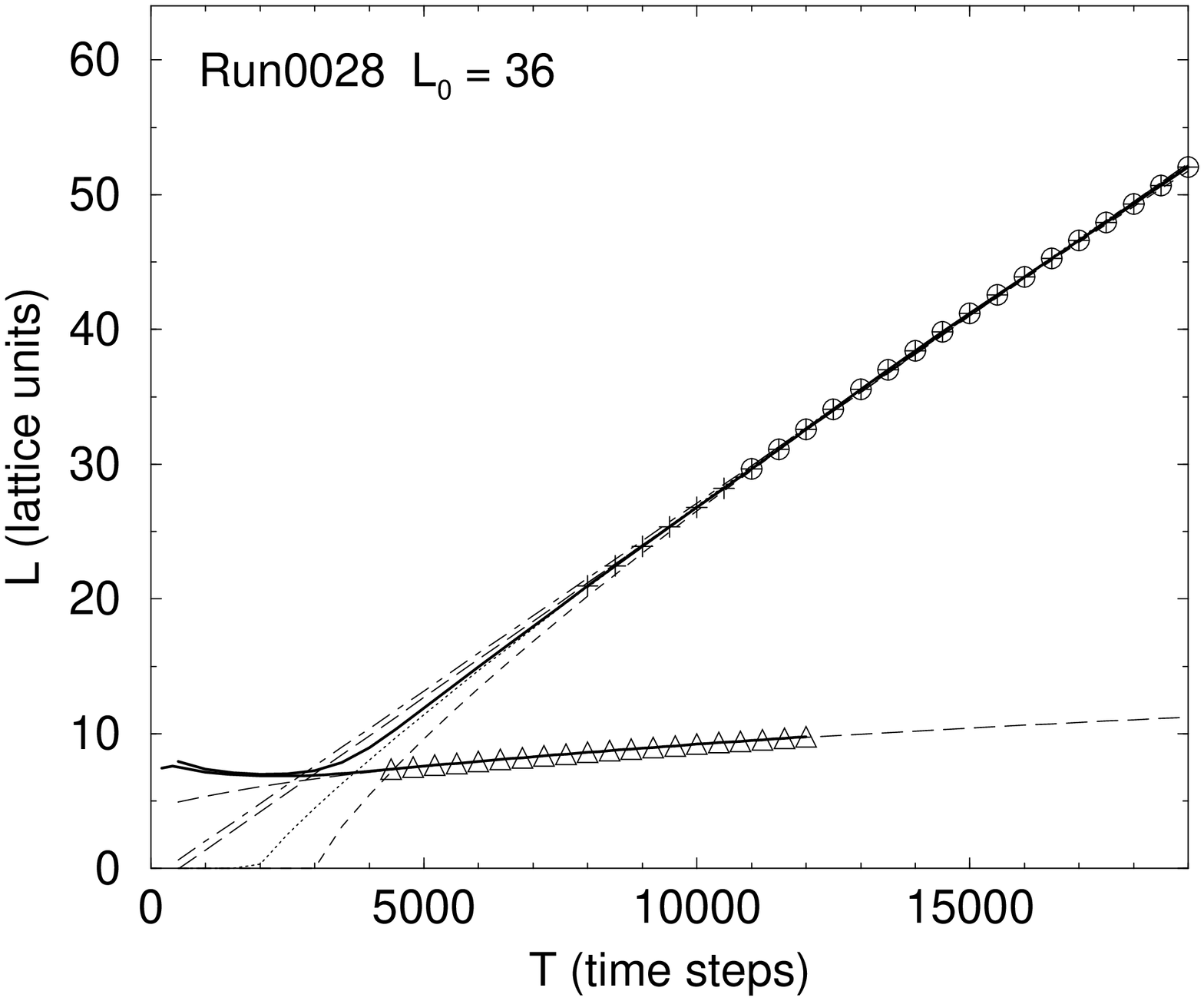}}}
    \end{minipage}
    \hfill
    \begin{minipage}{0.49\textwidth}
        \resizebox{\textwidth}{!}{\rotatebox{-90}{\includegraphics{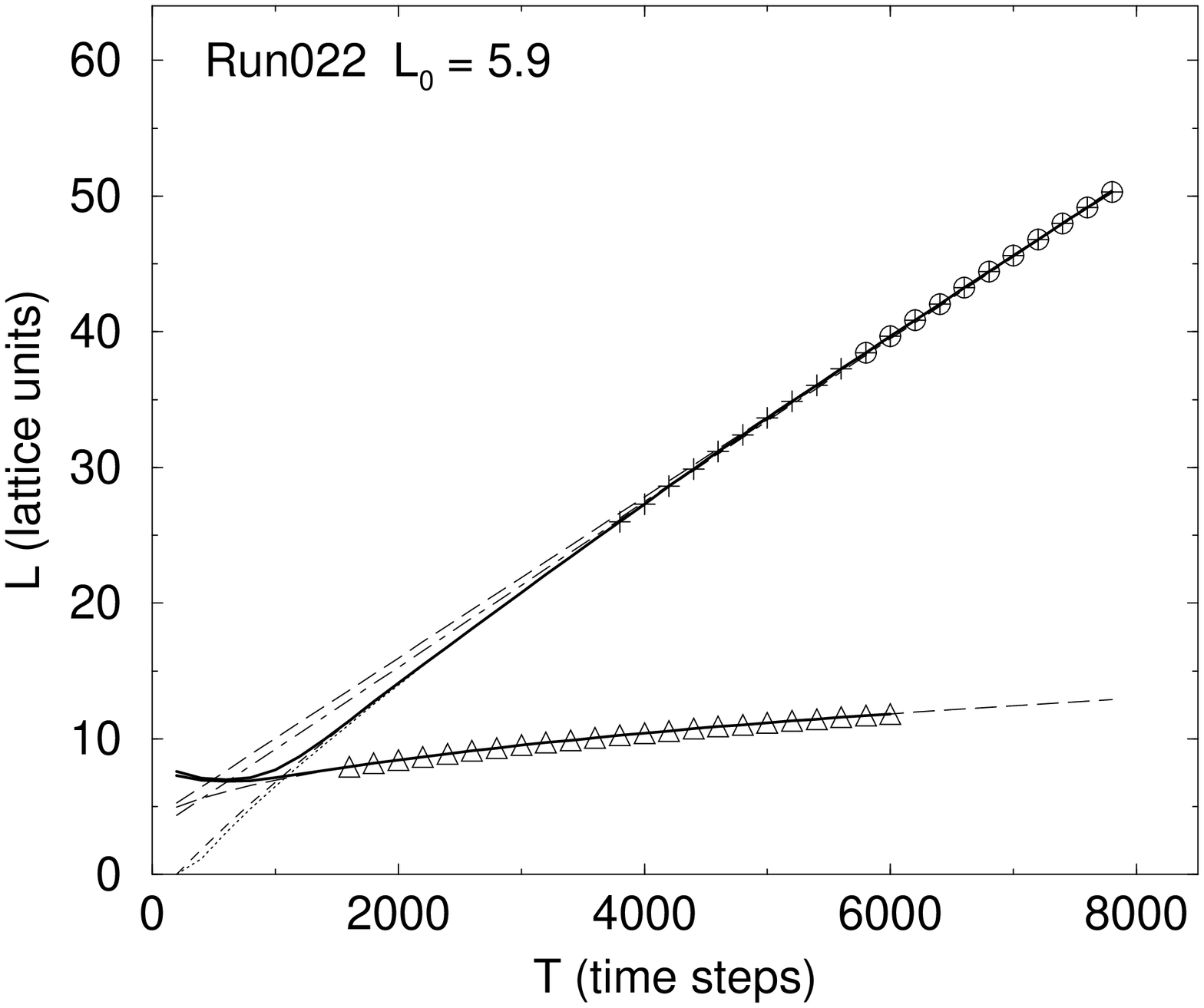}}}
    \end{minipage}
    \hfill \\
    \begin{minipage}{0.49\textwidth}
        \resizebox{\textwidth}{!}{\rotatebox{-90}{\includegraphics{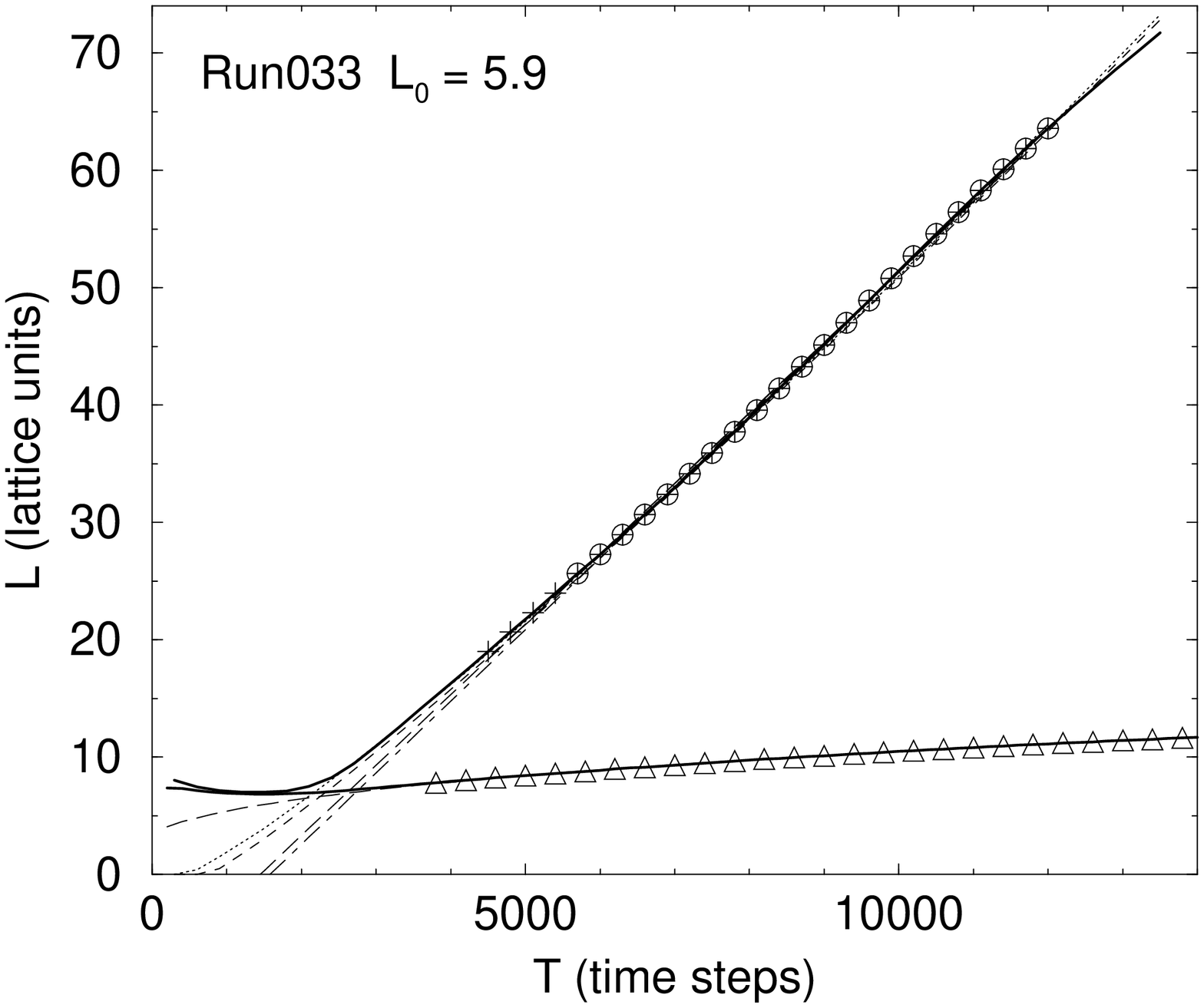}}}
    \end{minipage}
    \hfill
    \begin{minipage}{0.49\textwidth}
        \resizebox{\textwidth}{!}{\rotatebox{-90}{\includegraphics{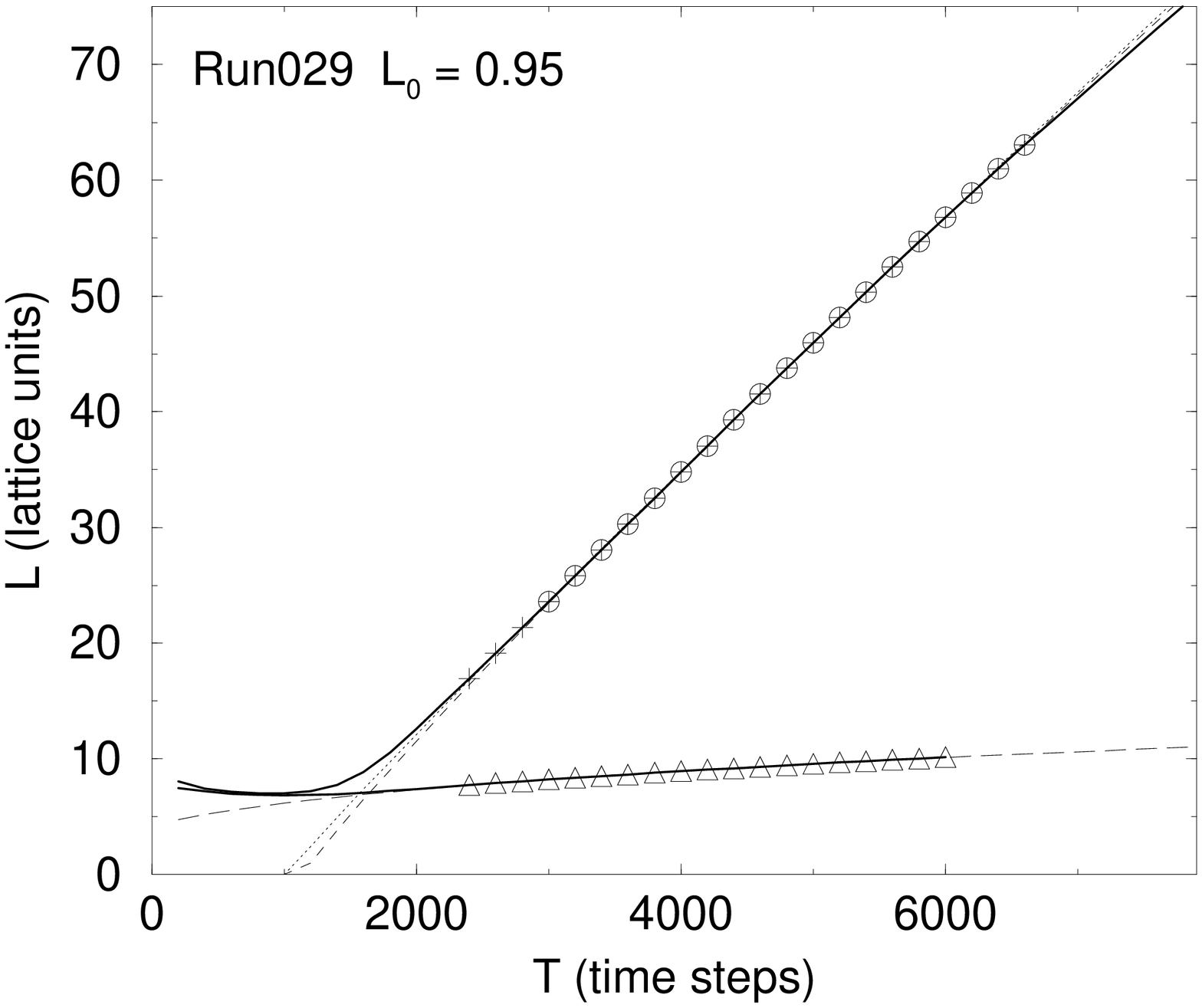}}}
    \end{minipage}
    \renewcommand{\baselinestretch}{1.0} \small\normalsize
    \caption[Fit $L(T)$: $L_0 = 36$, $L_0 = 5.9$ and $L_0 = 0.95$]
            {\textit{Fitting $L(T)$ and $L_D(T)$.
	     Upper left: Run028, $L_0 = 36$.
	     Upper right: Run022, $L_0 = 5.9$.
	     Lower left: Run033, $L_0 = 5.9$.
	     Lower right: Run029, $L_0 = 0.95$.
	     Solid lines indicate full set of recorded $L(T)$ data,
	     $+$ indicates data points used for fits
	     with $L_{min}$ set by 2\% diffusion,
	     $\bigcirc$ indicates data points  used for fits
	     with $L_{min}$ set by 1\% diffusion,
	     $\bigtriangleup$ indicates data points used for
	     fits to diffusion-only data.
	     table \protect\ref{table:L_min} summarises the main fit results.}}
    \protect\label{graph:l_t_fit1}
\end{minipage}
\end{figure}

\begin{figure}
\begin{minipage}{\textwidth}
    \raggedright
    \begin{minipage}{0.49\textwidth}
        \resizebox{\textwidth}{!}{\rotatebox{-90}{\includegraphics{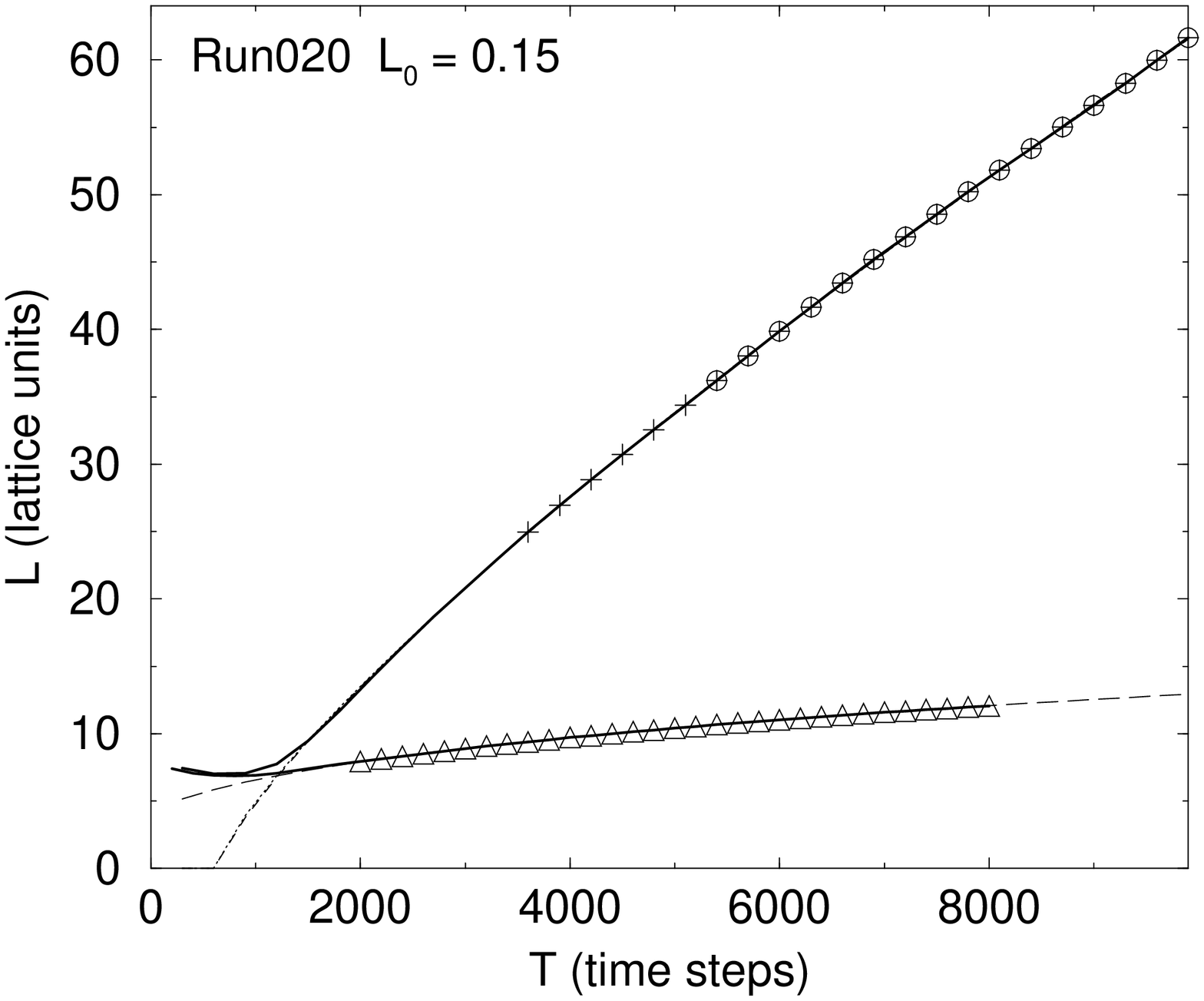}}}
    \end{minipage}
    \hfill
    \begin{minipage}{0.49\textwidth}
        \resizebox{\textwidth}{!}{\rotatebox{-90}{\includegraphics{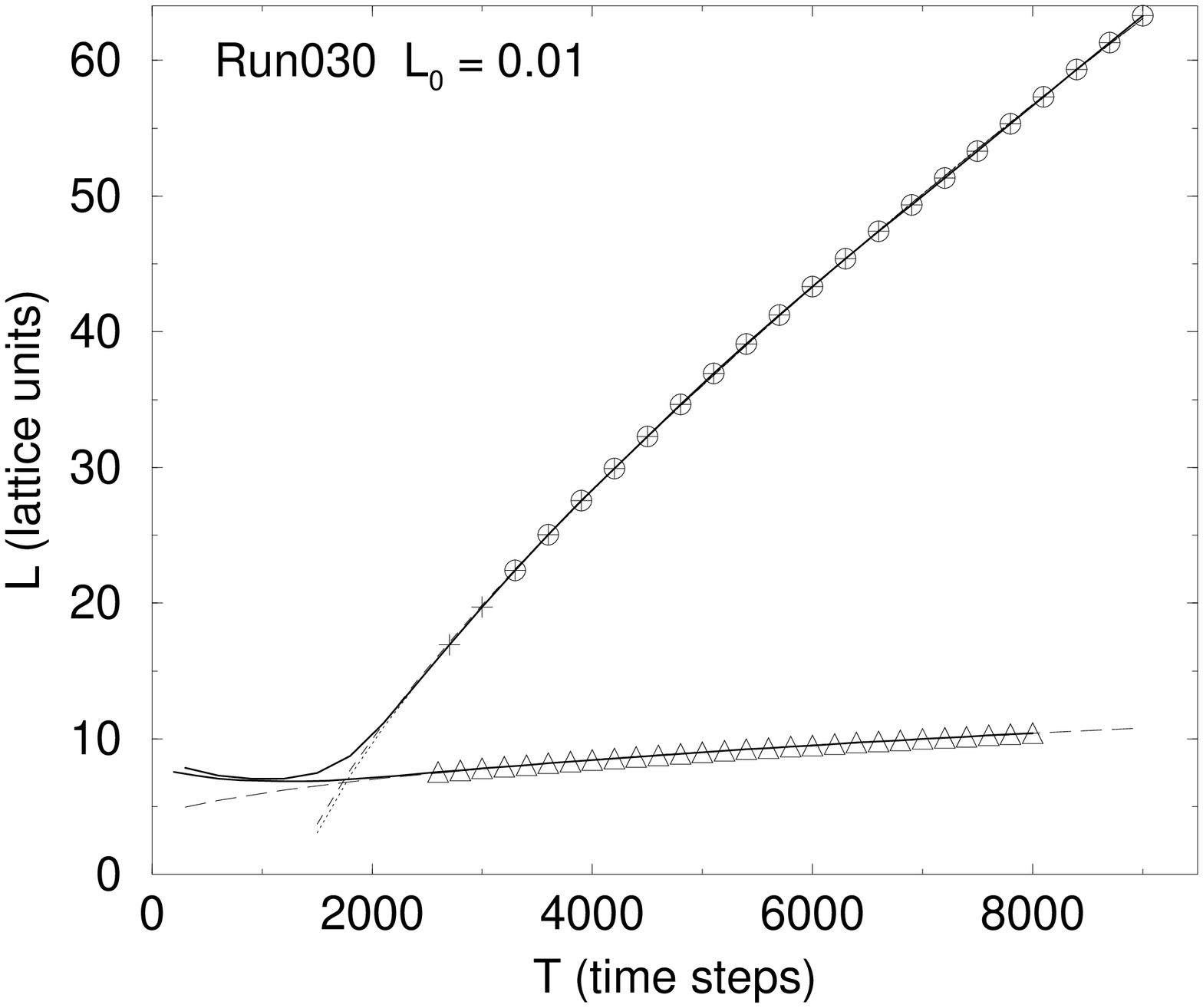}}}
    \end{minipage}
    \hfill \\
    \begin{minipage}{0.49\textwidth}
        \resizebox{\textwidth}{!}{\rotatebox{-90}{\includegraphics{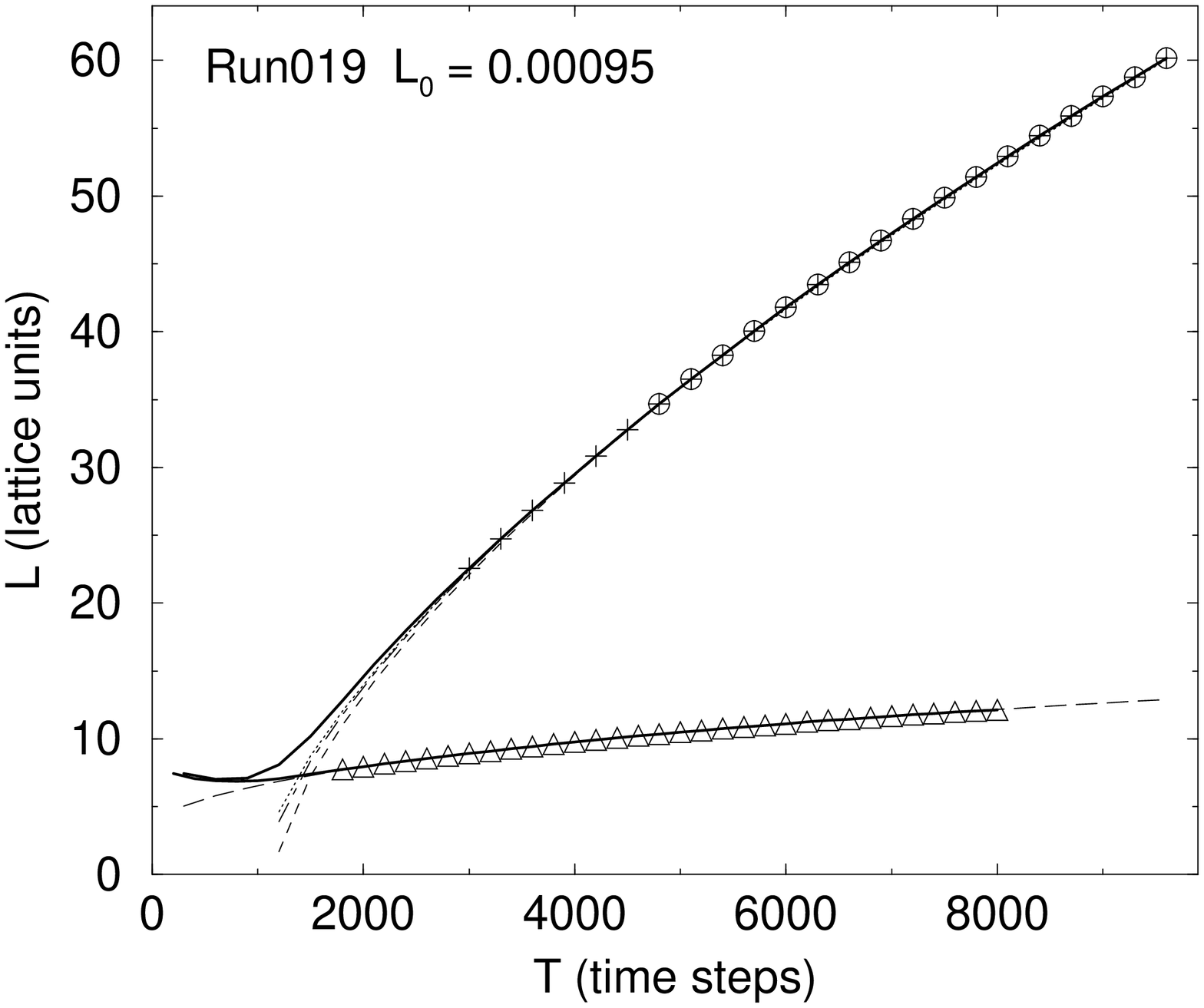}}}
    \end{minipage}
    \hfill
    \begin{minipage}{0.49\textwidth}
        \resizebox{\textwidth}{!}{\rotatebox{-90}{\includegraphics{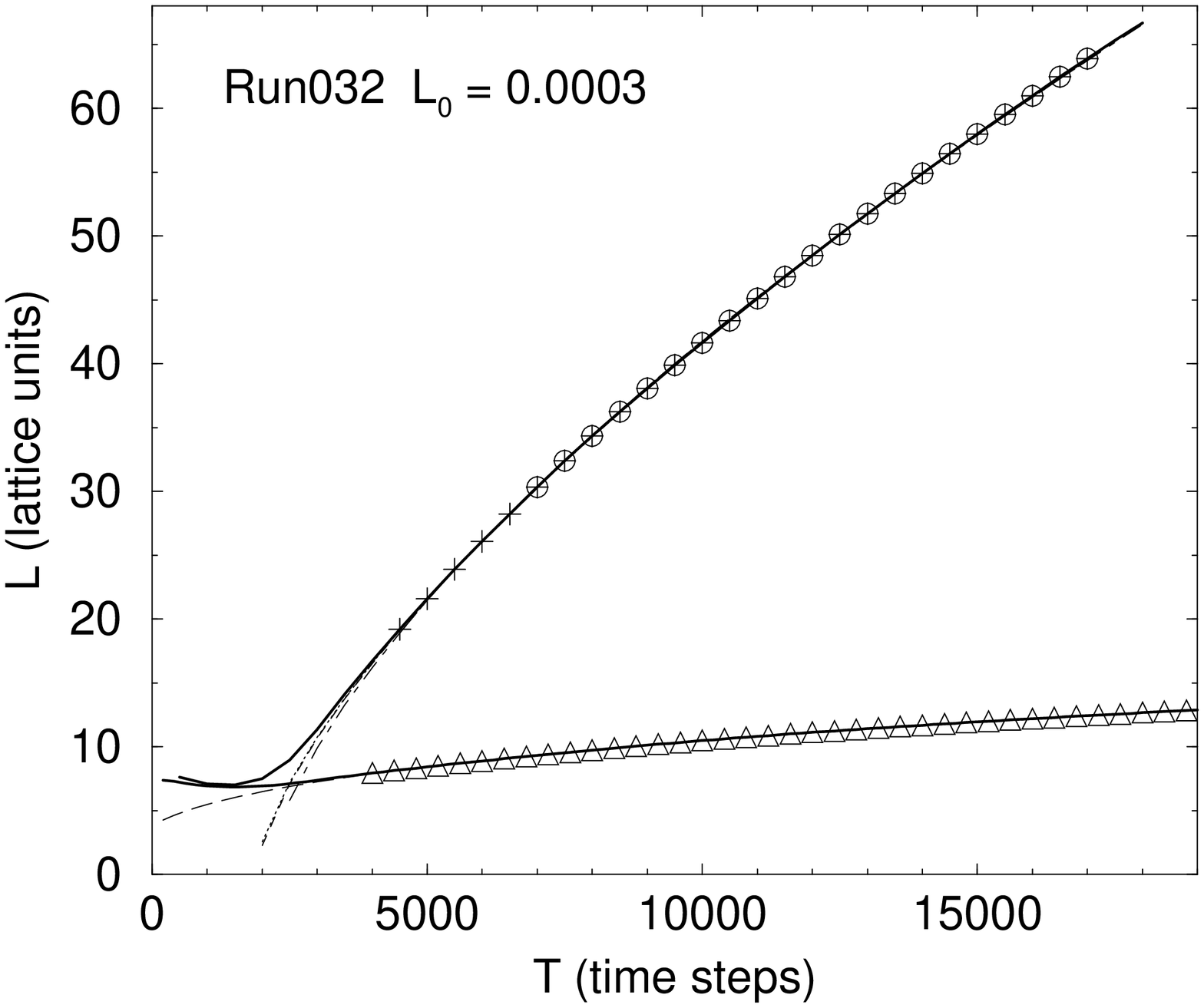}}}
    \end{minipage}
    \caption[Fit $L(T)$: $L_0 = 0.15$, $L_0 = 0.01$, $L_0 = 0.00095$ and
	     $L_0 = 0.0003$]
	    {\textit{Fitting $L(T)$ and $L_D(T)$.
	     Upper left:Run020, $L_0 = 0.15$.
	     Upper right: Run030, $L_0 = 0.01$.
	     Lower left:Run019, $L_0 = 0.00095$.
	     Lower right: Run032, $L_0 = 0.0003$.
	     Solid lines indicate full set of recorded $L(T)$ data,
	     $+$ indicates data points used for fits
	     with $L_{min}$ set by 2\% diffusion,
	     $\bigcirc$ indicates data points  used for fits
	     with $L_{min}$ set by 1\% diffusion,
	     $\bigtriangleup$ indicates data points used for
	     fits to diffusion-only data.
	     table \protect\ref{table:L_min} summarises the main fit results.}}
    \protect\label{graph:l_t_fit2}
\end{minipage}
\end{figure}

To parameterize the time dependence of $L(T)$, the `good data' was fitted, for each run separately, to the following form
\begin{equation}
L = v(T - T_{int})^{\alpha},
\label{eq:L_T}
\end{equation}
(equivalent to  (\ref{alphafit})) where $v$, $T_{int}$ and $\alpha$ are fitting parameters.
A nonlinear curve-fitting utility was used to create the fits,
which all fell within a specified tolerance of 1\%.
However, some trade-off is possible between the three fit parameters and a realistic uncertainty estimate for the exponent,
$\alpha$, is around 10\% for the first three runs in table \ref{table:pars_256},
and 5\% for the rest. 
The fits are shown in figures \ref{graph:l_t_fit1} and \ref{graph:l_t_fit2}, which also shows the diffusion-only data used to determine $L_{min}$ as described in \S\ \ref{sec:diffusion}. The fitted results are summarised in table \ref{table:L_min}.

%
\begin{table}
  \vspace{0.3cm}
  \begin{center}
  \begin{tabular*}{\textwidth}
{l@{\extracolsep{\fill}}
 r@{\extracolsep{0pt}}l@{\extracolsep{\fill}}
 r@{\extracolsep{0pt}}l@{\extracolsep{\fill}}
 r@{\extracolsep{0pt}}l@{\extracolsep{\fill}}
 r
 r@{\extracolsep{0pt}}l@{\extracolsep{\fill}}
 r@{\extracolsep{0pt}}l@{\extracolsep{\fill}}
 r@{\extracolsep{0pt}}l@{\extracolsep{\fill}}
 r@{\extracolsep{0pt}}l}
    & && \multicolumn{5}{c}{\textit{fits at 2\% diffusion}} &
         \multicolumn{2}{c}{\textit{fit}} &
         \multicolumn{4}{c}{\textit{$L_{min}$ at diffusion}} &
         \multicolumn{2}{c}{\textit{fit 1\%}} \\
    \textit{Run} & $L_0$  &&& $\alpha$ && $v$ & $T_{int}$ && $v_D$ &
        \multicolumn{2}{c}{2\%} &
        \multicolumn{2}{c}{1\%} &
        \multicolumn{2}{c}{$\alpha$}  \\[5pt]
    Run028 & 36&       & 0&.88 & 0&.0096 & 1948 & 0&.41 &\hspace{0.5em} 20&.0 &\hspace{0.5em} 28&.5 & 0&.81 \\
    \multicolumn{2}{r}{linear fit} &       & 1&.0  & 0&.00028&  516 &&&&&&&&\\
    Run022 &  5&.9     & 0&.86 & 0&.023  &  304 & 0&.64 & 26&.0 & 38&.0 & 0&.88 
\\
    \multicolumn{2}{r}{linear fit} &       & 1&.0  & 0&.00605& $-524$ &&&&&&&&\\
    Run033 &  5&.9     & 1&.16 & 0&.0012 &  442 & 0&.48 & 17&.5 & 24&.9 & 1&.12 
\\
    \multicolumn{2}{r}{linear fit} &       & 1&.0  & 0&.0060 & 1445 &&&&&&&&\\
    Run029 &  0&.95    & 0&.95 & 0&.0175 & 1020 & 0&.54 & 15&.3 & 21&.7 & 0&.92 
\\
    Run020 &  0&.15    & 0&.80 & 0&.0418 &  603 & 0&.60 & 23&.4 & 34&.9 & 0&.80 
\\
    Run030 &  0&.01    & 0&.75 & 0&.0747 & 1362 & 0&.51 & 14&.8 & 22&.4 & 0&.76 
\\
    Run019 &  0&.00095 & 0&.67 & 0&.134  & 1008 & 0&.60 & 21&.5 & 33&.8 & 0&.66 
\\
    Run032 &  0&.0003  & 0&.69 & 0&.0833 & 1855 & 0&.48 & 19&.0 & 29&.8 & 0&.69 
\\

  \end{tabular*}
  \end{center}
  \caption[Fits and lower cut-off, $L_{min}$, for $256^3$ runs]
          {Fits and lower cut-off, $L_{min}$, for $256^3$ runs. The parameter $v_D$ is the fit parameter corresponding to $v$ in the presence of diffusive growth only (see \S\ \ref{sec:diffusion}).}
  \protect\label{table:L_min}
  \vspace{0.1cm}
\end{table}

The data show $\alpha$ values ranging from 1.12 to 0.66 with a decreasing trend as $L_0$ is decreased. Certainly, an increasingly negative curvature of the $L(T)$ plots with decreasing $L_0$ is apparent from figures \ref{graph:l_t_fit1}, \ref{graph:l_t_fit2}. However, the resulting fit parameters were 
relatively erratic for the three runs of largest $L_0$ (expected to lie in the viscous regime). Indeed, we found $\alpha = 0.86, v = 0.023$ and $\alpha = 1.16, v = 0.0012$ for two runs with the same nominal $L_0$.
This was partly due to a relatively ill-conditioned fit as can be appreciated from figure \ref{graph:l_t_fit1}. (A second possible cause of the erratic fits
is the presence of long-range velocity fluctuations; see \S\ \ref{sec:finitesize}.)

Therefore it was decided to refit the data for the three most viscous runs, imposing $\alpha = 1$, the anticipated value. This yielded much better consistency among the fitted values of $v$, which with viscous scaling should obey $(T_0/L_0)v = b_1$, where $b_1$ is universal; with the forced linear fits this was indeed the case with $b_1$ extracted as $0.073, 0.072, 0.072 \pm 0.02$
for the three runs under discussion. Subject to this, we obtain a range
of values of $\alpha$ from 1.0 (Run028, Run022 and Run033) 
to 0.67 (Run019),
with intermediate exponents 0.95 (Run029), 0.80 (Run020)
and 0.75 (Run030) at intermediate $L_0$.  
This suggests that the simulations have indeed
covered the viscous, crossover and inertial regions.
However, the ultimate test of this is to convert to reduced physical units and construct the $l(t)$ curve explicitly.

\subsection{Universal scaling plot for $l(t)$}
\label{sec:scaled}

Our method for combining the data from different simulation
runs to give the $l(t)$ curve follows \cite{jury98a}.
As apparent from definitions  (\ref{defs}), the only fit parameter that is 
actually needed when converting $L(T)$ data to reduced physical units ($l(t)$)
is the intercept, $T_{int}$.  Then one uses the known density and viscosity, and the measured interfacial tension, to complete the conversion.

\begin{figure}
\begin{minipage}{\textwidth}
    \raggedright
    \resizebox{\textwidth}{!}{\rotatebox{-90}{\includegraphics{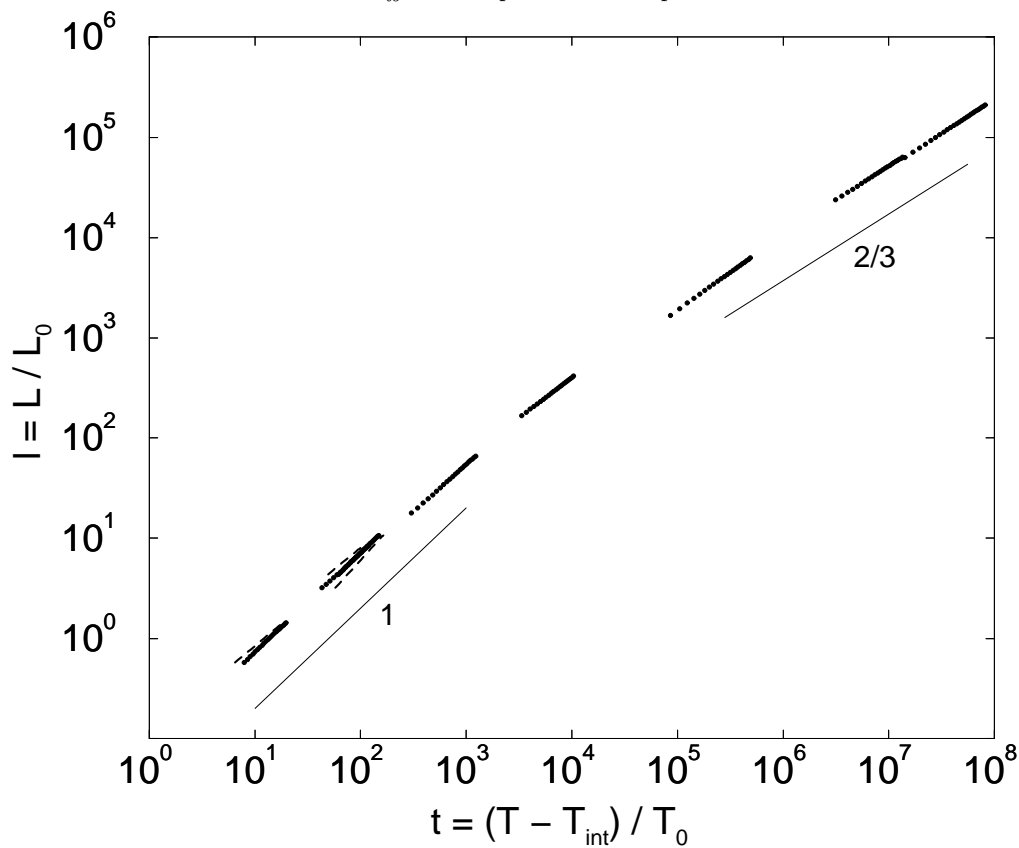}}}
    \caption[Scaling plot in reduced variables for $256^3$ LB data]
            {Scaling plot in reduced variables ($L/L_0$, $T/T_0$)
             for $256^3$ LB data.  Dots (left to right)
             are the runs in table \ref{table:pars_256} (top to bottom).
             Dashed lines show free exponent fits for the first three
             data sets in table \ref{table:pars_256} for comparison
             with linear fits (dots).}
    \protect\label{graph:l_t_fin}
\end{minipage}
\end{figure}

Figure \ref{graph:l_t_fin} shows
the $l(t)$ data from all the runs in table \ref{table:pars_256} combined on a single log-log plot. Note that for the two runs of $L_0 = 5.9$ the resulting data collapse is much improved by the forced linear fit (giving two very different values of the nonuniversal offset $T_{int}$ instead of two disparate values of $\alpha$). With the former, the two datasets overlie on the $l(t)$ plot but with the latter they do not; this helps to vindicate our choice of fit. Apart from a similar reservation about force fitting $\alpha$ for the most viscous run ($L_0 = 36$), the $l(t)$ curve is free of adjustable parameters. Although we did not have resources to cover the entire curve with data, there is no evidence for any breakdown of universality: the various runs do appear to lie on a smooth
underlying curve. (In particular, the two most inertial runs virtually join up.)

The apparently universal $l(t)$ curve shows scaling that is first linear ($l = b_1t$, with $b_1 = 0.072 \pm 0.02$), then passes through a broad crossover region before reaching $l = b_{2/3}t^{2/3}$ (with $b_{2/3} = 1.0 \pm 0.05$) at large $l$, $t$. The positions of the crossover and inertial runs on the graph are in keeping with the trend for the scaling exponent, $\alpha$, fitted directly from each run. This confirms that the exponents determined from our fitting
procedure do accurately reflect what is going on in these simulations.
The extreme breadth of the crossover regime, $10^2\lesssim t \lesssim 10^6$,
justifies the use of a single exponent to fit each run (\ref{alphafit}), (\ref{eq:L_T}) even in the crossover: no single run is long enough to see a change in exponent from beginning to end beyond the estimated errors.
There is no hint that the exponent is reducing
still further to $\alpha\le 1/2$, as predicted by \cite{grant99a},
although a further crossover beyond the range of $l,t$ reached in these simulations cannot be ruled out.  

Recall that intersection of asymptotes on the $l(t)$ plot defines $t^*$, the characteristic crossover time from viscous to inertial behaviour. As mentioned previously (\S\ \ref{sec:scaling1}), scaling theory says only that $t^*$ is `of order unity'. The measured value is close to $10^4$, a value that should raise no eyebrows in the turbulence community but may do so among workers in phase separation kinetics. Since $b_{2/3}$ is very close to $1$, the largeness of $t^*$ can be traced to the smallness of $b_1$ and to the relatively minor change in exponent on crossing from viscous to inertial scaling: for by its definition, $t^* = (b_{{i}}/b_{v})^{1/(\alpha_{v} - \alpha_{i})}$ where subscripts $i,v$ signify viscous and inertial values.

Note too, the huge range of scales covered by the combination of
eight simulation runs: five decades of length and seven decades of time.
This achievement is only possible by fully exploiting the expected scaling. This means that, although our work is capable of falsifying the scaling hypothesis (our $l(t)$ plot might not have joined up, and might yet not do so when more data is added), its non-falsification in our work may not represent persuasive proof that the scaling is true.  

For, as mentioned previously (\S\ \ref{sec:pars}), to navigate the $l(t)$ curve we are forced to correlate the simulation parameters in a systematic way. Hence if the coarsening rate was in fact dependent on $M$, say (for example by being pinchoff-limited, \cite{jury98a}), this would not \textit{necessarily} show up as bad data collapse in figure \ref{graph:l_t_fin}, since $M$ is strongly correlated with $L_0$ and/or $T_0$. In principle, however, our parameter steering has no effect on data within a run, so that any `steering-induced' data collapse could in principle be detected because curves would not quite line up with their neighbours on the plot, although their midpoints would lie on a smooth curve \citep{jury98a}. Although we believe this is not happening for our $l(t)$ data, we do not have enough results to entirely rule it out, especially as something similar \textit{does} occur in our own velocity derivative data
(\S\ \ref{sec:velocity_derivatives}).

Our results for the evolution of the interfacial structure are compared with those of previous authors in Appendix \ref{sec:other_work}.

\section{Results for the fluid velocity}
\label{sec:velocity}

\begin{figure}
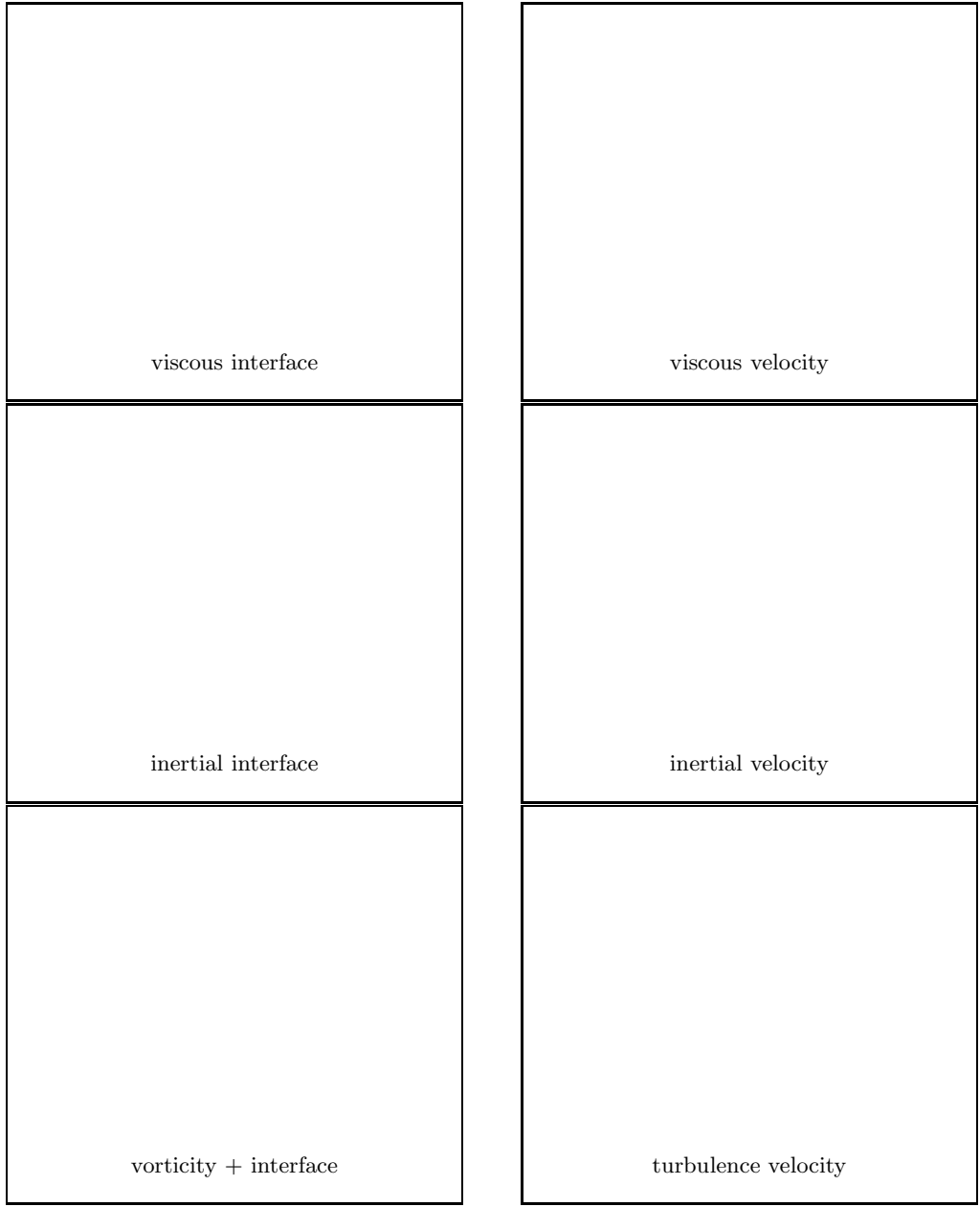

\begin{center}
    \framebox[0.47\textwidth]{\rule[-1em]{0em}{5.3cm}viscous interface}
    \hfill
    \framebox[0.47\textwidth]{\rule[-1em]{0em}{5.3cm}viscous velocity}
    \\
    \framebox[0.47\textwidth]{\rule[-1em]{0em}{5.3cm}inertial interface}
    \hfill
    \framebox[0.47\textwidth]{\rule[-1em]{0em}{5.3cm}inertial velocity}
    \\
    \framebox[0.47\textwidth]{\rule[-1em]{0em}{5.3cm}vorticity + interface}
    \hfill
    \framebox[0.47\textwidth]{\rule[-1em]{0em}{5.3cm}turbulence velocity}
\end{center}
    \caption{Pictures of interface and velocity maps in viscous and inertial
             regimes.  Top row: interface (left) and velocity field (right)
	     for Run027 in the viscous region at time step 8500, when the
	     domain size is about 21 lattice units.
	     Middle row: interface (left) and velocity field (right)
	     for Run031 in the inertial region, for time step 5000 when the
	     domain size is about 22 lattice units.
	     Only a $32^3$ section of the simulation is shown (the same
	     section for both interface and velocity).
	     The velocity is shown for the front two lattice planes only
	     for clarity.  Arrow colours indicate speed (redder is faster).
	     Bottom row: (left) interface (blue) for Run031 with the red
	     interface at a contour of 50\% of the
	     maximum vorticity and thus enclosing regions of high vorticity.
	     Velocity from single fluid turbulence (right).
             Turbulence has Re$_{\lambda} \simeq 10$,
	     matching that of the most inertial run, Run031.
	     The vorticity for single fluid turbulence (not shown) is
	     qualitatively similar to that shown for Run031 (left).}
    \protect\label{fig:vel_fields}
\end{figure}

Due to data storage limitations for the largest ($256^3$) runs,
our velocity analysis also made extensive use the $128^3$ runs listed in table \ref{table:pars_128}. 
The velocity field was analysed as a single, continuous field, filling
the whole simulation; there is no explicit information about
the location of the interface between the two fluid phases.  
However, visualisation of the velocity field was done using the AVS package
and examples (one viscous, one inertial) are shown in figure \ref{fig:vel_fields}, where the
flow patterns can be compared with the domain structure defined by
the interface.

There are almost no prior data on fluid mixtures with which to compare these results. A simulation using a pseudospectral method written
by \cite{young99a} was therefore used to generate a velocity field
for single fluid, freely decaying turbulence with a similar Reynolds number to those of the spinodal system in the inertial regime. A velocity map for this single fluid turbulence is also pictured in figure \ref{fig:vel_fields}.

A trend from locally laminar flow to more chaotic motion is
apparent in passing from the viscous to the inertial regime. 
The vorticity map in the latter case is comparable to the one for the turbulent single fluid (not shown).
However, the comparison is hindered by the fact that the interfacial motion at length scale $L$ introduces a `whorly' velocity pattern even in the purely viscous flow regime. A better discriminator between the two regimes, pursued elsewhere, comes from watching the time evolution of the interfacial structure itself, which is clearly underdamped in the inertial case, \cite{desplat00a}.

\subsection{Velocity structure factor}
\label{sec:s_k_v}

The velocity structure factor was introduced in \S\ \ref{sec:vel_comp}. For numerical purposes we define it (following  (\ref{eq:sk})) as
\begin{equation}
S_v(k) =  \frac{1}{n_k}\sum_{k-\frac{\upi}{\Lambda} < |\mathbf{k}| < k+\frac{\upi}{\Lambda}}
	  \mathbf{v}(\mathbf{k}).\mathbf{v}(\mathbf{-k})\;.
\label{eq:svk}
\end{equation}
The results were shown already in figure \ref{graph:vel_comp}, where $S_v(k)$ is depicted for three of the
runs in table \ref{table:pars_128}, alongside two calculations (LB and spectral) for single fluid turbulence. These structure factors are in unscaled units but in each case correspond to a point during the run where the domain size is around 30 lattice units.

The bumps on the $S_v(k)$ curves at high $k$ were discussed in \S\ \ref{sec:vel_comp}. But even apart from these, the velocity structure factors have very different shapes in the viscous (Run027), crossover (Run018), and inertial (Run031) regimes; these differences are much larger than for the order parameter structure factor (figure \ref{graph:s_k}). In other words, the geometry of the fluid flow is changing much more significantly, as one moves along the $l(t)$ curve, than the geometry of the interface. 

We return to this in \S\ \ref{sec:vel_len}, but first address an issue raised in \S\ \ref{sec:finitesize}, which is the apparent $k^{-2}$ divergence in $S_v(k)$ at low $k$ in the viscous regime. 
This can be qualitatively explained as follows. In a purely viscous approximation (Stokes flow) the NSE (\ref{eq:nse}) becomes in Fourier space
\begin{equation}
\eta k^2\mathbf{v(k)} = -i\mathbf{k}. \mbox{\boldmath{$\cal P$}}^{th}(\mathbf{k}).
\label{eq:nsestokes}
\end{equation}
Here $\mbox{\boldmath{$\cal P$}}^{th}$ contains the chemical term $\mbox{\boldmath{$\cal P$}}^{chem}$ which is mainly localised on the interface between the two fluids. We now argue that this term is strongly correlated at length scales up to the domain size $L$ but not larger ones: this means that, for the purposes of low wavenumbers ($k \ll \upi/L$) it is a random variable with short range correlation. Ignoring for simplicity all tensor indices, one thus has
$\langle |\mbox{\boldmath{$\cal P$}}^{th} (\mathbf{k})|^2\rangle \to \chi$, a constant, as $k\to 0$.
From  (\ref{eq:nsestokes}) we find immediately
\begin{equation}
S_v(k) \sim \frac{\chi}{\eta^2 k^2}.
\label{eq:divergence}
\end{equation} 
Thus the long range, Stokesian hydrodynamic propagation converts short range fluctuations in $\mbox{\boldmath{$\cal P$}}^{th}$ into long range fluctuations in the fluid velocity.
As mentioned in \S\ \ref{sec:finitesize}, the resulting divergence could lead to erratic coarsening rates and/or problems with finite size effects, throughout the viscous regime. This appears not to have been noticed by previous authors.

There is a somewhat related anomaly that arises in colloidal suspensions under gravity, although in that case the short range fluctuations are in the density, 
which is effectively a random body force, rather than in a random stress: \citet*{segre97a}. 

\subsection{Length scales from the velocity field}
\label{sec:vel_len}

The velocity structure factor, $S_v(k)$, can be used to calculate
a velocity length scale, $L_v(T)$ analagous to $L(T)$ (compare  (\ref{eq:Ldef})):
\begin{equation}
L_v(T) = 2\upi\frac{\int S_v(k,T) \mathrm{d}k}{\int k S_v(k,T) \mathrm{d}k}.
\label{eq:fmv}
\end{equation}

\begin{figure}
\begin{minipage}{\textwidth}
    \raggedright
    \begin{minipage}{0.48\textwidth}
        \resizebox{\textwidth}{!}{\rotatebox{-90}{\includegraphics{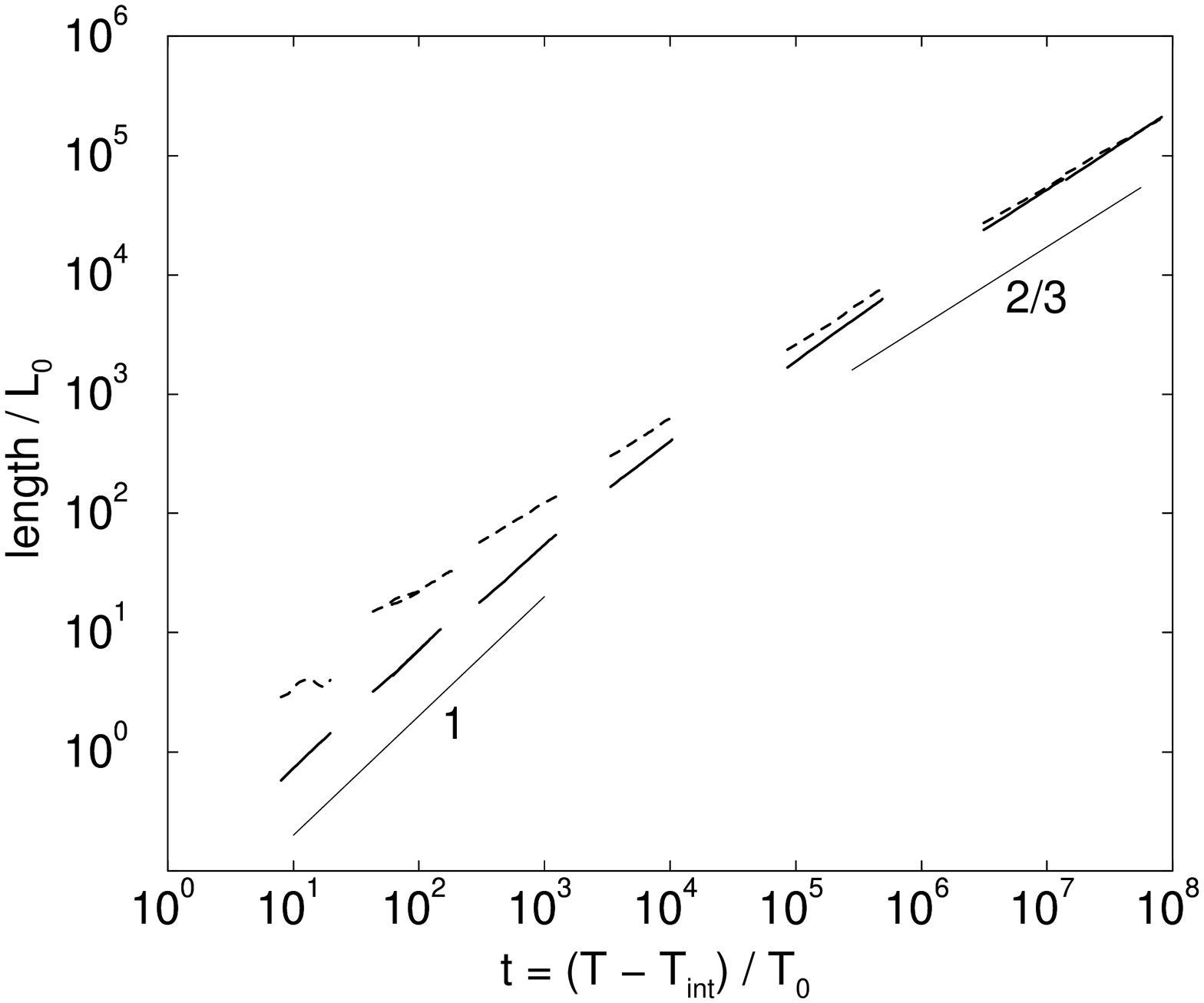}}}
    \end{minipage}
    \hfill
    \begin{minipage}{0.48\textwidth}
        \resizebox{\textwidth}{!}{\rotatebox{-90}{\includegraphics{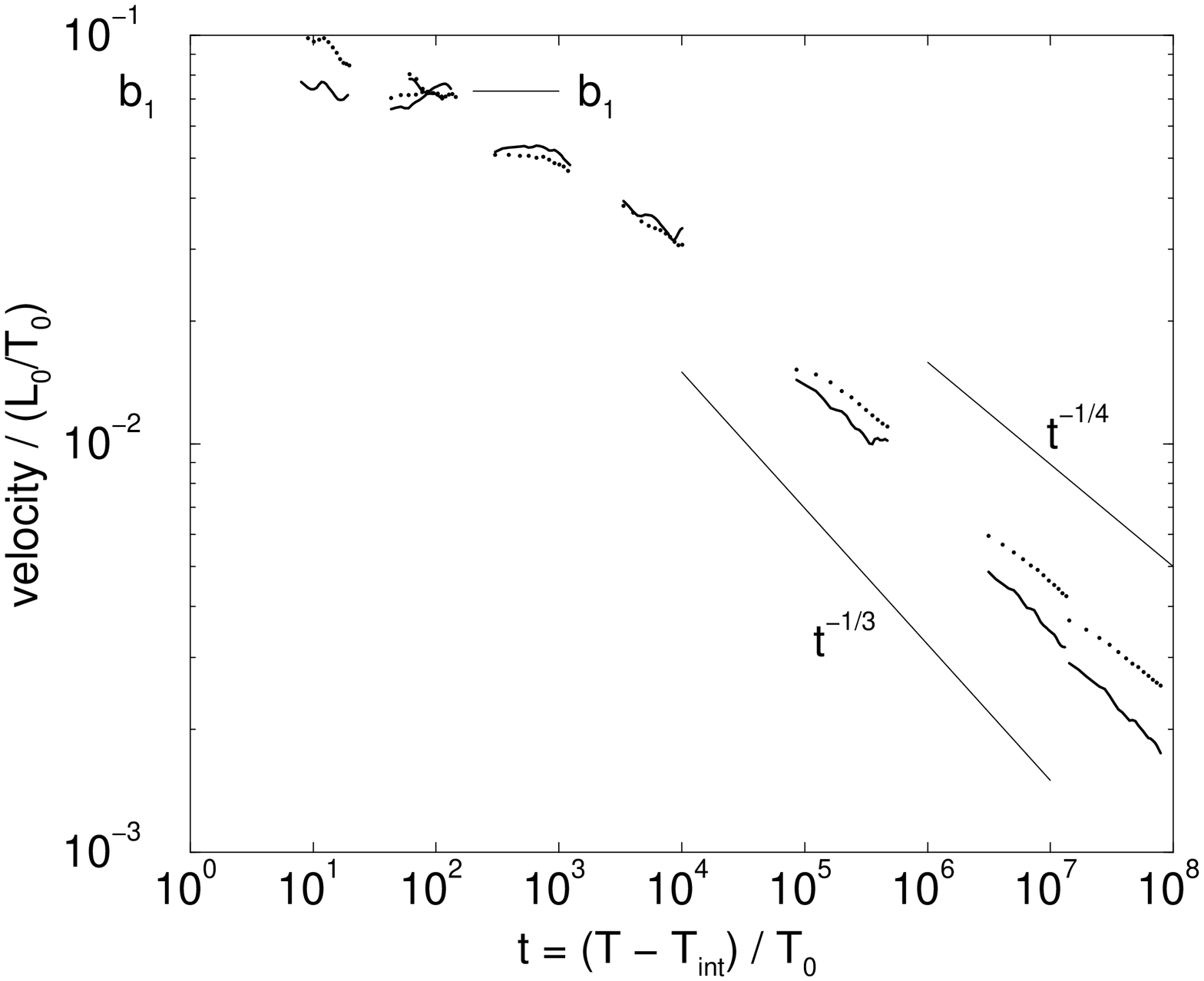}}}
    \end{minipage}
    \vspace{1ex}
    \caption {Left: $L_v(T)$ (dashed) compared with $L(T)$ (bold)
	     for runs in table \protect\ref{table:pars_256}.
             Right: For runs in table \protect\ref{table:pars_256},
	     reduced interface velocity, $\dot l(t)$ (bold)
	     compared with the {\sc rms} fluid velocity measured in
             reduced physical units (dotted).}
    \protect\label{graph:l_t_final_vel}
\end{minipage}
\end{figure}

This length measure was found to be insensitive to coarse-graining 
in nearly all cases. 
Data collected for $L_v(T)$ from the $256^3$ runs was converted to reduced physical units, using the values of $T_{int}$ already obtained from the $L(T)$ fits and given in table \ref{table:L_min}.
(Hence no further fitting was involved.)
The resulting scaling plot is shown, alongside the $l(t)$ data presented earlier, in figure \ref{graph:l_t_final_vel} (left).
The results in the inertial regime show a strong convergence between $l_v \equiv L_v/L_0$ and $l \equiv L/L_0$: the velocity length $l_v$ shows the same $t^{2/3}$ scaling as $l$, with a similar prefactor. This is not obvious \textit{a priori}, since, as mentioned above, the shapes of $S_v(k)$ and $S(k)$ are very different.

More surprisingly, we find that to a fairly good numerical approximation, $l_v(t)$ shows a $t^{2/3}$ growth throughout the crossover region, and that this even extends far into the viscous regime, within which $l_v$ exceeds the domain scale $l$ by a significant factor. However, as the viscosity is increased to access the bottom left corner of the plot, the data are
increasingly affected by finite size effects, since $L_v$ then is comparable to the system size $\Lambda$. These are especially pronounced for the most viscous run (with $L_v$ almost constant during that run). 
Allowing for these effects, the data is consistent with $l_v \sim t^{2/3}$ at all times; however we have no reason to expect this result in the viscous regime, where both the 
simple and the extended scaling analyses (\S\ \ref{sec:scaling1}, \ref{sec:scaling2}) predict instead $l_v \sim l \sim t$.
Note, though, that the velocity length measure chosen (\ref{eq:fmv}) is sensitive to the low $k$ divergence found above in $S_v(k)$,
which would give a contribution of order $\Lambda/\ln \Lambda$ from the lower limits of integration. 
It is possible that for the parameters and system sizes used here, this size-limited contribution combines with those from higher wavevector to give an apparent $2/3$ power in the viscous regime. 

Alternative length measures may be had by taking the ratio of two other successive moments of $S_v(k)$, to replace  (\ref{eq:fmv}). 
Adding one extra power of $k$ to the top and bottom integrand gives
a length measure that lies between $l_v$ and $l$ throughout the viscous regime, reducing the exponent discrepancy there, but without attaining the linear scaling of $l$ itself. For more than two extra powers of $k$ the scaling gets worse, not better, as the integral in the denominator becomes dominated by high $k$ contributions. 

\subsection{Average velocities}
\label{sec:vel_com}

The {\sc rms} fluid velocities (spatially averaged) were calculated for all the runs in table \ref{table:pars_256} and are plotted in reduced physical units in
figure \ref{graph:l_t_final_vel} (right), alongside the reduced interface velocity $\dot l(t)$ derived from the order parameter.

In the most viscous run, Run028, the {\sc rms} fluid velocity is 
larger than the interface velocity.  Both velocities are fluctuating
quite far from the expected constant behaviour in the linear region,
and the fluctuations are more or less in step. This may in part be a facet of the erratic, finite-size limited behaviour seen in the far viscous regime (\S\ \ref{sec:finitesize}). 

Otherwise we observe that the {\sc rms} fluid velocity 
matches the interface velocity in the viscous and early crossover region,
but grows larger than it in the
inertial region, by about 40\% at the largest $l,t$.
The two most inertial runs (Run019 and Run032) appear to have the {\sc rms} velocity 
scaling with a slightly different exponent than the interface velocity (approximately as $t^{-1/4}$ rather than $t^{-1/3}$), though this may not be significant. Such a deviation is foreseen by neither the simple nor the extended scaling theory, both of which have velocities scaling as $v \sim \dot l$ at all times. It would imply a buildup of kinetic energy in the fluid beyond that predicted by either scaling analysis. The excess may be caused by our approaching the limits of numerical accuracy in resolving velocity gradients with a consequent breakdown in energy conservation (see \S\ \ref{sec:diss_rate},
\ref{sec:nonscale} below). A similar breakdown, caused instead by having too high a viscosity in lattice units (as indicated in \S\ \ref{sec:pars}), may likewise contribute to the excess {\sc rms} velocity seen in the most viscous run.

\subsection{Velocity distributions}

The {\sc pdf} of the velocity components in a homogeneous, isotropic
turbulent fluid is known to be almost Gaussian,
and uncorrelated over large distances in both space and time.
Non-Gaussian behaviour 
is found only in velocity increments and derivatives, \citep[e.g.][]{monin75a}.

The spinodal system is crucially different: it has a structural length scale $L(T)$, and correlations and inhomogeneity can be expected at this scale.
Hence the velocity components themselves can show non-Gaussian {\sc pdf}s.
The departures are expected to show up through the fourth moment, since there is no preferred direction in space that would allow one to create a scalar from a third moment of velocity. (The third moment was checked and found to be close to zero.)
The fourth moment is characterised by
the flatness,
$\langle v_{\alpha}^4 \rangle/\langle v_{\alpha}^2\rangle^2 -3$, with $\alpha$ a Cartesian index; this
vanishes for a Gaussian distribution.

\begin{figure}
\begin{minipage}{\textwidth}
    \raggedright
    \begin{minipage}{0.49\textwidth}
        \resizebox{\textwidth}{!}{\rotatebox{-90}{\includegraphics{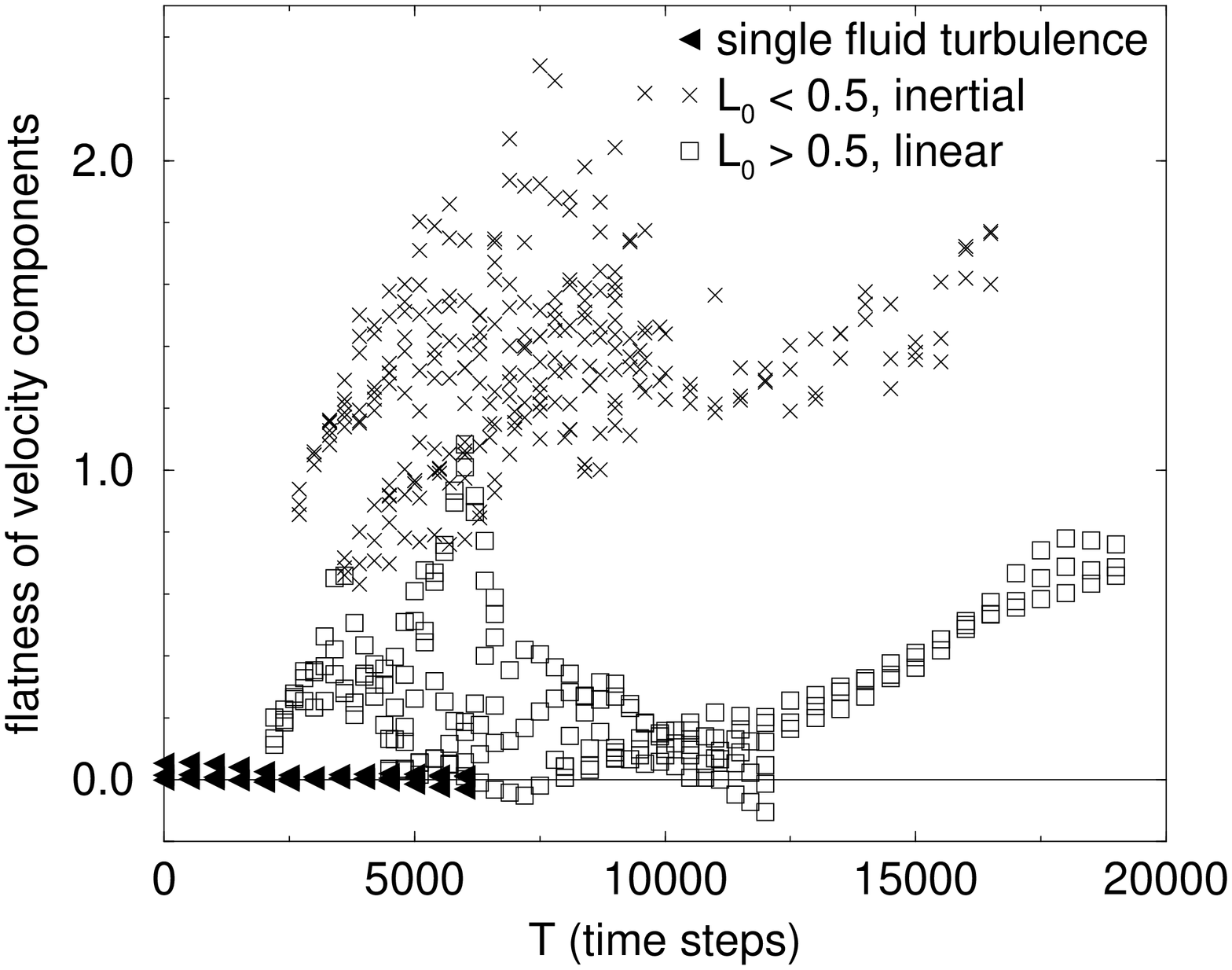}}}
    \end{minipage}
    \hfill
    \begin{minipage}{0.49\textwidth}
        \resizebox{\textwidth}{!}{\rotatebox{-90}{\includegraphics{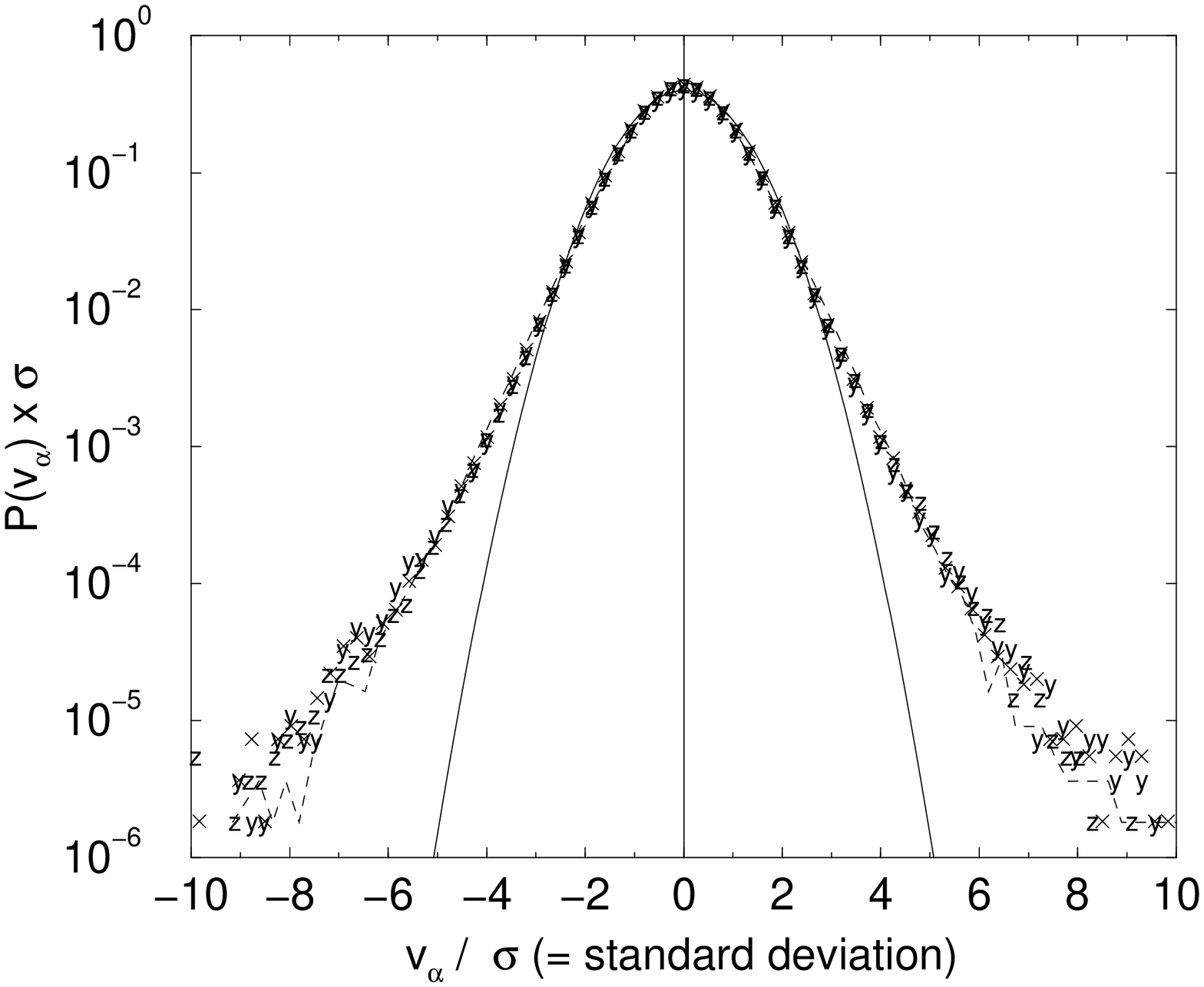}}}
    \end{minipage}
    \vspace{1ex}
    \caption[Velocity statistics, flatness and pdf]
            {Left: flatness of velocity components for runs in 
	     table \protect\ref{table:pars_256}; runs with 
	     $L_0>0.5$ (crosses) and with $L_0<0.5$ (squares). Single fluid
	     turbulence (filled triangles) are shown for comparison.
	     Right: {\sc pdf}s of velocity components for Run032 at time step 12000,
	     with Gaussian {\sc pdf} (solid) and transverse velocity derivative (dashed)
	     shown for comparison.}
    \protect\label{graph:vel_stats}
\end{minipage}
\end{figure}
Figure \ref{graph:vel_stats} (left) shows the flatness for runs in table
\ref{table:pars_128} as a function of time; a distinction is drawn between viscous runs (those with $L_0<0.5$)
and crossover/inertial runs (with $L_0>0.5$).  
It can be seen that the flatness is quite variable,
but as a general trend it grows slightly with time through each
run, and also grows with decreasing $L_0$ (increasing inertia).  
The velocity {\sc pdf}s show correspondingly wider tails and
narrower peaks than for a Gaussian; an example for Run032 is shown in 
figure \ref{graph:vel_stats} (right).  The shape is close to that found
in the transverse velocity derivatives in the same system
(shown dashed for comparison). These non-Gaussian effects are much more pronounced in
the inertial than in the viscous regime.

\section{Velocity derivatives}
\label{sec:velocity_derivatives}

We now turn to the analysis of spatial derivatives of the velocity field.
Velocity derivatives come in two types, longitudinal, e.g.,
$\mathrm{d}v_x/\mathrm{d}x$, and transverse, e.g.,
$\mathrm{d}v_x/\mathrm{d}y$ or $\mathrm{d}v_x/\mathrm{d}z$.
Representative velocity derivatives (in fact, $\mathrm{d}v_x/\mathrm{d}x$, $\mathrm{d}v_y/\mathrm{d}y$, and
$\mathrm{d}v_y/\mathrm{d}x$) were calculated. The differentiation was done by 
calculating $ik_xv_y(\mathbf{k})$ etc., and Fourier transforming back to real space. The derivatives are unambigous so long as the gradients of velocity are small on the scale of the lattice spacing, but otherwise the resulting gradient is
not the same as taking a lattice derivative (which we define via  (\ref{eq:grad_def}) with $\mathbf{v}$ replacing $\phi$). 
In practice we have found signigicant ($\sim 40\%$) discrepancies 
the two methods; these were investigated in the context of energy conservation 
and are discussed further in \S\ \ref{sec:diss_rate}.
For many purposes 
there is no particular reason to prefer the lattice to the Fourier definition
and we retain the latter for simplicity. 

\subsection{Skewness of velocity distribution}
\label{sec:vel_skew}

In fully-developed turbulence the skewness of the longitudinal velocity derivatives approaches $-0.5$, \citep[e.g.,][]{monin75a}. (The skewness of a variable $y$ is $\langle y^3 \rangle/\langle y^2 \rangle^{3/2}$.)
The transverse derivatives have zero skewness, by symmetry, but
positive flatness.

\begin{figure}
\begin{minipage}{\textwidth}
    \raggedright
    \begin{minipage}{0.49\textwidth}
        \resizebox{\textwidth}{!}{\rotatebox{-90}{\includegraphics{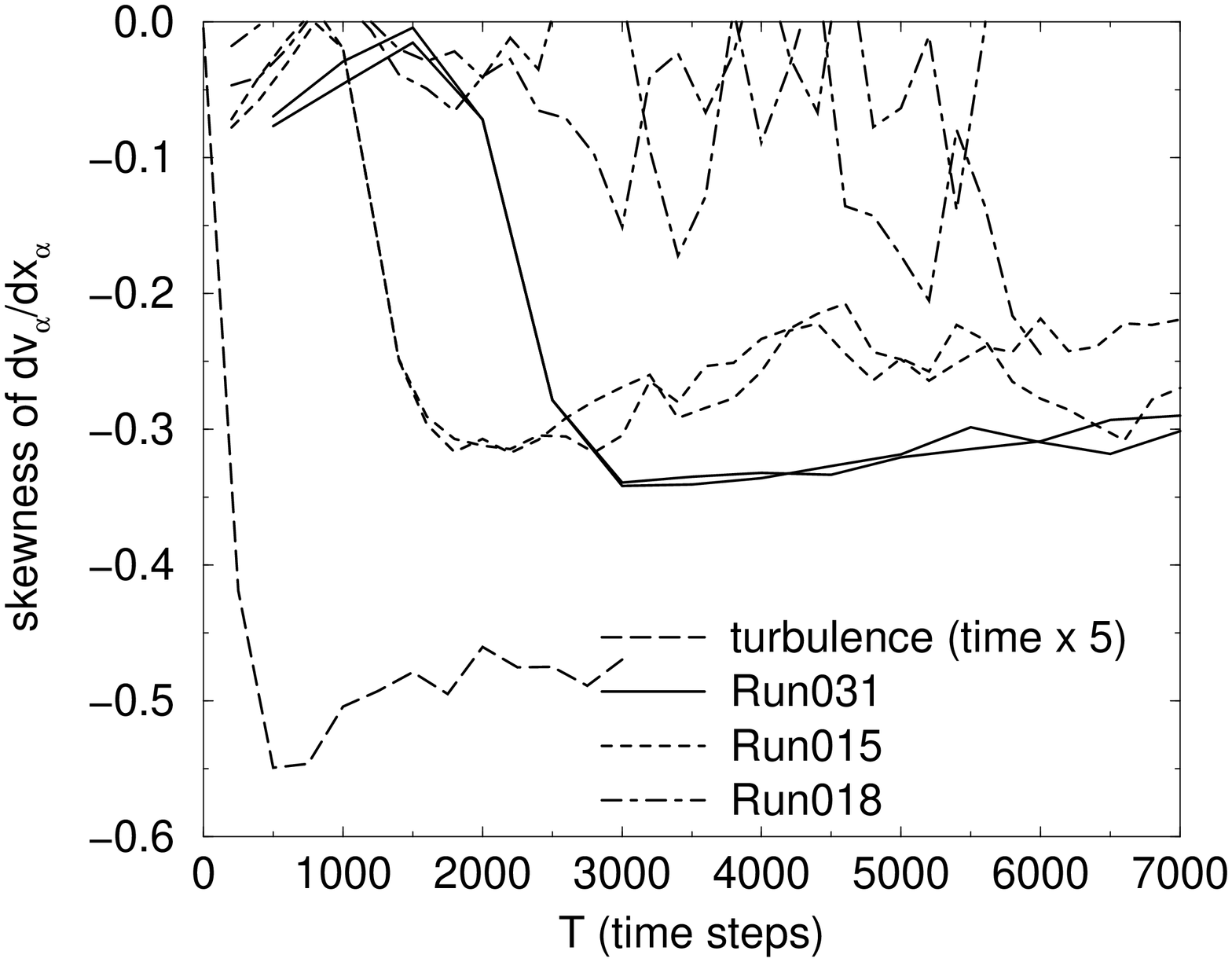}}}
    \end{minipage}
    \hfill
    \begin{minipage}{0.49\textwidth}
        \resizebox{\textwidth}{!}{\rotatebox{-90}{\includegraphics{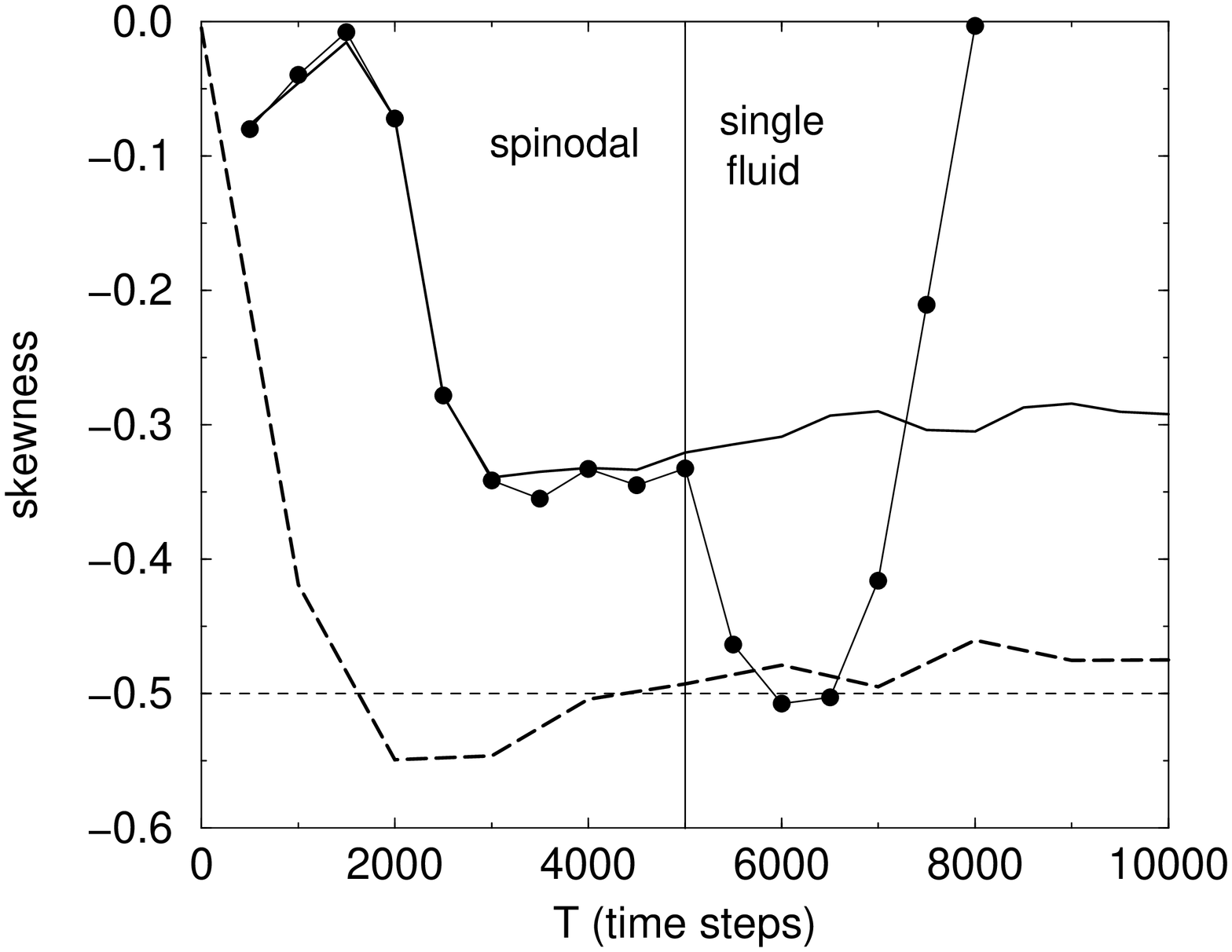}}}
    \end{minipage}
    \vspace{1ex}
    \caption[Velocity derivatives: skewness]
            {Left: Skewness of the longitudinal velocity derivatives for
	     Run031 (solid), Run015 (dashed), Run018 (dot dashed)
	     and single fluid decaying turbulence (long dashed).
	     Right: Skewness of a longitudinal velocity derivative for
	     Run031 (solid), a $96^3$ run with the same parameters in
	     which the interface was removed at time step 5000 (filled circles),
	     and single fluid turbulence (dashed).
	     The time scale for the turbulence data has been multiplied
	     by 5 (left) or 10 (right) to facilitate comparison with
	     the LB data.  (The time dependence is not of relevance here
	     once the initial stages are passed in both simulation methods.)}
    \protect\label{graph:skewness}
\end{minipage}
\end{figure}

Figure \ref{graph:skewness} (left) shows the skewness of the 
longitudinal velocity derivatives against time for the three most inertial runs
in table \ref{table:pars_128}.  Also shown for comparison
is the skewness from the freely decaying turbulence simulation.
In the two most inertial LB simulation runs,
the skewness of the longitudinal velocity derivative reaches around $-0.35$.
A plausible interpretation of this result is that patches of turbulence
arise, but do not fill the whole system; if the patches have skewness $-0.5$, the overall value is less.  From visualizations (see figure \ref{fig:vel_fields}) we know that the interface remains smooth so that any turbulent regions should be in the middle of the fluid domains. 

To test this idea further, a $96^3$ run with the same parameters
was done, and at timestep 5000 (when the domain size was about 21,
after residual diffusion had decayed to an acceptable level) the interface was suddenly removed by setting the
order parameter to 1 throughout the system. This converted it into a single
fluid with the same velocity field, which was then allowed to evolve.  The
longitudinal velocity derivative skewness for this run is shown in figure
\ref{graph:skewness} (right), with Run031 and the single fluid
turbulence data repeated for comparison.
Once the interface was removed, the skewness quickly jumped to around $-0.5$ from $-0.35$, providing
strong support for the `turbulence in patches' hypothesis; it appears that on removal
of the interface, the turbulence rapidly infects the whole system.
(Fairly soon after this, the system became numerically unstable causing the skewness to rise rapidly back toward zero as spuriously large
velocities were generated; this is visible in figure \ref{graph:skewness}, but was not investigated further.)

\subsection{Reynolds numbers}
\label{sec:spin_reyn}

In this work we choose to regard all Reynolds numbers as estimates of the {\sc rms} ratio, $R_2$, of the nonlinear term to the viscous term in the NSE (\ref{eq:nse}), as defined in  (\ref{eq:R_2t}). 

In practice, Reynolds numbers are usually constructed (via $\Rey=T_0\ell v/L_0^2$) from a characteristic length ($\ell$) and a characteristic velocity ($v$). 
In a homogeneously turbulent single fluid, 
the usual choice of velocity scale is $v_{\mbox{\sc rms}}$.
Also, the length measure $\ell$ must itself
be constructed from the velocity field; various choices then arise. 
One is the integral scale, defined (to within prefactor conventions)
as $L_{int}=(3\upi^2\rho/2E)\int kS_v(k)\mathrm{d}k$
where $E = 2\upi\rho\int k^2S_v(k)\mathrm{d}k$ is the total kinetic energy. (This a close relative of our $L_v$,  (\ref{eq:fmv}.)) A second is the Taylor microscale, $\lambda$, defined in Eq. (\ref{eq:microscale}).
A third length is the Kolmogorov microscale, $\lambda_d$,  (\ref{eq:k_d}),
but this is not normally used to form a Reynolds number.
In practice, most isotropic homogeneous turbulence simulation studies use as Reynolds number $\Rey_{\lambda} = \rho v_{\mbox{\sc rms}}\lambda/\eta$. This should give a reasonable estimate of the ratio $R_2$,
because as noted in \S\ \ref{sec:scaling2},
$\lambda$ is the length scale associated with the $\bnabla$ operator in 
$\bnabla\mathbf{v}$. But note that, according to the extended scaling analysis (table \ref{table:predict}), it is still not an accurate estimate asymptotically at late times: there one has the prediction that $\Rey_\lambda \sim \lambda v \sim T^{1/6}$ while $R_2 \sim T^0$. 

In the spinodal system an obvious alternative to $\Rey_\lambda$ arises from choosing $L(T)$, the domain size, 
as the length scale; either $v_{\mbox{\sc rms}}$ or the interface velocity $\dot L$ can then be used for the velocity scale. However, the resultings Reynolds numbers,
$\Rey_L = \rho L v_{\mbox{\sc rms}}/\eta$, 
and
$\Rey_\phi = \rho L\dot L/\eta$ (see  (\ref{eq:Re_phi})) are not directly comparable with ${Re}_\lambda$, which is always much smaller. 

\begin{figure}
\begin{minipage}{\textwidth}
    \raggedright
    \begin{minipage}{0.49\textwidth}
        \resizebox{\textwidth}{!}{\rotatebox{-90}{\includegraphics{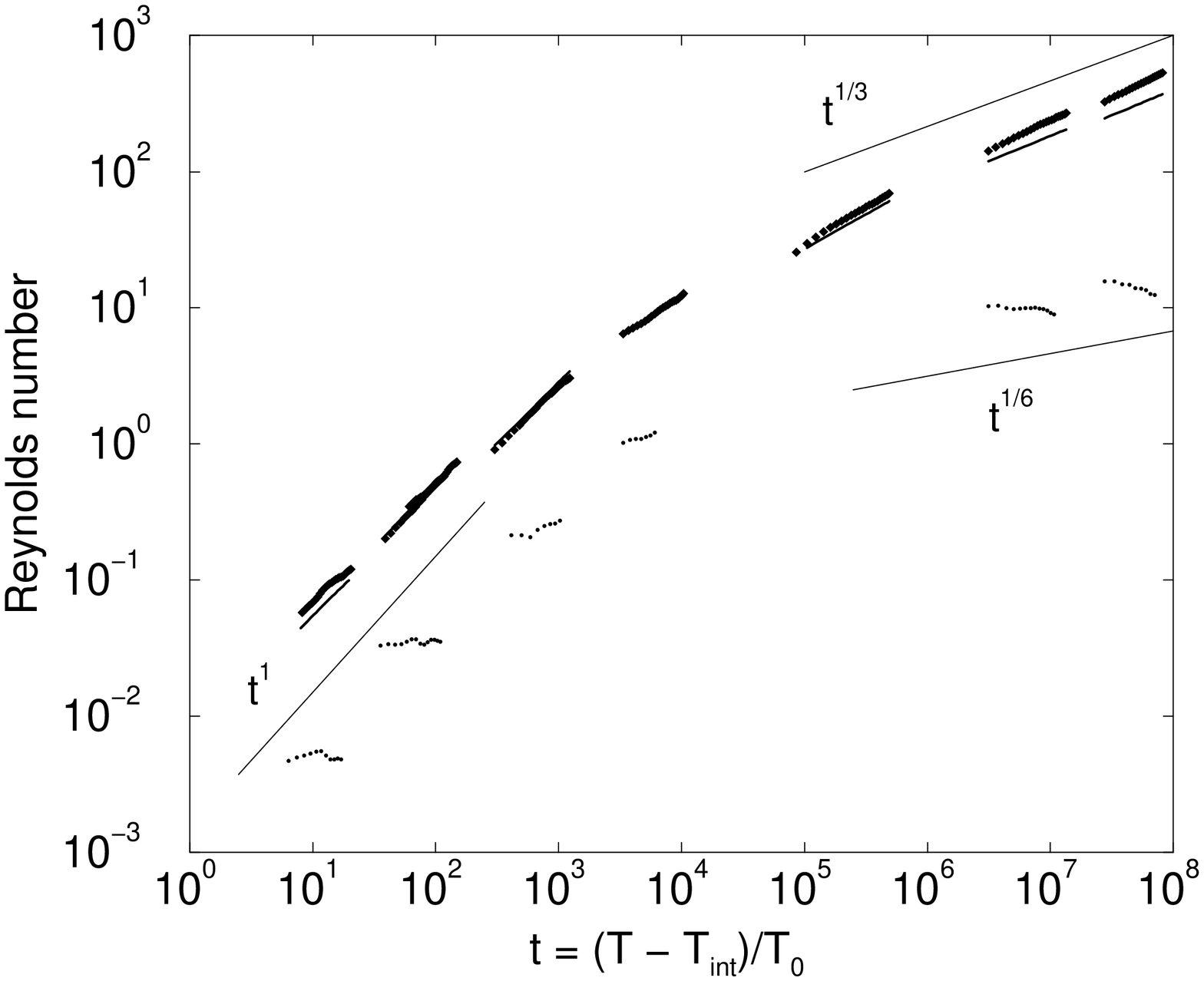}}}
    \end{minipage}
    \hfill
    \begin{minipage}{0.49\textwidth}
        \resizebox{\textwidth}{!}{\rotatebox{-90}{\includegraphics{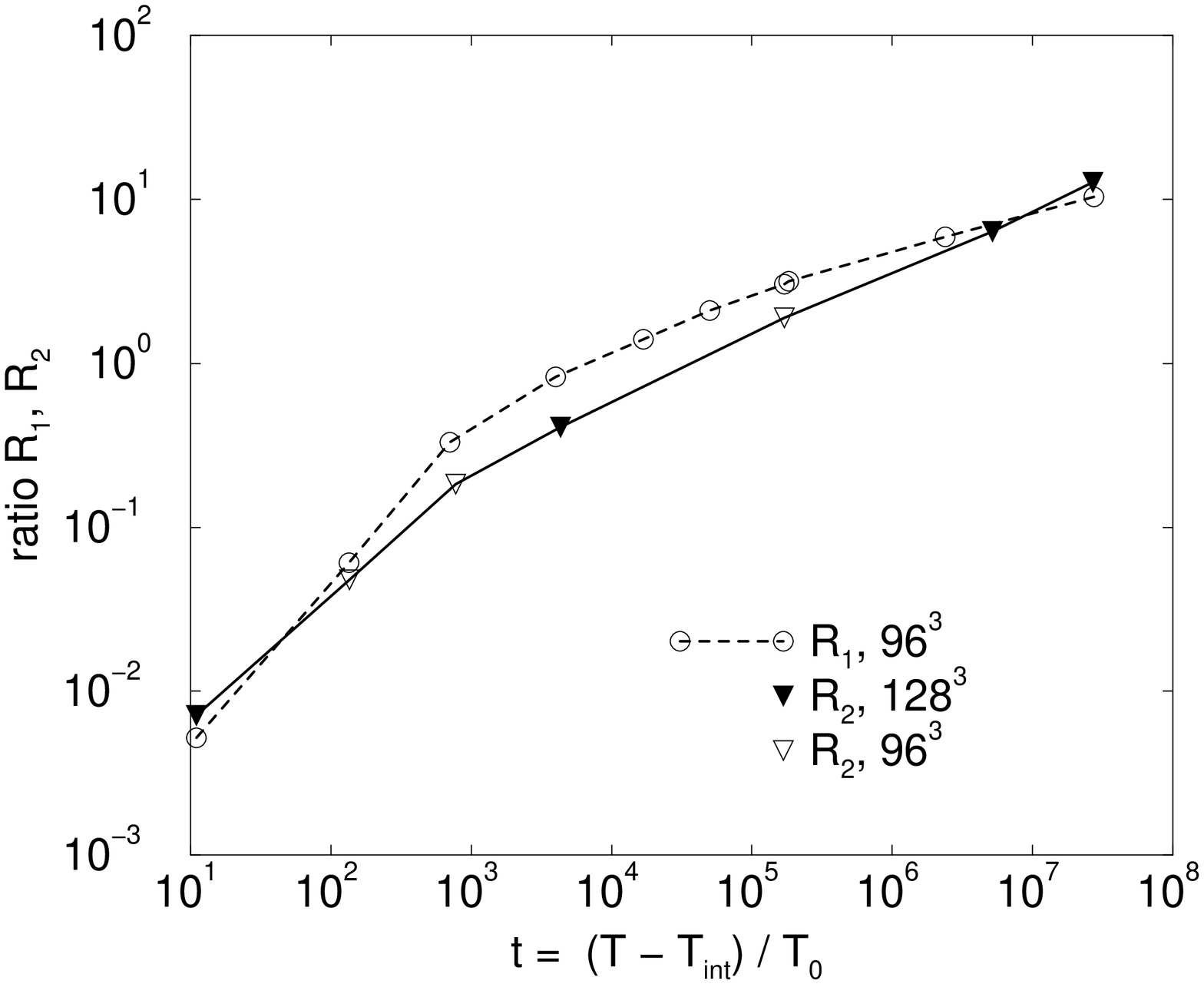}}}
    \end{minipage}
    \vspace{1ex}
    \caption[Terms in the Navier--Stokes equation]
	    {Left: 
For runs in table \protect\ref{table:pars_256},
             Re$_{\phi}$, (bold line), and $\Rey_L$ (diamonds)
	     and for Runs in table
	     \protect\ref{table:pars_128}, Re$_{\lambda}$ (dots).
	     Right:  Ratios $R_1$ and $R_2$ for runs with
	     $L_0=$ $36$, $2.9$, $0.59$, $0.15$, $0.054$, $0.024$,
	     $0.01$, $0.01$ 
	     (different parameters), $0.0016$, $0.00095$, $0.00039$, $0.0003$.
	     System sizes are  $96^3$ (open symbols) and
	     $128^3$ (filled symbols).}
    \protect\label{graph:reyn_t}
\end{minipage}
\end{figure}

All three quantities are shown as functions of the reduced time $t$ in figure \ref{graph:reyn_t} (left).
There is little difference between $\Rey_{\phi}$ and
$\Rey_L$; both show reasonable scaling, with the deviations in the most extreme runs stemming from those discussed already in \S\ \ref{sec:vel_com}.
The large $t$ asymptote for $\Rey_{\phi}$ in the inertial region is
approximately $\Rey_{\phi}\sim t^{1/3}$
as predicted by both the simple and the extended scaling analysis
\citep[but questioned by][]{grant99a}.
Our simulations extend from 
$0.1\lesssim \Rey_{\phi}\lesssim 350$, and the crossover
region occupies the range $1\lesssim\Rey_{\phi}\lesssim 100$.
Thus in terms of $\Rey_\phi$ (rather than $t$) the crossover is not, after all, quite so broad.

In contrast to $\Rey_L$ and $\Rey_\phi$, the data for
$\Rey_{\lambda}$ does not show good scaling behaviour.
The overall trend is of linear scaling in the viscous region
and slower growth
at around $t^{1/6}$ in the inertial region; both are broadly consistent with the extended
scaling theory (see table \ref{table:predict}). However, the individual runs
do not line up onto a single curve. 
This non-scaling behaviour of $\Rey_{\lambda}$ can be traced to that of $\lambda$ itself, which is examined in 
more detail in \S\ \ref{sec:nonscale}.

\subsection{Ratios of terms in the NSE}
\label{sec:NSE_r}

Shown in figure \ref{graph:reyn_t} (right) are the actual {\sc rms} ratios $R_1$ and $R_2$ defined in  (\ref{eq:R_1t}) and (\ref{eq:R_2t}) respectively. (Recall that the latter is the ratio of nonlinear to viscous terms in the NSE, which is what we believe a Reynolds number should estimate.) 
In order to form these ratios, one must first evaluate ({\sc rms} values of) the three relevant NSE terms,
$\eta\nabla^2\mathbf{v}$,
$\rho\dot{\mathbf{v}}$ and $\rho(\mathbf{v}.\bnabla)\mathbf{v}$,
which are vector quantities at each point in space.
An {\sc rms} value is calculated as, $|f|_{\mbox{\sc rms}} =
(\langle f_x^2\rangle + \langle f_y^2\rangle + \langle f_z^2\rangle)^{1/2}$,
where $f_x$, $f_y$, and $f_z$ are the Cartesian components of the vector
$\mathbf{f} = \eta\nabla^2\mathbf{v}$ (for example), and the average is taken over the whole
system. The first and second order spatial derivates of the velocity were found from $k$-space data (as in \S\ \ref{sec:vel_skew}) on $96^3$ and $128^2$ runs;
a further complication is that to evaluate $\dot{\mathbf{v}}$, velocity data from
consecutive time steps is required.  Due to data storage constraints, this was only collected for a set of $96^3$ runs. This is large enough to obtain $R_1$ values for a domain size just larger than
$L_{min}$, but not to determine its time dependence accurately within any particular run; checks were made for consistency by comparing other quantities with $128^3$ data.

The values of $R_1$ and $R_2$ shown in figure \ref{graph:reyn_t} (right)
lie between $\Rey_\phi$ and $\Rey_\lambda$ but are closer to the latter. The two ratios remain the same order of magnitude as each other throughout,
but vary by three orders of magnitude, from $R_2 \sim 10^{-2}$ at the viscous end (indicating the viscous term is dominant by two orders
of magnitude) to $R_2 \sim 10^1$ at the inertial end (indicating that the inertial terms
are dominant by one order of magnitude). Though we cannot calculate it directly, $R_2$ is presumably somewhat higher (about 20) towards the end of
our most inertial $256^3$ run.
This confirms the claim
made in \S\ \ref{sec:spin} that our simulations
have reached a regime where the inertial terms are dominant in 
the dynamics. 

Looking more closely at the behaviour of $R_1$ relative to $R_2$,
there is a significant difference in the crossover region,
by around a factor of two ($R_1 > R_2$).
Then, in the inertial regime, $R_1$ becomes less than $R_2$ by
about 50\% and appears to be heading for a lower growth rate.
This deviation
suggests that the asymptotic behaviour of these two ratios is perhaps
going to be different.  That would be consistent with the extended scaling 
predictions which are (table \ref{table:predict}) that
$R_1\sim t^{-1/6}$ while $R_2 \rightarrow$ constant; but if so neither
curve is close to its final asymptote yet.

That the LB simulations are still far from the final asymptotic behaviour in
our most inertial runs, even though inertial terms are clearly dominant there, is not unreasonable.
The largest ratio $R_2 \simeq 20$ actually achieved in our simulations is, by turbulence standards, still a fairly low Reynolds number. If the extended scaling theory of \cite{kendon99c} is correct, the final asymptotes for quantities relating to fluid velocity cannot be attained until an appreciable `inertial range' has developed between the interfacial driving scale $L(T)$, and the smaller scales ($\lambda$, $\lambda_d$) where
energy dissipation is taking place. 
However, to observe scaling of $l(t)$ for the structural (as opposed to velocity) data, it may not be necessary that the inertial cascade is fully established; one might only require that a reasonable degree of decoupling between interfacial motion and viscous dissipation has taken place. Our results for $l(t)$ suggest that this has already happened by the time $R_2 = 10$.

\subsection{Structure factors of the NSE terms}
\label{sec:NSE_k}

\begin{figure}
\begin{minipage}{\textwidth}
    \raggedright
    \begin{minipage}{0.49\textwidth}
        \resizebox{\textwidth}{!}{\rotatebox{-90}{\includegraphics{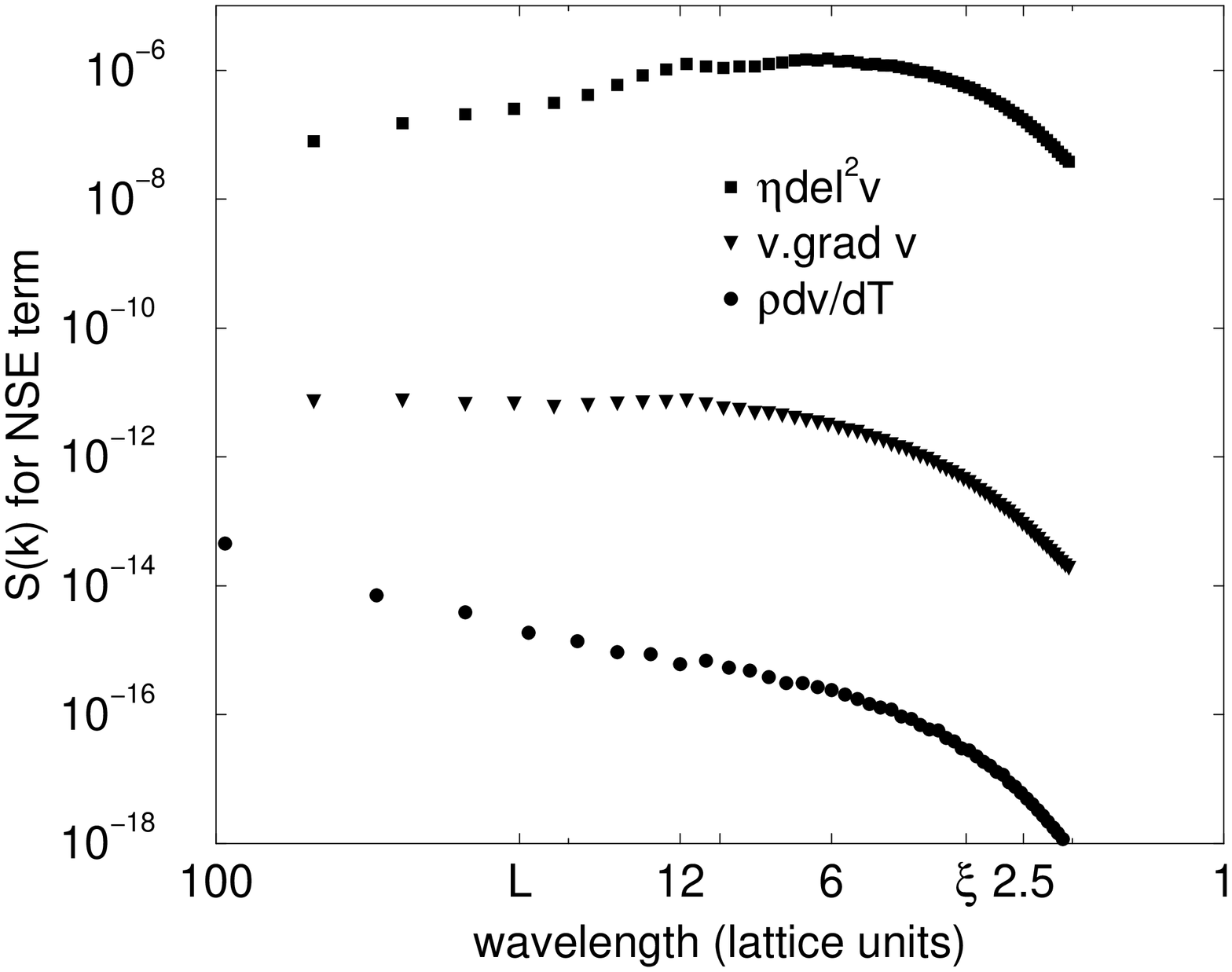}}}
    \end{minipage}
    \hfill
    \begin{minipage}{0.49\textwidth}
       \resizebox{\textwidth}{!}{\rotatebox{-90}{\includegraphics{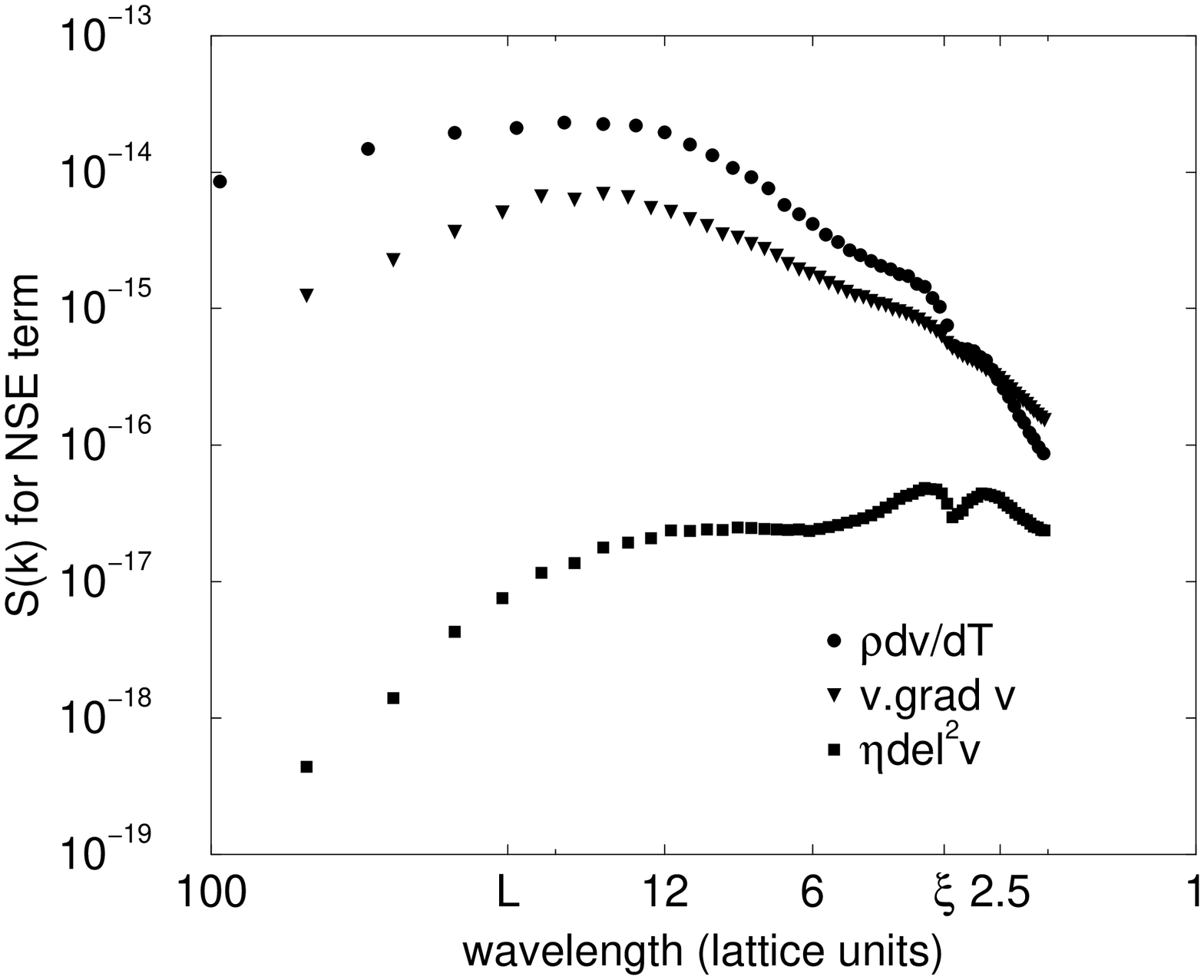}}}
    \end{minipage}
    \vspace{1ex}
    \caption[Structure factors of terms in the NSE]
	    {Structure factor of the viscous (squares) and nonlinear
	     (triangles) terms in the NSE.
	     Left: Run027 is in the viscous regime with $L_0=36$.
	     Right: Run031 is in the inertial regime with $L_0=0.0003$.
	     Both are shown on a log-log plot
	     for the timestep at which $L\simeq 25$
	     lattice units.  $L=25$ is marked on the x-axis along
	     with the interface width, $\xi\simeq 3$.
	     (The $\rho\partial\mathbf{v}/\partial T$ data is from the
	     corresponding $96^3$ runs.)}
    \protect\label{graph:NSE_k_027_031}
\end{minipage}
\end{figure}

Further information on the behaviour of the NSE terms can be obtained
by calculating the structure factor for each term. (These are $\langle|\mathbf{f}(\mathbf{k})|^2 \rangle$ where $\mathbf{f} = \eta \nabla^2 \mathbf{v}, \mathbf{v}. \bnabla \mathbf{v}$ or  $\rho\dot\mathbf{v}$.)
Results for one viscous and one inertial run are shown in 
figure \ref{graph:NSE_k_027_031}.
Looking first at the viscous run, 
the structure factor of the viscous 
term takes the form 
of a broad peak stretching from a small bump at 
wavelengths around 12 lattice units (which is around half the domain size, $L(T)=25$),
down to the interface width, $\xi\simeq 3$.
Thus the dissipation is taking place over the smaller
length scales in the system.  (The small bump is
a manifestation of the domain size in the dynamics; it is present
at around $L(T)/2$ throughout the run.)

In the inertial
regime, the viscous term is, as expected, smaller than the inertial 
terms.
The viscous term is similar in shape to that in the viscous regime,
with the
addition of two large peaks at high $k$ that presumably arise from the
presence of the interface (compare \S\ \ref{sec:vel_comp}) and/or lattice effects.
This suggests perhaps that the largest dissipative forces, and therefore most
dissipation, are happening close to the interface (or at least on that length scale).
The acceleration term has a broad peak in the structure factor at wavelengths around $L=25$, and tails off quite sharply below
10 lattice units.
The nonlinear term has a structure factor with a broad peak
centred around 15 lattice units, intermediate
between the length scale of the interface, $L(T)=25$, and
the dissipation length scales. 

The overall picture, though not quantitative (in view of the numerous sources of uncertainty, especially in the velocity derivatives) is qualitatively consistent
with the extended scaling picture, in which
the nonlinear term transports energy from large to small scales, where it is then dissipated. But in these runs there is still considerable overlap between the length scales for each term, as expected at relatively low Reynolds number.

\subsection{Energy conservation, dissipation rate}
\label{sec:diss_rate}

The dissipation rate is crucial to the energy balance in the simulation.
The LB algorithm is isothermal, and therefore does not strictly conserve energy, in that the energy dissipated as heat by the viscous stresses is taken out of the system locally rather than having to diffuse thermally to the boundaries. Nonetheless, the simulation should faithfully
recreate the energetics of the isothermal Navier--Stokes equation, driven by interfacial motion (as in (\ref{eq:nse})), which 
makes precisely the same assumption. 
In the inertial regime, the interfacial energy should first be transformed into kinetic energy of the fluid; the observation of $l\sim t^{2/3}$ (see
\S\ \ref{sec:scaled}) shows that this is happening at roughly the expected rate.
To complete the energy balance, energy must also be removed from
the simulation at the correct rate through viscous dissipation.  
This requires accurate modelling of the smaller length scales ($\lambda - \lambda_d$) at which such dissipation is taking place.

The dissipation rate can be calculated directly from the velocity data as
\begin{equation}
\varepsilon = \eta \langle (\nabla_{\alpha}v_{\beta})(\nabla_{\alpha}v_{\beta}) \rangle = \eta \int k^2 S_v(\mathbf{k}) \mathrm{d}^3\mathbf{k}
\label{eq:diss_rate}           
\end{equation}
where the angle brackets are spatial averages and both the incompressibility ($\nabla_{\alpha}v_{\alpha} = 0$) and homogeneity ($\langle v_{\alpha} v_{\beta} \rangle = 
(1/3)\delta_{\alpha\beta}S_v(k)$) of the binary fluid system have been exploited. 
The second expression is tantamount to using the Fourier definition of spatial derivatives, and as mentioned previously, this need not
coincide with lattice derivatives when gradients are not small. Therefore the dissipation rates were calculated first using the Fourier space expression in  (\ref{eq:diss_rate}) and then, for some but not all runs, using the real space expression with the lattice gradient operator as defined in Eq. (\ref{eq:grad_def}). 

\begin{figure}
\begin{minipage}{\textwidth}
    \begin{center}
    \begin{minipage}{0.7\textwidth}
        \resizebox{\textwidth}{!}{\rotatebox{-90}{\includegraphics{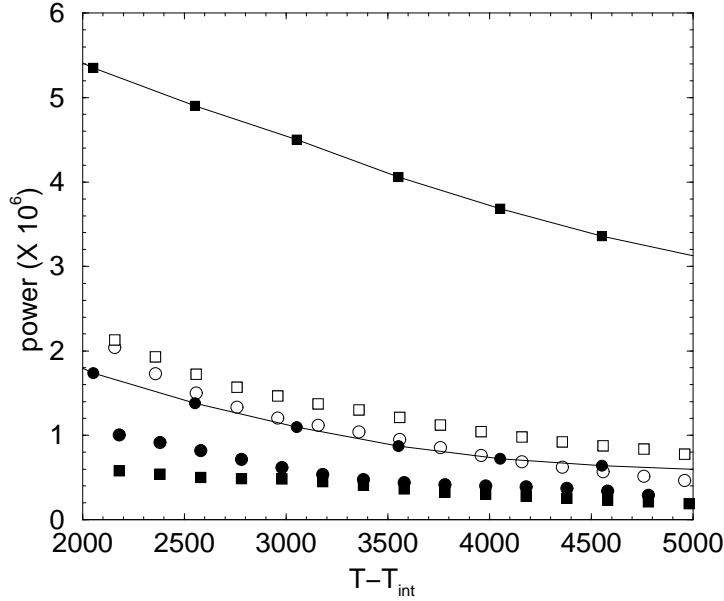}}}
    \end{minipage}
    \end{center}
    \vspace{1ex}
    \caption{Energy conservation in viscous region.
	    $\varepsilon_{in}$ (circles), $\varepsilon$ (squares),
	     for Run008 (filled), Run014 (open), and
	     Run027 (solid line).  In these runs
	     $\dot{\mathbf{v}}$ is negligible on the scale of the graph.}
    \protect\label{graph:plot_energyconsv}
\end{minipage}
\end{figure}

The two estimates (Fourier and lattice) of the dissipation rate were compared with
the rate of decrease of the sum of the interfacial energy and the fluid kinetic energy in the system: these should balance, once the interfaces are well formed and residual diffusion has become negligible. (Note that diffusion introduces its own contribution to the dissipation, not accounted for in the NSE, which may dominate early in each run.)
We found that the lattice derivatives gave better agreement than Fourier ones in this comparison. However, even with these, the results were never more than satisfactory; best in the crossover region (within 25\% of the expected value) but giving dissipation rates well above those required by energy conservation in the viscous regime, somewhat below in the inertial. At the viscous extreme the discrepancies were a factor three or more. Even in the crossover region, there was a tendency for the computed dissipation rate to drift below that required for energy conservation.
Examples are shown in figure \ref{graph:plot_energyconsv}.
Use of Fourier derivatives to estimate the dissipation rate instead gave values that were consistently too high, through all regimes, by a factor 2 or so for runs in the crossover and inertial regimes and much more than this in the viscous regime. 

A check was made to ensure that these discrepancies in the dissipation rate did not arise from compressibility effects, by computing the full expression for $\varepsilon$, \textit{without} assuming fluid incompressibility as was done in  (\ref{eq:diss_rate}). (See \cite{landau59f} for the relevant expression.) This check was done with lattice derivatives; the results did not differ appreciably from the lattice version of Eq. (\ref{eq:diss_rate}), so that compressibility is not responsible here.

The fact that the two methods of finding derivatives differ, shows that dissipation is primarily taking place at short scales (of order a few lattice spacings). The results do not imply that LB is failing to conserve energy, but do show that the viscous dissipation actually generated by the algorithm is not accurately estimated from the Fourier derivatives; the lattice
difference estimates for $\varepsilon$ are better, but still not accurate. 
In principle even these need not give the true dissipation since the LB algorithm actually dissipates energy by relaxing the velocity distribution functions (\ref{eq:tau_1}), and not by calculating lattice gradients. 
Note, in any case, that for qualitiative comparisons between runs (requiring log-log plots covering many decades) the factor two difference between Fourier and lattice derivatives is barely detectable, and we ignore it, for those purposes, below.

\begin{figure}
\begin{minipage}{\textwidth}
    \raggedright
    \begin{minipage}{0.48\textwidth}
        \resizebox{\textwidth}{!}{\rotatebox{-90}{\includegraphics{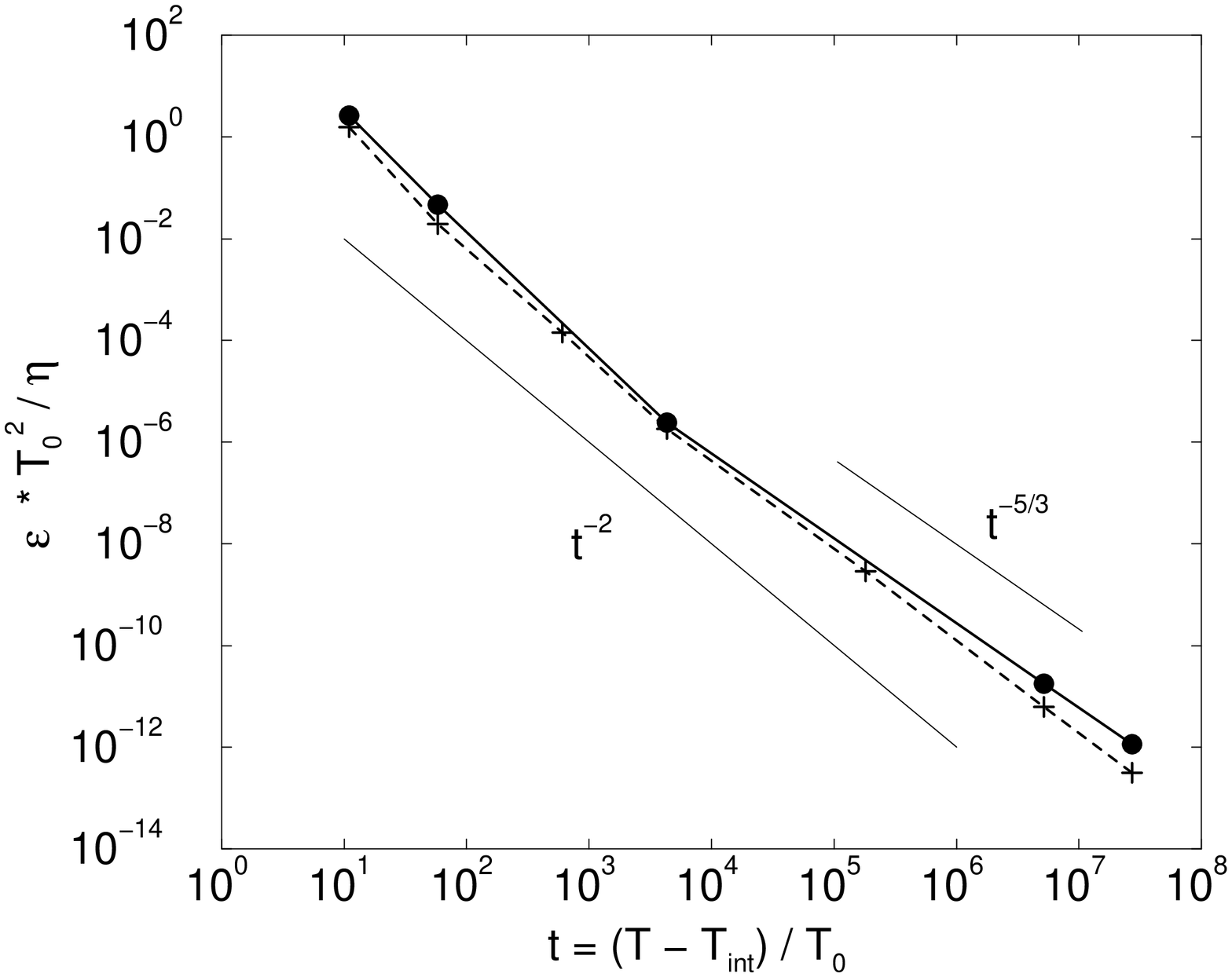}}}
    \end{minipage}
    \hfill
    \begin{minipage}{0.48\textwidth}
        \resizebox{\textwidth}{!}{\rotatebox{-90}{\includegraphics{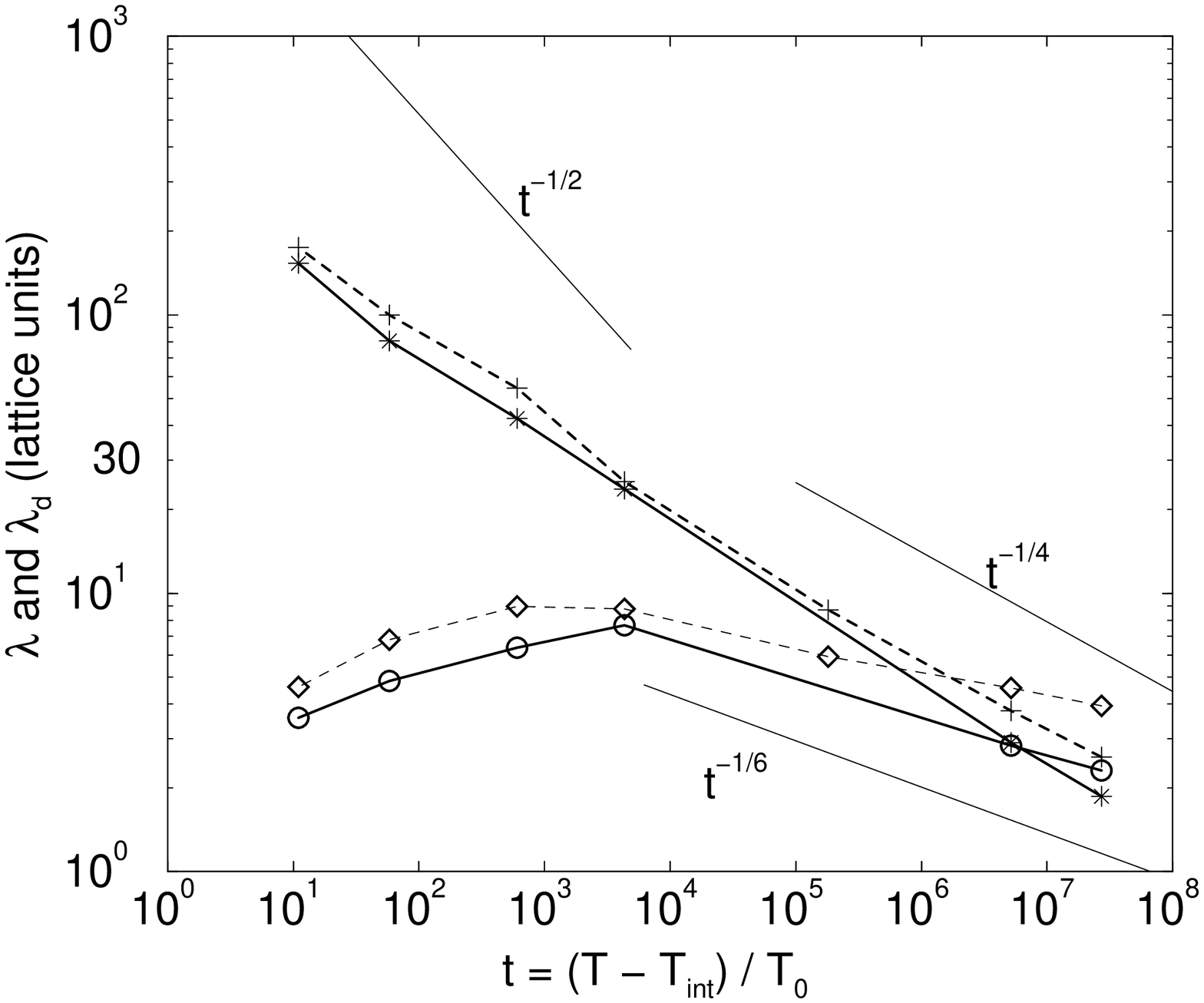}}}
    \end{minipage}
    \vspace{1ex}
    \caption{Left: Dissipation rate
	     at time step when domain size is $L=30$ (in simulation units) for
	     Runs in table \protect\ref{table:pars_128} (circles, solid line)
	     and Runs in table \protect\ref{table:pars_256} (plus, dotted).
             Right: Dissipation scale, $\lambda_d$, for
	     runs in table \protect\ref{table:pars_256} (plus, dashed), and
	     runs in table \protect\ref{table:pars_128} (stars, solid);
	     Taylor microscale, $\lambda$ for same runs (diamond, dashed; 
	     circles, solid respectively).
	     All in lattice units with domain size $L=30$ lattice units.}
    \protect\label{graph:dr_30}
\end{minipage}
\end{figure}

For the dissipation rate itself, simple scaling theory predicts that $\varepsilon$ will always 
scale as $t^{-2}$.  \cite{kendon99c} 
predicts the scaling to be $\varepsilon \sim t^{-2}$ in the viscous region, but $\varepsilon \sim t^{-5/3}$ in the inertial region (table \ref{table:predict}).
A scaling plot of the (Fourier) dissipation rate (in reduced physical units, as usual) is shown in figure \ref{graph:dr_30} (left). 
To the accuracy we are working, these data barely discriminate between the
two scaling predictions though the upward curvature may slightly favour that of \cite{kendon99c}. There is some sign in the viscous regime that the rate increases
too fast; this might be connected with the velocity anomalies discused
previously (\S\ \ref{sec:s_k_v}).

\subsection{Taylor and Kolmogorov microscales}
\label{sec:vder_len}

The Taylor and Kolmogorov scales were defined in 
 (\ref{eq:microscale}) and (\ref{eq:k_d}) respectively.
Figure \ref{graph:dr_30} (right) shows $\lambda,\lambda_d$ \textit{in simulation units}
for each run in table \ref{table:pars_256} and table \ref{table:pars_128}. Values are evaluated mid-run at time $T$ such that the domain size $L(T)= 30$,
but plotted against the reduced physical time.
The curves thus have the same shape as a scaling plot of $\lambda/L(T)$ and $\lambda_d/L(T)$ (since
$L(T)$ is here fixed in simulation units) but also allows the $\lambda,\lambda_d$ values to be compared with the lattice spacing (unity) and the linear system size (128 or 256).
Given the expected effects of coarse-graining on the $256^3$ data at short length scales, the agreement between the two lattice sizes is satisfactory
but only the $128^3$ data is discussed in what follows.

As can be seen in figure \ref{graph:dr_30} (right), $\lambda/L(T)$ does not vary by much between runs, although it is roughly a factor of two larger
in the crossover region than in the viscous and inertial extremes.
The extended scaling analysis (table \ref{table:predict}) predicts that $\lambda/L(T) \sim t^{-1/6}$ in
the inertial region, and the data is broadly consistent with this; an asymptotically constant ratio, predicted by simple scaling, is marginally less plausible though not ruled out. Likewise, $\lambda_d/L(T)$ is not far from the prediction of the new scaling theory ($t^{-1/4}$), and perhaps somewhat further from that of simple scaling ($t^{-1/6}$). For the viscous regime, however, $\lambda_d/L(T)$ does not appear to scale as $t^{-1/2}$, nor does $\lambda/L(T)$ approach a constant,
as predicted by both scaling theories. Once again the low $k$ velocity anomaly (\S\ \ref{sec:s_k_v}) in the viscous regime could be to blame for this, but so could inaccuracies in the estimated dissipation rate (\S\ \ref{sec:diss_rate}).

\subsection{Resolution of the energy cascade}
\label{sec:vder_res}

The Kolmogorov microscale has a further significance for any attempt at numerical simulation in inertial fluids.
In a fully turbulent fluid, $\lambda_d$ is expected to be smaller than 
$\lambda$, and to mark the smallest length scale relevant for dissipation.  
This should, if the results are to be relied upon, be `resolved', that is, $\lambda_d$ should lie above the discretization scale set by the lattice. 
There is some debate over exactly how small $\lambda_d$ can be in relation
to the lattice spacing, but
a factor of 1.0 to 1.6 has been put forward; see for example,
\cite{eswaran88a,yeung91a}. Our best estimate for $\lambda_d$ in the most inertial ($256^3$) run is around $\lambda_d=1.8$ lattice units; values are higher in all other runs. However, in view of the quantitative uncertainty in our velocity derivative estimates (see \S\ \ref{sec:diss_rate}), this estimate is good to only a factor of 2 or so. Therefore, the most inertial of our $256^3$ runs is certainly
close to the resolution limit, and we would not wish to proceed to higher $l,t$ without a larger system size. Note that only in that run is $\lambda_d$ appreciably smaller than $\lambda$ as
expected asymptotically in a turbulent flow; but the Taylor scale $\lambda$ is well-resolved, by the same criterion, in all runs.

However, estimates of how well the cascade is resolved using these two
lengths could be misleading, in that they assume local as well as global homogeneity. So if, for example, most of the dissipation were to occur in a thin layer around the 
interfaces in the system, the globally averaged $\lambda_d$ might suggest
that the dissipation was fully resolved whereas in fact it was not.
This would be consistent with our findings (\S\ \ref{sec:diss_rate}) that
the actual dissipation rate in LB is poorly estimated from the Fourier velocity derivatives, and imperfectly even from lattice ones.

On balance, we believe that the energy cascade is adequately resolved in most
of our LB simulations, at least for the purposes of getting the correct evolution of the structural length scale $l(t)$. (As noted in \S\ \ref{sec:NSE_r}, a fully resolved cascade might not be needed for this, so long as there is sufficient decoupling of interfacial motion and viscous dissipation.)
However, the most inertial runs may well be marginal in terms of resolution; in common with several other aspects of the simulations (such as the
residial diffusion, and anisotropy), these runs are at the limit of what is possible using this simulation method with
current computational resources.

\subsection{Apparent scaling violations}
\label{sec:nonscale}

In our various analyses of the velocity derivatives, we focused mainly on
scaling plots made by taking a representative `mid-run' data point from each LB simulation run (e.g. figure \ref{graph:dr_30}, left). These plausibly connect to form a smooth curve.
However, one test of scaling is whether the full datasets (rather than a representative point) from different runs appear to join up smoothly when plotted in reduced physical units; this was satisfactorily the case for the structural data $l(t)$ (figure \ref{graph:l_t_fin}) and reasonable also (though not by any means perfect) for the velocity data $l_v(t)$ (figure \ref{graph:l_t_final_vel}, left), $\Rey_L$ and $\Rey_\phi$ (figure \ref{graph:reyn_t}, left). 

\begin{figure}
\begin{minipage}{\textwidth}
    \raggedright
    \begin{minipage}{0.48\textwidth}
        \resizebox{\textwidth}{!}{\rotatebox{-90}{\includegraphics{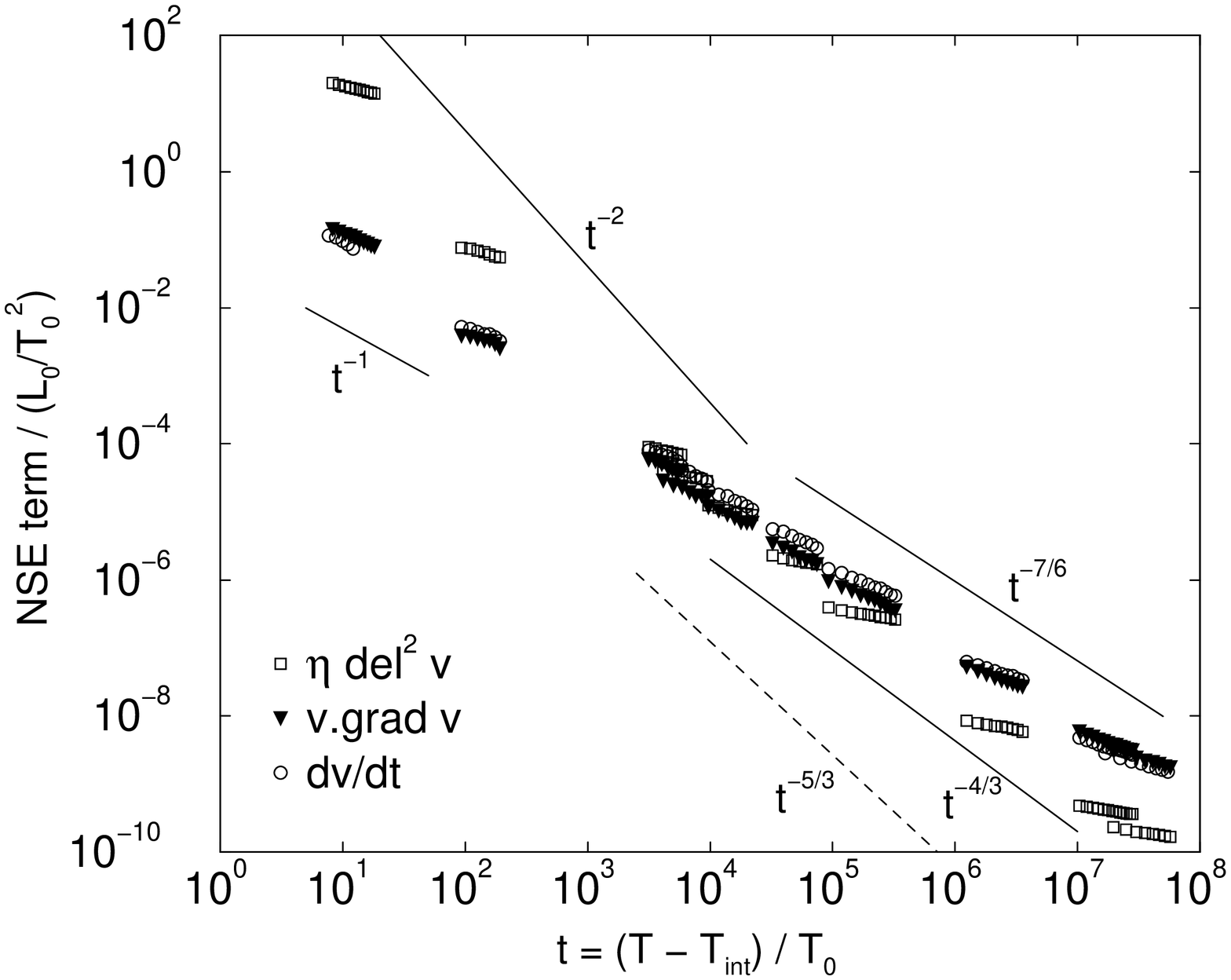}}}
    \end{minipage}
    \hfill
    \begin{minipage}{0.48\textwidth}
        \resizebox{\textwidth}{!}{\rotatebox{-90}{\includegraphics{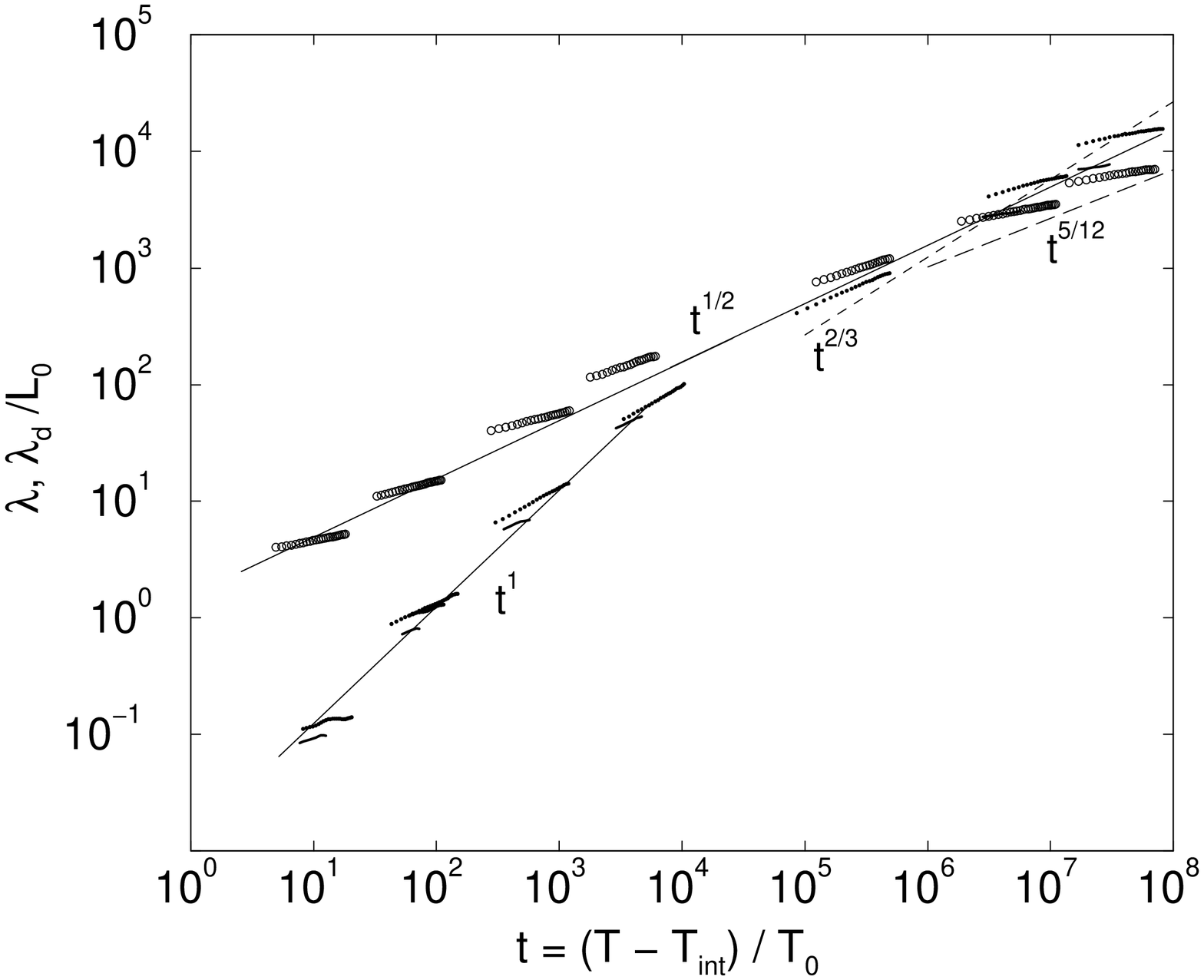}}}
    \end{minipage}
    \vspace{1ex}
    \caption{Left: NSE terms for all runs scaled. Most of
             this data is for $96^3$ systems, with $128^3$ data
             where available.
             Right: Dissipation scale, $\lambda_d$, for
	     runs in table \protect\ref{table:pars_128} (open circles),
	     Taylor microscale, $\lambda$ for same runs (shorter solid lines)
	     and runs in table \protect\ref{table:pars_256} (filled circles).
	     All in reduced units.}
    \protect\label{graph:NSE_t}
\end{minipage}
\end{figure}

However, the expected scaling within runs failed completely for $\Rey_\lambda$ (also in figure \ref{graph:reyn_t}, left). This quantity differs from the others just mentioned in that its definition involves $\lambda$, which in turn involves the dissipation rate $\varepsilon$ (see  (\ref{eq:microscale})) and therefore
depends on velocity derivatives. More generally, we find similar
`scaling violations' in \textit{all} the quantities we have looked at that  involve calculating velocity derivatives.  Figure \ref{graph:NSE_t} shows this
for (left) the various {\sc rms} terms in the NSE and (right) the Taylor and Kolmogorov scales.
The dissipation rate (not shown) shows similar features.

Apart from the failure of the runs to join up, the data for the NSE terms
is in reasonable accord with the scaling predictions
(table \ref{table:predict}) for the viscous term of
$t^{-2}$ in the linear region and $t^{-5/3}$ or
$t^{-7/6}$ in the inertial region, and for the inertial terms of
$t^{-4/3}$ or $t^{-7/6}$ in the inertial region.
Even the scaling of $t^{-1}$ for the nonlinear term in the viscous 
region is as predicted by the new scaling theory \citep{kendon99c}.
However, within each run the quantities are falling more slowly than 
all these predictions. Similarly $\lambda$ and $\lambda_d$ do not scale within single runs in the
same way as they scale between runs, except perhaps in the middle of the
crossover region.  (The same is true of $\varepsilon$, the dissipation rate,
from which $\lambda_d$ is derived.) 

At present, we have no simple explanation of these apparent scaling violations.
They could perhaps be a sign of subtle nonuniversalities of the type suggested by \cite{jury98a}, though it would certainly be premature to conclude this without similar (though perhaps less severe) effects being detected in $l(t)$. Another possibility, noted previously (\S\ \ref{sec:diss_rate}, \ref{sec:vder_res})
is that the we are not able, using LB with these system sizes, to fully resolve the velocity gradients that arise. 

On the other hand, our velocity derivatives do not appear to be completely lattice controlled, since for example, in (\ref{eq:diss_rate}) one could then set $\bnabla \mathbf{v} \sim v$. This would give 
$\varepsilon \sim \dot L^2$, giving in the inertial regime $\lambda_d \sim T^{1/6}$ within each run
(which is about right) but in the viscous regime $\lambda_d\sim T^0$, and also  
$\lambda \sim T^0$ throughout both regimes, neither of which is plausible 
for our data (see figure \ref{graph:NSE_t}, right). 
We have made various further attempts to explain the data by assuming `composite' 
scalings, based for example on the idea that dissipation occurs mainly near the interface where velocity gradients might be anomalously large ($\bnabla v \sim \dot L/\xi$)\footnote{We note that in a very recent analysis, \cite{solis00a} developed an heuristic alternative to that of \cite{kendon99c} in which dissipation takes place in an (asymptotically) thin layer around the interface rather than in the bulk. They argue that coarsening is limited by the damping rate of capillary waves of wavelength $L$, and that this damping rate is unaffected by nonlinearity; this gives $l\sim t^{4/7}$. We find no evidence for this in our data. Using the oscillation frequency instead of the linear damping rate recovers $l\sim t^{2/3}$.}. In that case, one would have $\varepsilon \sim (\xi/L)\dot L^2/\xi^2$, where the factor $\xi/L$ is the volume fraction of interface. Although none of these attempts was successful, an explanation along these lines is not completely ruled out. 

One further complicating factor is that, within LB, the fluid viscosity is frequency dependent: the Navier Stokes limit requires that velocities are not only small (in lattice units) but also evolving sufficiently slowly. For very low viscosities, $\tau_1$ approaches $1/2$ which means that the distribution function $f_i$ is over-relaxed (nearly reversing sign at each time step), see Equation \ref{eq:tau_2}. (In effect, the `natural' viscosity scale in LB is of order unity and the over-relaxation is used to create an unnaturally small value.) This works successfully if accelerations are small but could result in a higher dissipation rate than would be the case for a purely newtonian fluid, especially in the most inertial runs.

\section{Conclusions}
\label{sec:conc}


Our LB results for symmetric binary fluids, undergoing spinodal decomposition after a deep quench, appear to confirm the dynamical scaling hypothesis, which requires a universal time dependence of the structural length scale $L$
when expressed in reduced physical units as a function of time, $l(t)$ (figure \ref{graph:l_t_fin}). We found
no signs of nonuniversality here, although some weak breakdown of it, as suggested by \cite{jury98a}, cannot be entirely ruled out.
By exploitation of the expected scaling, and careful parameter steering
and validation tests to eliminate residual diffusion and other unwanted effects, we achieved an $l(t)$ curve spanning seven decades of reduced time $t$ and five of reduced length $l$. The $l(t)$ curve 
asymptotes to $l = b_1 t$ at $t \ll t^*$ (the viscous regime), with $b_1 \simeq 0.072$, and $l = b_{2/3} t^{2/3}$ at $t \gg t^*$ (the inertial regime), with $b_{2/3} \simeq 1.0$.
On our reading of \cite{guenoun87a}, $b_1$ is within 10\% of the most careful experimental measurements (albeit not for a deep quench) although others \citep{laradji96a} extract an estimate about twice as large
from the same experimental results.

The crossover time, $t^*$ in reduced physical units, as defined by the interception of the viscous and inertial asymptotes, is surprisingly large ($t^*\simeq 10^4$). So is the width of the crossover region (four decades). 
This may explain why the $t^{2/3}$ region has never yet been confirmed in laboratory experiments. The default assumption that $t^*$ is `of order unity' would mean that for fluid pairs typically studied (short chain alcohols and water for example), the inertial regime could be accessed at length scales of a few microns, easily studied by light scattering. But in fact, to exit the crossover region (at $t^*\sim 10^6$) one needs $l = L/L_0 > 10^4$ which for typical fluids gives $L$ of order centimetres (and larger still for the near-critical quenches often used, see \cite{guenoun87a}). This requires direct visualisation experiments, not light scattering, and more importantly  requires rigorous exclusion of thermal convection and gravitational effects. The latter is likely to be possible only with extremely careful density matching, or in microgravity, \cite{cambon97a}.

Our findings are less extreme when translated into Reynolds numbers. The conventional definition of the Reynolds numbers for the spinodal problem is $\Rey_\phi = l\dot l$; the crossover regime then spans $1\lesssim \Rey_\phi
\lesssim 100$ and the Reynolds number range we actually achieved was $0.1 \lesssim \Rey_\phi\lesssim 350$.
(Note that we see no sign of a saturating $\Rey_\phi$ as predicted by \cite{grant99a}, but cannot rule this out, at some value well above 350. )

These figures are reduced further if the actual ratio $R_2$ of nonlinear to viscous terms is used as the Reynolds number; our results span $0.01 \lesssim R_2 \lesssim 20$. 
At the upper end of this range, we have described in detail the first unambiguous simulations of a regime in which the inertial terms in the Navier--Stokes equation are actually large compared to the viscous ones; here we observed $l\sim t^{2/3}$. 
However these simulations may still be far from the asymptotic regime
addressed by the scaling analysis of \cite{kendon99c}, in which ultimately one expects $\lambda_d \ll \lambda \ll L$ for the Kolmogorov, Taylor and structural length scales. But in fact, although $\lambda_d$ is only slightly smaller than $\lambda$ even for our most inertial run, our data for the inertial regime is already somewhat more consistent with Kendon's analysis than with the simpler (single length scale) scaling predictions of \cite{furukawa85a}. 

In the viscous regime, we recovered the expected ($l\sim t$) scaling though this is somewhat erratic and accompanied by an unexpected behaviour of the velocity correlation length $l_v = L_v/L_0$ (figure \ref{graph:l_t_final_vel}, left) and possibly also the dissipation rate (figure \ref{graph:dr_30}, left). The first of these, at least, may be related to the presence of anomalous long-range correlations of the velocity over length scales much larger than the domain size $L(T)$. These were detected numerically in our most viscous runs, and can be explained simply by assuming that the interfaces contribute a random stress with local correlations on scale $L$ (see \S\ \ref{sec:s_k_v}).  It is quite possible that such effects have
arisen undetected in previous simulations of the viscous regime; in principle they could lead to finite size problems arising long before the domain size $L$ approaches the size $\Lambda$ of the simulation cell. Fortunately, the same does not appear to happen in the inertial regime.

As well as through their differing domain growth laws ($l\sim t$ or $l\sim t^{2/3}$) the viscous and inertial regimes can be distinguished by rather subtle changes in interface geometry (evident through the structure factor and by visualisation, with more thin necks in the inertial case) and by much larger changes in the velocity statistics. For example, the velocity {\sc pdf} is significantly flatter for inertial than for viscous runs and also there is significant skewness for longitudinal velocity derivatives in the inertial regime.
In that regime, we achieve results that show some of the characteristics of a turbulent fluid, although non-Gaussian features are seen directly in the velocity distribution as well in that of derivatives. This reflects the presence of microstructure (interfaces), as well as some turbulence, in the fluid mixture.

The scaling of quantities involving velocity derivatives was found to give
reasonably continuous curves when mid-run values were plotted in reduced physical units (e.g. figure \ref{graph:dr_30}, left) but showed apparently systematic violations of the scaling behaviour which that would imply, when analysed in detail within each run. This could indicate some physical nonuniversality, or some shortcoming of the LB code; but equally it can be attributed to uncertainties over how to accurately estimate derivatives that are not small on a lattice scale. The LB code is implicitly aware of differences in the local velocity distributions but does not itself calculate velocity derivatives.
Therefore, when these derivatives are not small, there can be discrepancies between their `effective' values (as defined, e.g., through the actual dissipation rate generated by the algorithm) and any estimate based on Fourier, or lattice difference, data.

The fact that such ambiguities arise at all is an indication that our LB simulations are close to the limits of accuracy acceptable for the fluid motion, since ideally all gradients encountered in LB should be small, not large, on the lattice scale. It is interesting that they arise even when the Taylor and Kolmogorov microscales, $\lambda$ and $\lambda_d$, are both significantly larger the lattice spacing; this perhaps suggests 
heterogeneity of the velocity field that is stronger than in single-fluid homogeneous turbulence. It may mean that, despite our best endeavours, we have
not adequately resolved the fluid motion at short length scales.
Fortunately, according to the analysis of \cite{kendon99c}, the main requirement for observation of $l\sim t^{2/3}$ in the inertial regime is that the interface motion is decoupled (by fluid convection) from the dissipation scale, and not that the latter is modelled with complete accuracy. The reason is that, once the interface is decoupled, its motion is controlled by how fast it feeds kinetic energy into the fluid at large scales; the precise details of the dissipation further down the cascade has no further influence on the interfacial dynamics. If so, statistics based on the domain growth law $l(t)$ may be much more robust than those for the fluid velocity and (especially) velocity derivatives, in the inertial regime.

Very similar accuracy limits were also reached in the thermodynamic sector, where we required narrow interfaces, with large composition gradients, 
to satisfy the conflicting requirements of rapid interfacial equilibration and low residual diffusion. (This gave interfacial tensions 10-15\% different from nominal, with measurable lattice anisotropy.) 
We have taken unusual care to analyse (and maintain under reasonable control) the various sources of error. The reader has, we hope, enough information to decide for herself how much confidence to place in our various results. 

It is interesting to ask how simulations might be taken further into the inertial regime. If the analysis of \cite{kendon99c} is any guide, the ultimate asymptotic state entails, computationally, resolution of the following scale hierarchy:
\begin{equation}
a \ll \xi \ll \lambda_d \ll \lambda \ll L \ll \Lambda
\end{equation}
with $a$ the lattice spacing, or, more generally, $a^{-3}$ the density of degrees of freedom in the simulation.
A factor 3 between each of these length scales is roughly what we achieve at $\Lambda = 256$. But a worthwhile improvement to give, say, a factor 10 between each would require $\Lambda = 10^5$.
For a three dimensional system, this lies beyond any foreseeable innovation
in computational hardware if any current methodology is used. Possibly the way forward would be to couple a coarse-grained algorithm (such as large eddy simulation) for the turbulent fluid to a moving interface; we leave this to others to explore.

\begin{acknowledgments}

We would like to thank Alan Bray, Anatoly Malevanets,
Alexander Wagner, Patrick Warren, Julia Yeomans and Alistair Young
for helpful discussions.
Work funded in part under EPSRC GR/M56234.

\end{acknowledgments}

\begin{appendix}
\section{Analysis of hydrodynamic modes}
\label{app:hydro_modes}
In this appendix we analyse the hydrodynamic modes for the LB hydrodynamic equations (\ref{eq:NS_LB}-\ref{eq:diff_LB}). We take 
as our reference state one in which the fluid is at rest with uniform concentration. In 
this case, to linear order 
in the perturbation of the hydrodynamic fields the Navier-Stokes equation is satisfied,  and only the spurious terms in the balance equation for the order parameter remain. 
Specifically, the equations we are going to analyse are
\begin{eqnarray}
\dot{\rho}+\nabla.(\rho{\bf v}) &=& 0 \nonumber\\
\dot{\bf v} &=& -\frac{1}{\rho}\nabla.\mbox{\boldmath{$\cal P$}}
^{th}+\nu\nabla^2{\bf v}+
\zeta\nabla(\nabla.{\bf v})\nonumber\\
\dot{\phi}+\nabla.(\phi{\bf v}) &=& (\tau_2-\frac{1}{2})
\left[\tilde{M}\nabla^2\mu-\nabla.(\frac{\phi}{\rho}\nabla.{\cal P}^{th})\right]
\label{eq:linear}
\end{eqnarray}
where $\nu=\eta/\rho$ and $\zeta=(\xi+2\eta/3)/\rho$ are the kinematic shear and bulk viscosities, respectively.
Using eq.(2.9), the last term in the third equation 
can be rewritten as

\begin{equation}
\nabla.
\mbox{\boldmath{$\cal P$}}^{th} =c_s^2\nabla\rho+\phi\nabla\mu
\label{eq:divp}
\end{equation}
where $c_s$ is the speed of sound and $\mu$ the chemical potential. This expression 
shows that there is a spurious coupling between the order parameter and gradients 
of the density field. Such a coupling can only be relevant when the binary mixture 
exhibits compressibility. In terms of the deviations of the hydrodynamic fields with respect to their 
equilibrium values, $\rho_0=1, \phi_0, {\bf v}=0$, eqs.(\ref{eq:linear}) become

\begin{eqnarray}
\dot{\delta \rho}+\rho_0\nabla.\delta{\bf v} &=& 0 \nonumber\\
\frac{\partial \delta{\bf v}}{\partial t} &=& -\frac{c_s^2}{\rho_0}\nabla\delta\rho-(A+3B\phi_0^2)\nabla\delta\phi+\nu\nabla^2\delta{\bf v}+
\zeta\nabla(\nabla.\delta{\bf v})\nonumber\\
\dot{\delta\phi}+\phi_0\nabla.\delta{\bf v} &=& \omega_2
\left[(\tilde{M}-\phi_0^2)(A+3B\phi_0^2)\nabla^2\delta\phi-c_s^2\phi\nabla^2\delta\rho\right]
\end{eqnarray}
where we have defined $\omega_2=\tau_2-\frac{1}{2}$. In Fourier space, this set of equations can be 
expressed in matrix form, setting $\mathbf{X} = (\delta\rho, \mathbf{k}. \mathbf{v}, \delta\phi, \delta \mathbf{v} - \mathbf{k k}. \delta\mathbf{v}/k^2)$ as
\begin{equation}
\dot{\mathbf{X}}
= \left(\begin{array}{cccc}
               0 & -i & 0 & 0\\
               -i c_s^2 k^2 & -(\nu+\zeta)k^2 & -i\phi_0 a k^2 & 0\\
               \omega_2 \phi_0 c_s^2 k^2 & -i \phi & -\omega_2 (\tilde{M}-\phi_0^2) a k^2 & 0\\
               0 & 0 & 0 & -\nu k^2
               \end{array}
\right).
\mathbf{X}
\end{equation}
where $a = A + 3B\phi_0^2$. The hydrodynamic mode eigenvalues are correspondingly
\begin{eqnarray}
\lambda_1 &=& -\nu k^2 \\
\lambda_{2,3} &=& \pm i c_s k\sqrt{1+\frac{a\phi_0^2}{c_s^2}}\nonumber\\
&-&\frac{k^2}{2}\left[\nu+\zeta+a\omega_2\phi_0^2\left(-1+\frac{a(\tilde{M}+2\phi_0^2)}{3 (c_s^2+a\phi_0^2)}\right)\right]\\
\lambda_4 &=& -k^2 \frac{a\omega_2 \tilde{M}}{1+a\phi_0^2/c_s^2}
\end{eqnarray}

The first mode corresponds to the propagation of shear waves, the subsequent two modes 
are related to the propagation and damping of compressible momentum waves, and the final 
mode is related to the diffusion of the order parameter. One can see that the presence of the 
spurious term in the convection-diffusion equation
does not qualitatively modify the propagation of the order parameter: it remains diffusive. 
However, its eigenvalue depends on the compressibility of the the fluid, and for a compressible fluid there is then a reduction of diffusivity
coming from the coupling to sound waves. Such a coupling also modifies the propagation 
and damping of longitudinal waves in the medium, so that these become a function of the order 
parameter. 

To whatever extent the compressibility of the fluid can be neglected, the additional coupling between the order parameter and the density becomes irrelevant, and the expected 
diffusion coefficient for the convection-diffusion equation of the order parameter is recovered.


\section{Comparisons with other work}
\label{sec:other_work}

\begin{figure}
\begin{minipage}{\textwidth}
    \begin{center}
    \resizebox{\textwidth}{!}{\rotatebox{-90}{\includegraphics{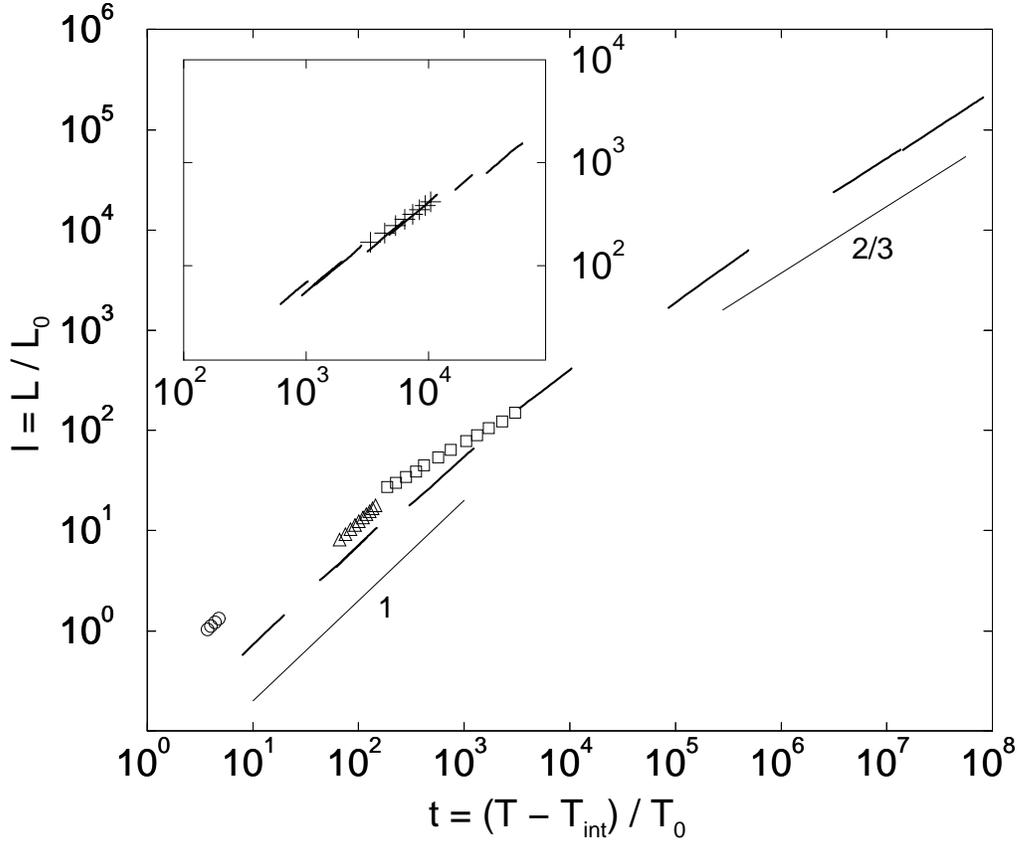}}}
    \end{center}
    \caption[Scaling plot for $256^3$ LB data compared with other published work]
	    {Scaling plot in reduced variables ($L/L_0$, $T/T_0$)
	     Bold lines (left to right) are LB $256^3$ data from
             table \ref{table:pars_256} (top to bottom).
             Also shown are results from other published work:
             squares \protect\cite{appert95a},
             triangles \protect\cite{laradji96a},
             circles \protect\cite{bastea97a},
             Inset: DPD data of \protect\cite{jury98a} (solid lines)
             with one LB
             data set ($L_0 = 0.15$, pluses) repeated for comparison.}
    \protect\label{graph:l_t_others}
\end{minipage}
\end{figure}

In order to compare our results with others' published work, a similar scaling
procedure must be applied to place their data onto the universal scaling
plot in figure \ref{graph:l_t_fin}.  This is only possible for work
reported in sufficient detail to enable values for $L_0$ and $T_0$
to be calculated.  Recent three-dimensional work where comparison is possible
includes that of \cite{bastea97a}, \cite{laradji96a},
\cite{appert95a}, and \cite{jury98a}.  These four sets of
results are shown along with the LB data in figure \ref{graph:l_t_others}.
Simulations of three-dimensional spinodal decomposition with hydrodynamics for which
quantitative comparisons were not possible include, \cite{koga91a},
\cite{puri92a}, and \cite{alexander93a}, all of whom claimed to have
simulated the linear regime; \cite{shinozaki91a}, \cite{ma92a},
\cite{lookman96a}.  The last two claimed to have simulated the inertial
regime but offered no evidence beyond their fitted exponent values
for definitely having inertial rather than diffusive effects.

The four studies for which detailed numerical comparisons are possible
will now be considered in turn.

\subsection{\protect\citet*{bastea97a}}
\label{sec:bastea97a}

\cite{bastea97a} carried out a three-dimensional simulation containing
about $1.4\times10^6$ particles whose motion was described mesoscopically by
Boltzmann-Vlasov equations.  They combined direct simulation 
Monte Carlo methods for the short range interaction with
particle-in-cell methods for long range interactions.  The fluid system is
relatively low-density; they describe it as a gas-gas phase separation.
The necessary fitting and scaling for
the results reported by \citeauthor{bastea97a} was done by \cite{jury98a}.

\begin{figure}
\begin{minipage}{\textwidth}
    \raggedright
    \begin{minipage}{0.49\textwidth}
        \resizebox{\textwidth}{!}{\rotatebox{-90}{\includegraphics{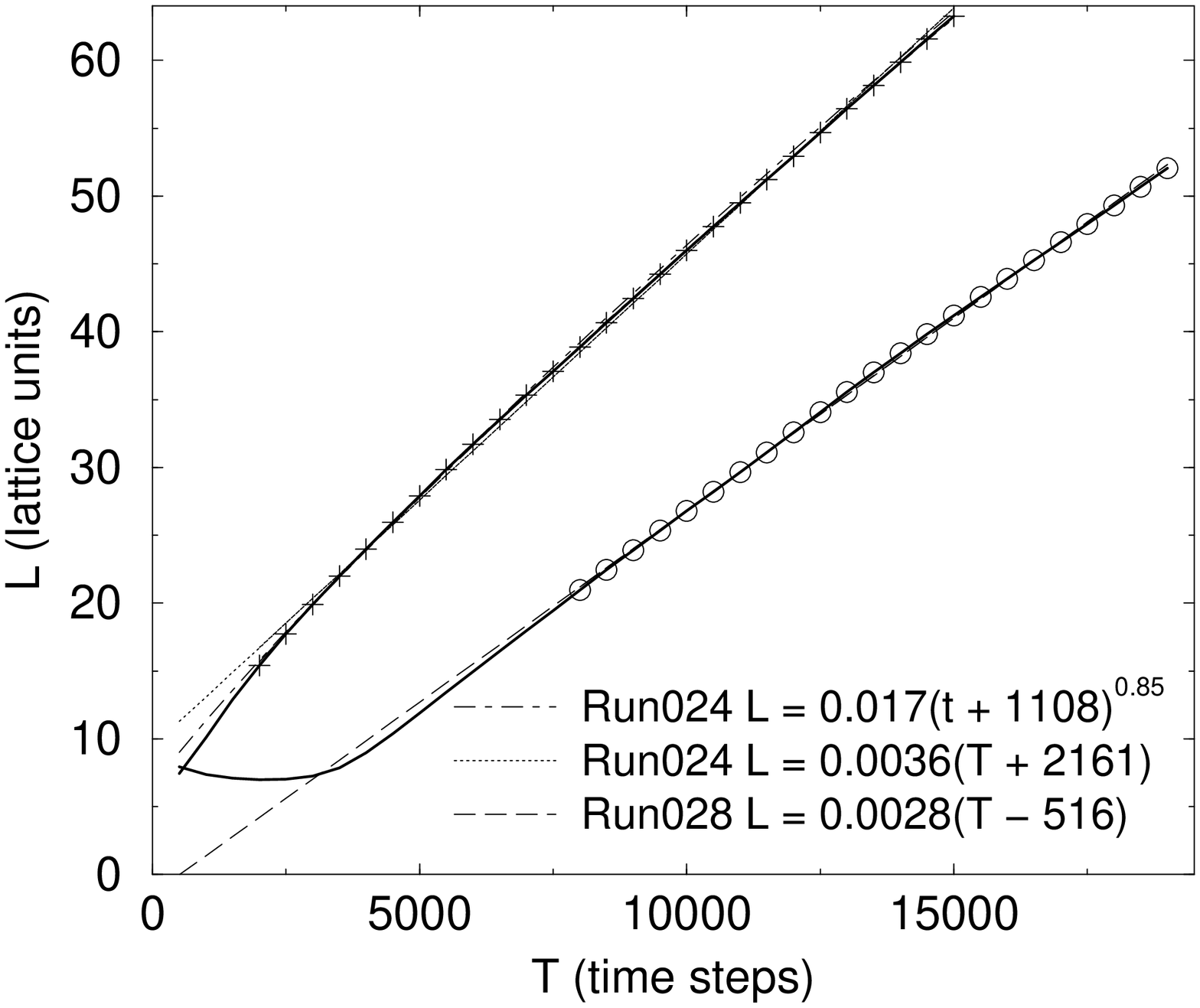}}}
    \end{minipage}
    \hfill
    \begin{minipage}{0.49\textwidth}
        \resizebox{\textwidth}{!}{\rotatebox{-90}{\includegraphics{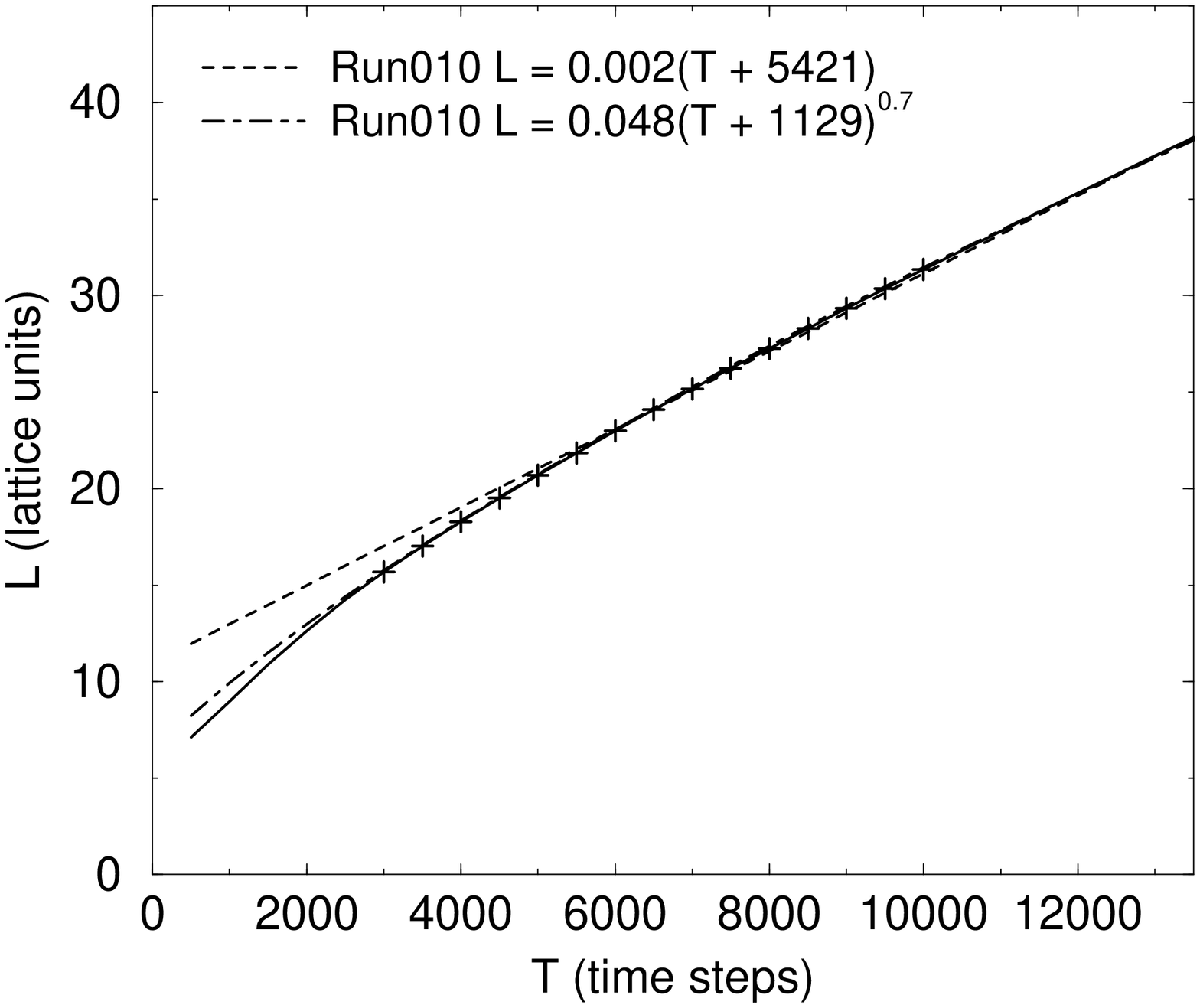}}}
    \end{minipage}
    \renewcommand{\baselinestretch}{1.0} \small\normalsize
    \caption[Excessive diffusion: Run024 and Run010]
	    {Left: $256^3$, $L_0 = 36$,
	     Run024 with $M=0.8$ compared with Run028 ($M=0.1$).
             Right: Run010, $128^3$, $L_0=381$, $M=0.5$.}
    \protect\label{graph:large_M}
\end{minipage}
\end{figure}

On the universal scaling plot, figure \ref{graph:l_t_others}, the
supposedly viscous-regime  portion of the data from \citeauthor{bastea97a}
is shown as circles.  It lies
in the correct (lower left) region of the graph, but well to the left of any of the LB results.
The scaled value of the fit parameter, $b_1=v/(L_0/T_0)$, from the
refitting by \citeauthor{jury98a} is $b_1 = 0.3$.
This compares with the considerably smaller values from the LB results in the linear region. The dynamical scaling hypothesis requires however
that $b_1$ should be universal to all systems in the viscous regime.
The high reported $b_1$ leads us to suspect residual diffusion.
\citeauthor{bastea97a} do not report the diffusion rate in their system
in a form that can be used to apply the analysis of \S\ \ref{sec:diffusion},
but test LB runs have been done with high diffusion rates that produce
LB data sets very similar to those of \citeauthor{bastea97a}.
Two such test runs, Run024 and Run010, are shown in figure \ref{graph:large_M}.

On the left, Run024 is
compared with Run028, one of the runs in table \ref{table:pars_256}.
Run024 has the same parameters as Run028 except for the mobility,
which is eight times higher.  Run024 fits a free
exponent of $\alpha=0.85$, which is similar to Run022 (also in the
linear region).  However, the lack of an initial flat diffusive region
(also absent in \cite{bastea97a}) strongly suggests the diffusion is too strong for uncontaminated linear growth to
be observed within the attainable system size.  A linear fit to
the upper part of the data produces a scaled value of $b_1 = 0.082$,
compared with 0.073 for Run028.

On the right, Run010 has a value of $L_0=381$, larger than for
any of the runs finally used by us, see table \ref{table:pars_128}.
This should put it even further into the linear region than the
rest of the runs.  However, a fit with
a free exponent produces $\alpha=0.7$.  A linear fit to the upper part
of the data produces a value for the scaled fit parameter
of $b_1 = 0.32$, i.e. about the same as the data from \citeauthor{bastea97a}, which this curve resembles.
It seems likely, therefore, that the data from \cite{bastea97a}
has strong residual diffusion,
and the results they present are a mixture of linear and diffusive growth.

\subsection{\protect\citet*{laradji96a}}
\label{sec:laradji96a}

\cite{laradji96a} used a large-scale molecular dynamics simulation
of a Lennard-Jones model with 343000 particles with a deep quench. 
The necessary fitting and scaling for
the data of \citeauthor{laradji96a} was done by \cite{jury98a}.
The results are shown on figure \ref{graph:l_t_others} as triangles lying,
like those of \cite{bastea97a}, to the left of
the LB data towards the lower left corner, again in the viscous regime
according to the LB analysis.
\citeauthor{laradji96a} claimed their results confirmed the linear
scaling, but their value of the scaled 
fit parameter, $b_1 = 0.13$, is again higher than ours. However, it is
in well within the range of values spanned by the the over-diffusive LB runs
shown in figure \ref{graph:large_M} ($b_1 = 0.082, 0.32$). 
The shape of their $L(T)$ curve
matches that of Run024 in figure \ref{graph:large_M} (left), i.e.
there is no initial diffusive plateau before domain growth begins.  
In LB data this was always a sign that the run had too high a mobility
and significant contamination by residual diffusion. 

\subsection{\protect\citet*{appert95a}}
\label{sec:appert95a}

\cite{appert95a} used a three-dimensional lattice gas simulation to simulate
spinodal decomposition.  
Their largest system size was $128^3$ and
they also used $64^3$ to test for finite size effects; they rejected
data with $L > \Lambda/2$, whereas we saw significant finite size effects by this stage (and applied the stricter condition, $L < \Lambda/4$).
They claimed a fitted exponent of $\alpha \simeq 2/3$, which,
if correct, would put their results in the inertial regime.
Taking the relevant interval of their data, refitting it by our method (giving an exponent $\alpha=0.62$) and converting into reduced physical units gives
the dataset (squares) on figure \ref{graph:l_t_others}. This
shows that, if our LB data is correct, the data of \citeauthor{appert95a} is actually in the crossover region. It appears to be
asymptoting onto the LB data from above, which again suggests significant residual diffusion, giving too low a value for the fitted exponent.
As with the data of \cite{bastea97a} and \cite{laradji96a}, the $L(T)$ plot
is lacking the initial diffusive plateau, an absence that we found was invariably associated with strong residual diffusion within LB.

\subsection{\protect\citet*{jury98a}}
\label{sec:jury98a}

\cite{jury98a} carried out a series of simulations of a symmetric, binary
fluid mixture using the DPD (dissipative particle dynamics) method with
$10^6$ particles and a deep quench. (The DPD algorithm combines soft interparticle repulsions with pairwise damping of
interparticle velocities and pairwise random forces.) In terms of the range of domain scales that can be probed, as a multiple of the interfacial width, these 
are roughly equivalent to our $128^3$ runs using LB. 
These authors found that each data set was well fitted by a linear scaling,
$L=v(T-T_{int})$,
but with a systematic increase of $b_1=v/(L_0/T_0)$ upon moving from
upper right to lower left in the universal scaling plot,
see figure \ref{graph:l_t_others} (inset).
These seven data sets have a range of $L_0$ values, $0.29\le L_0\le 0.013$,
which places them all firmly in the crossover region found in the
LB results, between Run029 ($L_0=0.95$) and Run030 ($L_0=0.01$).

\citeauthor{jury98a} suggested two alternative interpretations
of 
their own data and that of \cite{laradji96a} and \cite{bastea97a},
to explain the observed linear scaling within each run, but lack of 
consistency in the prefactor, $b_1$. The first was a possible nonuniversality
arising from the physics of pinchoff. The second was that all data sets formed  part of an extremely broad crossover region.  This explains the observed trend
$b_1 \sim t^{-0.2}$, but not the linear scaling within
each run. However, \citeauthor{jury98a} noted the possibility of `dangerous' finite size effects within their data, which was truncated at $L_{max} = \Lambda/2$. Our LB results support the latter explanation for the DPD data, with a separate or additional reason (residual diffusion) for rejecting the other datasets.
Unlike these, all the data sets of \citeauthor{jury98a}
do lie very close to the LB results, see figure \ref{graph:l_t_others} (inset).
Moreover, the DPD show a diffusive plateau prior to the onset of the apparent linear regime, consistent with low residual diffusion levels.

Since the DPD simulation method is very different from LB,
the correspondence of the two sets of results
lends broad support to the idea of a universal scaling,
although the fact that each DPD run is best fit by a locally linear growth law
is not consistent with this. Based on our LB results, we are inclined attribute the latter to finite size effects, which would certainly be large enough to spoil our nearest equivalent runs ($128^3$ with $L_{max} = 64$). 
However, we cannot rule out other explanations
as offered by \citeauthor{jury98a}. Larger DPD runs, and/or a series of LB runs, giving strongly overlapping datasets in a narrow window of the $l(t)$ like the DPD results (figure \ref{graph:l_t_others}, inset), might shed light on this.
\end{appendix}

\bibliography{../bibs/back,../bibs/dpd,../bibs/lb,../bibs/expt,../bibs/persist,../bibs/shear,../bibs/stats,../bibs/turb,../bibs/mike}

\end{document}